\documentclass[12pt]{article}
\usepackage{latexsym}
\usepackage{amsmath,amsfonts}
\usepackage{times}
\allowdisplaybreaks[4]

\catcode`\@=11

\newcount\hour
\newcount\minute
\newtoks\amorpm \hour=\time\divide\hour by 60\minute
=\time{\multiply\hour by 60 \global\advance\minute by-\hour}
\edef\standardtime{{\ifnum\hour<12 \global\amorpm={am}%
        \else\global\amorpm={pm}\advance\hour by-12 \fi
        \ifnum\hour=0 \hour=12 \fi
        \number\hour:\ifnum\minute<10
        0\fi\number\minute\the\amorpm}}
\edef\militarytime{\number\hour:\ifnum\minute<10 0\fi\number\minute}

\def\draftlabel#1{{\@bsphack\if@filesw {\let\thepage\relax
   \xdef\@gtempa{\write\@auxout{\string
      \newlabel{#1}{{\@currentlabel}{\thepage}}}}}\@gtempa
   \if@nobreak \ifvmode\nobreak\fi\fi\fi\@esphack}
        \gdef\@eqnlabel{#1}}
\def\@eqnlabel{}
\def\@vacuum{}
\def\marginnote#1{}
\def\draftmarginnote#1{\marginpar{\raggedright\scriptsize\tt#1}}
\def\draft{
        \pagestyle{plain}
        \overfullrule=2pt
        \oddsidemargin -.5truein
        \def\@oddhead{\sl \phantom{\today\quad\militarytime} \hfil
        \smash{\Large\sl DRAFT} \hfil \today\quad\militarytime}
        \let\@evenhead\@oddhead
        \let\label=\draftlabel
        \let\marginnote=\draftmarginnote
        \def\ps@empty{\let\@mkboth\@gobbletwo
        \def\@oddfoot{\hfil \smash{\Large\sl DRAFT} \hfil}
        \let\@evenfoot\@oddhead}
        \def\@eqnnum{(\theequation)\rlap{\kern\marginparsep\tt\@eqnlabel}%
        \global\let\@eqnlabel\@vacuum}  }

\newcommand{\rf}[1]{(\ref{#1})}
\renewcommand{\theequation}{\thesection.\arabic{equation}}
\renewcommand{\thefootnote}{\fnsymbol{footnote}}
\newcommand{\newsection}{   
\setcounter{equation}{0}\section}

\def\appendix#1{\addtocounter{section}{1}\setcounter{equation}{0}
\renewcommand{\thesection}{\Alph{section}}
\section*{Appendix \thesection\protect\indent \parbox[t]{11.15cm}{#1}}
\addcontentsline{toc}{section}{Appendix \thesection\ \ \ #1}}

\def\be{\begin{equation}}
\def\ee{\end{equation}}
\def\beq{\begin{eqnarray}}
\def\eeq{\end{eqnarray}}

\def\parline{\,\partial\kern -0.55em /\,\,}

\def\half{{\frac{1}{2}}}

\def\LL{{\cal L}}

\def\RR{{\cal R}}
\def\VV{{\cal V}}

\def\phik{|\phi\rangle}
\def\phibr{\langle\phi|}

\def\phikst#1{|\phi_{#1}\rangle}

\def\psikst#1{|\psi_{#1}\rangle}

\def\psik{|\psi\rangle}
\def\psibr{\langle\psi|}

\def\xik{|\xi\rangle}

\def\smG{{\scriptscriptstyle G}}

\def\smzero{{\scriptscriptstyle (0)}}
\def\smone{{\scriptscriptstyle (1)}}
\def\smtwo{{\scriptscriptstyle (2)}}

\def\smn{{\scriptscriptstyle (n)}}

\def\oplussm{{\scriptscriptstyle \oplus}}
\def\ominussm{{\scriptscriptstyle \ominus}}

\def\smE{{\scriptscriptstyle E}}
\def\smGE{{\scriptscriptstyle GE}}
\def\smGamma{{\scriptscriptstyle \Gamma}}
\def\smG{{\scriptscriptstyle G}}

\def\oplussm{{\scriptscriptstyle \oplus}}
\def\ominussm{{\scriptscriptstyle \ominus}}

\def\plushat{{\scriptstyle \hat{+}}}
\def\minushat{{\scriptstyle \hat{-}}}

\def\bwt{\widetilde{b}}

\def\ewt{\widetilde{e}}

\def\fwt{\widetilde{f}}

\def\rwt{\widetilde{r}}

\def\mwt{\widetilde{m}}

\def\ebwt{\widetilde{\bar{e}}}
\def\mbwt{\widetilde{\bar{m}}}
\def\rbwt{\widetilde{\bar{r}}}
\def\Yb{\bar{Y}}

\def\Gwh{\widehat{G}}

\def\gaal{\gamma\alpha}
\def\gaalb{\gamma\bar\alpha}
\def\alpar{\alpha\partial}
\def\albpar{\bar\alpha\partial}

\def\Cb{\bar{C}}

\def\eb{\bar{e}}
\def\mb{\bar{m}}
\def\rb{\bar{r}}

\def\nbf{{\bf n }}

\def\i{{\rm i}}
\def\sym{{\rm sym}}
\def\as{{\rm as}}
\def\f{{\rm f}}
\def\u{{\rm u}}
\def\d{{\rm d}}
\def\lin{{\rm lin}}

\def\EH{{\scriptscriptstyle\rm EH}}
\def\W{{\scriptscriptstyle\rm W}}
\def\Max{{\scriptscriptstyle\rm Max}}

\def\hcwh{\widehat{h.c.}}

\def\smminone{{\scriptscriptstyle -1}}
\def\smmintwo{{\scriptscriptstyle -2}}
\def\smminthree{{\scriptscriptstyle -3}}

\begin{document}


\begin{flushright}
FIAN-TD-2007-10 \hspace{1.7cm}{}~\\
arXiv: 0707.4437 [hep-th] \hspace{0.4cm}{}~\\
Modified, November 2011\hspace{0.5cm}{}~
\end{flushright}

\vspace{1cm}

\begin{center}

{\Large \bf Ordinary-derivative formulation

\bigskip of conformal low-spin fields}

\vspace{2.5cm}

R.R. Metsaev\footnote{ E-mail: metsaev@lpi.ru }

\vspace{1cm}

{\it Department of Theoretical Physics, P.N. Lebedev Physical
Institute, \\ Leninsky prospect 53,  Moscow 119991, Russia }

\vspace{3.5cm}

{\bf Abstract}

\end{center}

Conformal fields in flat space-time of even dimension greater than or equal
to four are studied. Second-derivative formulation for spin 0,\,1,\,2
conformal bosonic fields and first-derivative formulation for spin 1/2,\,3/2
conformal fermionic fields are developed. For the spin 1,\,3/2,\,2 conformal
fields, we obtain gauge invariant Lagrangians and the corresponding gauge
transformations. Gauge symmetries are realized by involving Stueckelberg
fields and auxiliary fields. Realization of global conformal boost symmetries
is obtained. Modified Lorentz and de Donder gauge conditions are introduced.
Ordinary-derivative Lagrangian of interacting Weyl gravity in $4d$ is
obtained. In our approach, the field content of Weyl gravity, in addition to
conformal graviton field, includes one auxiliary rank-2 symmetric tensor
field and one Stueckelberg vector field. With respect to the auxiliary tensor
field, the Lagrangian contains, in addition to other terms, the Pauli-Fierz
mass term. Using the ordinary-derivative Lagrangian of Weyl gravity, we
discuss interrelation of Einstein AdS gravity and Weyl gravity via breaking
conformal gauge symmetries. Also, we demonstrate use of the light-cone gauge
for counting on-shell degrees of freedom in higher-derivative conformal field
theories.

\newpage
\renewcommand{\thefootnote}{\arabic{footnote}}
\setcounter{footnote}{0}

\section{Introduction}

In view of the aesthetic features of conformal field theory  a interest in
this theory was periodically renewed (see Ref.\cite{Fradkin:1985am} and
references therein). Conjectured duality \cite{Maldacena:1997re} of large $N$
conformal ${\cal N}=4$ SYM theory and type IIB superstring theory in $AdS_5
\times S^5$ background has triggered intensive and in-depth study of various
aspects of conformal fields. Conformal fields in space-time of dimension
$d\geq 4$ can be separated into two groups: fundamental conformal fields and
shadow fields. This is to say that field having Lorentz algebra spin $s$ and
conformal dimension $\Delta = s+d-2$
is referred to as fundamental field,%
\footnote{We note that fundamental conformal fields with $s=1$, $\Delta= d-1$
and $s=2$, $\Delta = d$, correspond to conserved vector current and conserved
traceless rank-2 tensor field (energy-momentum tensor) respectively.
Conserved conformal currents can be built from massless scalar, spinor and
spin-1 fields (see e.g. Ref.\cite{Konstein:2000bi}).}
while field having Lorentz algebra spin $s$ and dual conformal dimension
$\hat\Delta = d - \Delta$ is sometimes referred to as shadow field. It is the
shadow fields that are used to discuss conformal equations of motion and
Lagrangian formulations (see e.g.
Refs.\cite{Fradkin:1985am},\cite{Erdmenger:1997gy},\cite{Segal:2002gd}%
\footnote{ Discussion of equations for mixed-symmetry conformal fields with
discrete $\Delta$ may be found in Ref.\cite{Shaynkman:2004vu}.}). In the
framework of AdS/CFT correspondence, the shadow fields manifest themselves in
two related ways at least. First, they appear as boundary values of
non-normalazible solution of equations of motion for bulk fields in AdS
background (see e.g. Refs.\cite{Balasubramanian:1998sn}-\cite{Metsaev:2005ws}%
\footnote{In earlier literature, discussion of shadow field dualities may be
found in Refs.\cite{Petkou:1994ad,Petkou:1996jc}.}).
It turns out that the action of IIB supergravity expanded over $AdS_5\times S^5$ background
and evaluated over Dirichlet problem reproduces
the action of $N=4$ conformal supergravity \cite{Liu:1998bu}.%
\footnote{Also, conformal symmetries manifest themselves in the tensionless
limit of strings \cite{Isberg:1992ia} (see also
Refs.\cite{Bonelli:2003kh,Bonelli:2003zu}).}
Second, shadow fields appearing in the conformal graviton supermultiplet of
$N=4$ superconformal algebra constitute the field content of $N=4$ conformal
supergravity and these shadow fields couple to operators constructed out of
the fields of ${\cal N}=4$ supersymmetric YM theory. Note also that $N=4$
conformal supergravity has the same global supersymmetries, though realized
in a different way, as supergravity/superstring theory in $AdS_5 \times S^5$
background. In view of these relations to IIB supergravity/superstring and
supersymmetric YM theories, we think that various alternative formulations of
shadow fields will be useful to understand string/gauge theory dualities
better.  In this paper, we deal with the shadow fields. These fields will be
referred to as conformal fields in what follows.

Lagrangian formulation of most conformal fields involves the higher
derivatives. The purpose of this paper is to develop ordinary-derivative,
gauge invariant, and Lagrangian formulation for free conformal fields. Our
approach can be summarized as follows.

\noindent {\bf i}) We introduce additional field degrees of freedom (D.o.F),
i.e., we extend the space of fields entering the standard higher-derivative
theories. Some of the additional fields turn out to be Stueckelberg fields,
while the remaining
additional fields turn out to be auxiliary fields.%
\footnote{ To develop ordinary-derivative approach one can use the
Ostrogradsky method. The conventional use of this method leads to appearance
only the auxiliary fields. Our method leads automatically to the appearance
of both auxiliary and Stueckelberg fields. The method involves two steps.
First, we use light-cone gauge to classify on-shell D.o.F of the standard
higher-derivative theories according to irreps of the $so(d-2)$ algebra.
Second, we replace irreps of the $so(d-2)$ algebra by the corresponding
representations of the Lorentz algebra $so(d-1,1)$ .}

\noindent {\bf ii}) Our Lagrangian for bosonic (fermionic) conformal field
does not contain higher than second (first) order terms in derivatives.
Two-derivative (one-derivative) contributions to the Lagrangian take the form
of the standard kinetic terms of bosonic (fermionic) fields.

\noindent {\bf iii}) All vector, tensor and vector-spinor fields are
supplemented by appropriate gauge symmetries.%
\footnote{ To realize those additional gauge symmetries we adopt the approach
of Refs.\cite{Zinoviev:2001dt,Metsaev:2006zy} which turns out to be the most
useful for our purposes.}
Gauge transformations of conformal fields do not involve higher than first
order terms in derivatives. One-derivative contributions to the gauge
transformations take the form of the standard gradient gauge transformations
of the vector, tensor, and vector-spinor fields.

\noindent {\bf iv}) The gauge symmetries of our Lagrangian make it possible
to match  our approach with  the higher-derivative one, i.e., by gauging away
the Stueckelberg fields and by solving constraints for the auxiliary fields
we obtain the standard higher-derivative formulation of conformal fields.
This implies that our approach retain propagating D.o.F of the
higher-derivative conformal field theory, i.e., our approach is equivalent to
the higher-derivative one, at least at the classical level.

Stueckelberg approach turned out to be successful for the study of theories
involving massive fields (see e.g. Ref.\cite{Siegel:1985tw}). Recall that all
Lorentz covariant formulations of string theories are realized by using
Stueckelberg fields. Therefore we expect that the use of Stueckelberg fields
might be useful for developing new interesting formulations of conformal
field theory.

This paper is devoted to ordinary-derivative formulation of free low-spin
conformal fields in space-time of even dimension $d \geq 4$. Note however
that, for the case of spin-2 field in $4d$, we also discuss the
ordinary-derivative formulation of interacting conformal field theory which
is related to Weyl conformal gravity in $4d$.

The rest of the paper  is organized as follows.

In Sec. \ref{sec-03}, we discuss conformal scalar field. In Sec.
\ref{sec-03-sub01} we start with brief review of the higher-derivative
formulation of conformal scalar field, while, in Sec. \ref{sec-03-sub02}, we
review the well-known ordinary-derivative formulation of conformal scalar
field. We use conformal oscillators to demonstrate how those oscillators
allow us to simplify the presentation of the well-known ordinary-derivative
Lagrangian for the conformal scalar field. We review realization of global
conformal algebra symmetries on space of fields entering the
ordinary-derivative formulation.

Sec. \ref{sec-04} is devoted to the discussion of spin-1 conformal field. In
Sec. \ref{sec-04-sub01} we start with brief review of the higher-derivative
formulation of spin-1 conformal field. Using light-cone gauge, we describe
on-shell D.o.F appearing in the standard higher-derivative approach to spin-1
conformal field. After this, in Sec. \ref{sec-04-sub02}, we describe our
ordinary-derivative Lorentz covariant and gauge invariant formulation of the
spin-1 conformal field. Also we discuss modified Lorentz gauge.

Sec. \ref{sec-05} is devoted to the discussion of spin-2 conformal field. In
this Section, we generalize results of Sec. \ref{sec-04} to the case of
spin-2 conformal field. Also, we describe modified de Donder gauge for the spin-2
conformal field.

In Sec. \ref{sec-06}, we describe the ordinary-derivative formulation of
interacting theory of $4d$ conformal gravity, i.e., we provide the
ordinary-derivative formulation of $4d$ Weyl gravity. We discuss
ordinary-derivative gauge invariant Lagrangian and its gauge symmetries. Also
we discuss interrelation of Einstein AdS gravity and Weyl gravity via
breaking conformal gauge symmetries.

In Sec. \ref{sec-07}, we discuss spin-$\half$ conformal field. In Sec.
\ref{sec-07-sub01} we review the higher-derivative formulation of
spin-$\half$ conformal field.  After this, in Sec. \ref{sec-07-sub02}, we
describe the ordinary-derivative formulation of spin-$\half$ conformal field.

Sec. \ref{sec-08} is devoted to spin-$\frac{3}{2}$ conformal field. In Sec.
\ref{sec-08-sub01}, we briefly review the higher-derivative formulation of
spin-$\frac{3}{2}$ conformal field. Also we present two new representations
for higher-derivative Lagrangian which to our knowledge have not been
discussed in literature. Using the light-cone gauge, we describe on-shell
D.o.F of the standard higher-derivative Lagrangian. After this, in Sec.
\ref{sec-08-sub02}, we develop our ordinary-derivative formulation of
spin-$\frac{3}{2}$ conformal field.

Section \ref{conl-sec-01} suggests directions for future research.

Technical details are collected in Appendices. In Appendices A,C,E,
we derive on-shell D.o.F. of the standard higher-derivative theories for the
respective spin-1, spin-2, and spin-$\frac{3}{2}$ conformal fields. In
Appendices B,D,F we present details of the derivation of ordinary-derivative gauge
invariant Lagrangian and gauge transformations for the respective spin-1,
spin-2, and  spin-$\frac{3}{2}$ conformal fields.

\section{Preliminaries}

\subsection{Notation}

Our conventions are as follows. $x^a$ denotes coordinates in $d$-dimensional
flat space-time, while $\partial_a$ denotes derivatives with respect to
$x^a$, $\partial_a \equiv \partial / \partial x^a$. Vector indices of the
Lorentz algebra $so(d-1,1)$ take the values $a,b,c,e=0,1,\ldots ,d-1$.  We
use the mostly positive flat metric tensor $\eta^{ab}$. To simplify our
expressions we drop $\eta_{ab}$ in scalar products, i.e., we use $X^aY^a
\equiv \eta_{ab}X^a Y^b$.

We use a set of the creation operators $\alpha^a$, $\zeta$,
$\upsilon^\oplussm$, $\upsilon^\ominussm$, and the respective set of
annihilation operators $\bar{\alpha}^a$, $\bar{\zeta}$,
$\bar\upsilon^\ominussm$, $\bar\upsilon^\oplussm$. These operators, to be
referred to as oscillators in what follows, satisfy the commutation relations%
\footnote{ We use oscillator formulation
\cite{Lopatin:1987hz,Vasiliev:1987tk,Labastida:1987kw} to handle the many
indices appearing for tensor fields. It can also be reformulated as an
algebra acting on the symmetric-spinor bundle on the manifold $M$
\cite{Hallowell:2005np}. Note that the scalar oscillators $\zeta$,
$\bar\zeta$, which appeared in gauge invariant formulation of massive fields,
arise naturally by a dimensional reduction
\cite{Biswas:2002nk,Hallowell:2005np} from flat space. It is natural to
expect that `conformal' oscillators $\upsilon^\oplussm$,
$\upsilon^\ominussm$, $\bar\upsilon^\oplussm$, $\bar\upsilon^\ominussm$ also
allow certain interpretation via dimensional reduction.
}
\beq
\label{manold-31102011-01}
&&{} [\bar{\alpha}^a,\alpha^b]=\eta^{ab}\,, \qquad [\bar\zeta,\zeta]=1\,,
\qquad [\bar{\upsilon}^\oplussm,\, \upsilon^\ominussm ]=1\,, \qquad\quad
[\bar{\upsilon}^\ominussm,\, \upsilon^\oplussm]=1\,,
\\
&& \bar\alpha^a |0\rangle = 0\,,\qquad\quad  \bar\zeta|0\rangle =
0\,,\qquad\quad \bar\upsilon^\oplussm |0\rangle = 0\,,\qquad\quad \quad
\bar\upsilon^\ominussm |0\rangle = 0\,.
\eeq
The oscillators $\alpha^a$, $\bar\alpha^a$ and $\zeta$, $\bar\zeta$,
$\upsilon^\oplussm$, $\upsilon^\ominussm$, $\bar\upsilon^\oplussm$,
$\bar\upsilon^\ominussm$ transform in the respective vector and scalar
representations of the Lorentz algebra $so(d-1,1)$. We use $2^{[d/2]}\times
2^{[d/2]}$ Dirac gamma matrices $\gamma^a$ in $d$-dimensions, $ \{
\gamma^a,\gamma^b\} = 2\eta^{ab}$, and adapt the following hermitian
conjugation rules for the derivatives, oscillators, and $\gamma$-matrices:
\be \label{03082011-01} \partial^{a\dagger} = - \partial^a, \qquad
\gamma^{a\dagger} = \gamma^0 \gamma^a\gamma^0\,,\qquad
\alpha^{a\dagger} = \bar\alpha^a\,, \qquad \zeta^\dagger = \bar\zeta \,,
\qquad
\upsilon^{\oplussm\dagger} = \bar\upsilon^\oplussm\,,\qquad
\upsilon^{\ominussm \dagger} = \bar\upsilon^\ominus \,.
\ee
We use operators constructed out of the derivatives, oscillators, and
$\gamma$-matrices,
\beq \label{manold-31102011-02}
&& \Box \equiv \partial^a\partial^a\,, \qquad\quad
\parline\equiv \gamma^a\partial^a\,,\qquad\quad
\alpha\partial \equiv \alpha^a\partial^a\,,\qquad\quad \bar\alpha\partial
\equiv \bar\alpha^a\partial^a\,,\qquad
\\
\label{manold-31102011-03} && \gamma\alpha \equiv \gamma^a\alpha^a\,,\qquad\
\ \gamma\bar\alpha \equiv \gamma^a\bar\alpha^a\,,\qquad \ \ \
\alpha^2 \equiv \alpha^a\alpha^a\,, \qquad\quad   \bar\alpha^2 \equiv
\bar\alpha^a\bar\alpha^a\,,\qquad
\\
\label{manold-31102011-05} && N_\alpha \equiv \alpha^a \bar\alpha^a \,,
\qquad \
N_\zeta \equiv \zeta \bar\zeta \,, \qquad\quad \ \ \
N_{\upsilon^\oplussm} \equiv \upsilon^\oplussm \bar\upsilon^\ominussm\,,
\qquad \
N_{\upsilon^\ominussm} \equiv \upsilon^\ominussm \bar\upsilon^\oplussm\,,
\\
\label{manold-31102011-06} && N_\upsilon \equiv N_{\upsilon^\oplussm} +
N_{\upsilon^\ominussm} \,, \qquad\qquad\qquad \quad \ \
\Delta' \equiv N_{\upsilon^\oplussm} - N_{\upsilon^\ominussm} \,.
\eeq
The $2\times 2$ matrices and antisymmetric products of $\gamma$-matrices are
defined as
\beq \label{manold-03112011-01}
&& \sigma_+  = \left(
\begin{array}{ll}
0 & 1
\\
0 & 0
\end{array}\right),
\quad
\sigma_-  = \left(
\begin{array}{ll}
0 & 0
\\
1 & 0
\end{array}\right),
\quad
\pi_+  = \left(
\begin{array}{ll}
1 & 0
\\
0 & 0
\end{array}\right),
\quad
\pi_-  = \left(
\begin{array}{ll}
0 & 0
\\
0 & 1
\end{array}\right)\,,
\\
\label{manold-31102011-04} && \hspace{1cm} \gamma^{ab} = \half (\gamma^a \gamma^b -
\gamma^b \gamma^a)\,,\qquad \gamma^{abc} =
\frac{1}{3!}(\gamma^a\gamma^b\gamma^c \pm 5 \hbox{ terms})\,.
\eeq
Throughout the paper the notation  $k' \in [n]_1$ implies that
$k' = -n,-n+1,-n+2,\ldots,n-2, n-1,n$, while the notation $k' \in [n]_2$
implies that $k' =-n,-n+2,-n+4,\ldots,n-4, n-2,n$:
\beq
\label{sumnot01} && k' \in [n]_1 \quad \Longrightarrow \quad k'
=-n,-n+1,-n+2,\ldots,n-2, n-1,n\,,
\\
\label{sumnot02} && k' \in [n]_2 \quad \Longrightarrow \quad k'
=-n,-n+2,-n+4,\ldots,n-4, n-2,n\,.
\eeq

\subsection{Global conformal symmetries }

The conformal algebra $so(d,2)$ of $d$ dimensional space-time taken to be in
basis of the Lorentz algebra $so(d-1,1)$ consists of translation generators
$P^a$, conformal boost generators $K^a$, and generators $J^{ab}$ which span
the Lorentz algebra $so(d-1,1)$. We assume the following normalization for
commutators of the conformal algebra:
\beq
\label{ppkk}
&& {}[D,P^a]=-P^a\,, \hspace{2cm}  {}[P^a,J^{bc}]=\eta^{ab}P^c -\eta^{ac}P^b
\,,
\\
&& [D,K^a]=K^a\,, \hspace{2.2cm} [K^a,J^{bc}]=\eta^{ab}K^c - \eta^{ac}K^b\,,
\\
\label{pkjj} && \hspace{2.5cm} {}[P^a,K^b]=\eta^{ab}D-J^{ab}\,,
\\
&& \hspace{2.5cm} [J^{ab},J^{ce}]=\eta^{bc}J^{ae}+3\hbox{ terms} \,.
\eeq

Let $\phik$ denotes free field propagating in flat space-time of dimension
$d\geq 4$. Let Lagrangian for the field $\phik$ be conformal invariant. This
implies that Lagrangian is invariant (up to total derivative) under the
transformations
\be \label{20072011-03} \delta_{\hat{G}} \phik  = \hat{G} \phik \,, \ee
where the realization of the conformal algebra generators $\hat{G}$ on space
of $\phik$ takes the form
\beq
\label{conalggenlis01} && P^a = \partial^a \,,
\\
\label{JIJdef}
\label{conalggenlis02} && J^{ab} = x^a\partial^b -  x^b\partial^a + M^{ab}\,,
\\
\label{conalggenlis03} && D = x^a\partial^a  + \Delta\,,
\\
\label{conalggenlis04}  && K^a = K_{\Delta,M}^a + R^a\,,
\\
&& \hspace{1cm} K_{\Delta,M}^a \equiv -\frac{1}{2}x^bx^b \partial^a + x^a D +
M^{ab}x^b\,.
\eeq
In \rf{conalggenlis02}-\rf{conalggenlis04}, $\Delta$ is operator of conformal
dimension, $M^{ab}$ is spin operator of the Lorentz algebra,
\be  [M^{ab},M^{ce}]=\eta^{bc}M^{ae}+3\hbox{ terms} \,, \ee
and $R^a$ is operator depending on the derivatives with respect to space-time
coordinates and not depending on the space-time coordinates $x^a$, $[P^a,R^b]=0$.%
\footnote{For the case of conformal currents and shadow fields studied in
Refs.\cite{Metsaev:2008fs,Metsaev:2009ym,Metsaev:2010zu}, the operator $R^a$
does not depend on the derivatives.}
The spin operator of the Lorentz algebra is well known for arbitrary spin
conformal field. In standard higher-derivative formulations of conformal
fields, the operator $R^a$ is often equal to zero, while in
ordinary-derivative approach, we develop in this paper, the operator $R^a$ is
non-trivial. This implies that, in the ordinary-derivative approach, complete
description of the conformal fields requires finding not only gauge invariant
Lagrangian but also the operator $R^a$. It turns out that requiring
Lagrangian to be invariant under gauge symmetries and global conformal
algebra symmetries allows us to fix both the Lagrangian and the operator
$R^a$ uniquely.

\newsection{ Conformal scalar field}\label{sec-03}

As a warm up let us consider spin-0 field (scalar field). To make contact
with studies in earlier literature we start with the presentation of the
standard higher-derivative formulation for the scalar field.

\subsection{ Higher-derivative formulation of conformal scalar field}\label{sec-03-sub01}

In the framework of the standard approach, conformal scalar field $\phi$
propagating in flat space of arbitrary dimension $d$ is described by the
Lagrangian
\be\label{standlag01} \LL = \half \phi \Box^{1+k}\phi\,, \ee
where $k$ is arbitrary positive integer independent of $d$. For $k=0$,
Lagrangian \rf{standlag01} describes field associated with unitary
representation of the conformal algebra $so(d,2)$, while, for $k\geq 1$,
Lagrangian \rf{standlag01} describes field related to non-unitary
representation of the conformal algebra.%
\footnote{ By now, unitary representations of (super)conformal algebras that
are relevant for elementary particles are well understood (for discussion of
conformal algebras see e.g. Refs.\cite{Evans}-\cite{Metsaev:1995jp} and
superconformal algebras Refs.\cite{Dobrev:1985qv,Gunaydin:1998sw}). In
contrast to this, non-unitary representations deserve to be understood
better.}
The field $\phi$ has the conformal dimension
\be \label{condim0001} \Delta_\phi = \frac{d-2}{2} - k\,.\ee

\subsection{ Ordinary-derivative formulation of conformal scalar field}\label{sec-03-sub02}

In the framework of ordinary-derivative approach, a dynamical system
that on-shell equivalent to the conformal scalar field $\phi$ with
Lagrangian \rf{standlag01} involves $k+1$ scalar fields
\be \label{scafiecol01}
\phi_{k'}\,,\qquad k' \in [k]_2\,,
\qquad
k - \hbox{ arbitrary positive integer}\,,
\ee
where we use the notation as in \rf{sumnot02}.
Conformal dimensions of the scalar fields $\phi_{k'}$ are given by
\be \label{delscalcomdef01} \Delta_{\phi_{k'}} = \frac{d-2}{2} + k'\,.\ee

To handle the many scalar fields \rf{scafiecol01} we use oscillators
$\upsilon^\oplussm$, $\upsilon^\ominussm$ \rf{manold-31102011-01} and
introduce the ket-vector defined by
\be
\label{phiscadef01}
\phik \equiv \sum_{k'\in [k]_2} \frac{1}{ (\frac{k+k'}{2})!}
(\upsilon^\oplussm)^{^{\frac{k+k'}{2}}}
(\upsilon^\ominussm)^{^{\frac{k-k'}{2}}} \, \phi_{k'} |0\rangle\,,
\ee
where a bra-vector $\phibr$ is defined according the rule $\phibr =
(\phik)^\dagger $. We note that ket-vector $\phik$ \rf{phiscadef01} is
degree-$k$ homogeneous polynomial in the oscillators $\upsilon^\oplussm$,
$\upsilon^\ominussm$,
\beq
&& (N_\upsilon - k ) \phik = 0 \,.
\eeq
Ordinary-derivative Lagrangian can entirely be expressed in terms of
the ket-vector $\phik$. This is to say that Lagrangian we found takes
the form%
\footnote{ We introduce mass-like term $m^2$ to show formal
similarity between our Lagrangian and the one for massive field.}
\be
\label{scafielm2def01x}    \LL = \half \phibr E \phik\,,
\qquad
E  \equiv \Box - m^2 \,, \qquad \quad m^2 \equiv \upsilon^\ominussm
\bar\upsilon^\ominussm \,.
\ee

{\bf Component form of Lagrangian}. For the reader convenience, we present
component form of the Lagrangian. The component form of Lagrangian
\rf{scafielm2def01x} can easily be obtained by plugging ket-vector
\rf{phiscadef01} into \rf{scafielm2def01x}. Doing so, we obtain
\be \label{13022011-oldman01}
\LL =  \sum_{k' \in [k]_2} \LL_{k'}\,,
\qquad\quad
\LL_{k'} = \half \phi_{-k'} \Box \phi_{k'} - \half \phi_{-k'}\phi_{k'+2}\,.
\ee
To illustrate \rf{13022011-oldman01}, we note that, for $k=1,2,3$, Lagrangian
\rf{13022011-oldman01} takes the form
\beq
\LL|_{k=1} & = & \phi_\smminone\Box \phi_{_1} -\half \phi_{_1}\phi_{_1} \,,
\nonumber\\
\LL|_{k=2} & =  &\phi_\smmintwo\Box \phi_{_2} + \half \phi_{_0}\Box \phi_{_0}
- \phi_{_0}\phi_{_2} \,,
\\
\LL|_{k=3} & = & \phi_\smminthree\Box \phi_{_3} +  \phi_\smminone\Box
\phi_{_1} - \phi_\smminone\phi_{_3} - \half \phi_{_1}\phi_{_1} \,.
\nonumber
\eeq

We now make comment on the interrelation of ordinary-derivative Lagrangian
\rf{13022011-oldman01} and higher-derivative Lagrangian \rf{standlag01}. To
this end we note that the scalar field $\phi_{-k}$ has the same conformal
dimension as the scalar field $\phi$ entering the higher-derivative approach
\rf{condim0001}. Therefore, we can use the identification $\phi_{-k} = \phi$.
The remaining scalar fields $\phi_{k'}$, $k'=-k+2,-k+4,\ldots k-2,k$, are
auxiliary fields. Using equations of motion for auxiliary fields obtained from
Lagrangian \rf{13022011-oldman01}, $\Box\phi_{k'-2}-\phi_{k'}=0$, we can
express all auxiliary fields in terms of $\phi_{-k}\equiv \phi$,
\be \label{manold-02112011-01} \phi_{k'} = \Box^{^{\frac{k+k'}{2}}}
\phi\,,\qquad k'=-k+2,-k+4,\ldots, k-2,k\,. \ee
Plugging \rf{manold-02112011-01} into \rf{13022011-oldman01}, we obtain
higher-derivative Lagrangian \rf{standlag01}, i.e., our approach and
higher-derivative approach are equivalent.

{\bf Conformal symmetries}. To complete ordinary-derivative description of
the scalar field we provide realization of the conformal algebra symmetries
on space of the ket-vector $\phik$. All that is required is to find
realization of the operators $M^{ab}$, $\Delta$ and $R^a$
\rf{conalggenlis01}-\rf{conalggenlis04} on space of $\phik$. For the case of
scalar field, the spin operator of the Lorentz algebra is trivial,
$M^{ab}=0$, while the realization of conformal dimension operator $\Delta$
and operator $R^a$ on space of $\phik$ \rf{phiscadef01} is given by
\beq
\label{manold-02112011-02} && \Delta  =  \frac{d-2}{2}+\Delta'\,,
\qquad \quad  \Delta' \equiv  N_{\upsilon^\oplussm} - N_{\upsilon^\ominussm} \,,
\nonumber\\[-12pt]
&&
\\[-12pt]
&& R^a =  -2\upsilon^\oplussm \bar\upsilon^\oplussm
\partial^a\,,
\nonumber
\eeq
where realization of $\Delta$ on space of $\phik$ \rf{manold-02112011-02} can
be read from \rf{delscalcomdef01}. Plugging $R^a$ \rf{manold-02112011-02}
into \rf{conalggenlis04}, we check that the commutator $[K^a,K^b]=0$ is
satisfied. In terms of component fields \rf{scafiecol01}, $R^a$
transformations take the form
\be
\delta_{R^a} \phi_{k'} =  - \half (k+k')(k+2-k')\partial^a \phi_{k'-2}\,.
\ee

Finally, we note that the  Lagrangian and operator $R^a$ are fixed
uniquely by requiring that%
\footnote{ Various alternative discussions of higher-derivative theories may
be found in Refs.\cite{Buchbinder:1987vp,de Urries:2001ns,Villasenor:2002kw}.}\\
i) Lagrangian does not contain higher than second order terms in
derivatives;\\
ii) operator $R^a$ does not contain higher than first order terms
in derivatives;\\
iii) Lagrangian is invariant under conformal algebra
symmetries $so(d,2)$.

\newsection{ Spin-1 conformal  field} \label{sec-04}

We proceed with discussion of conformal field theory for spin-1 field which
is the simplest example allowing us to demonstrate how gauge symmetries are
realized in the ordinary-derivative approach. We start with the presentation
of the standard higher-derivative approach to the spin-1 field.

\subsection{ Higher-derivative formulation of spin-1 conformal field}\label{sec-04-sub01}

In the framework of the standard approach, spin-1 conformal field $\phi^a$
propagating in flat space of arbitrary dimension $d\geq 4$ is described by
the Lagrangian
\be \label{hihgderLag01}
\LL = -\frac{1}{4}F^{ab}\Box^k F^{ab}\,, \qquad  F^{ab} \equiv
\partial^a \phi^b - \partial^b \phi^a\,,\qquad k \equiv \frac{d-4}{2} \,. \ee
We see that, for the case of spin-1 field, the integer $k$ turns out to be
fixed by dimension of space-time.  For $k=0$ (i.e., $d=4$), Lagrangian
\rf{hihgderLag01} describes Maxwell vector field associated with unitary
representation of the conformal algebra $so(4,2)$. For $k\geq 1$ (i.e., $d
\geq 6$), Lagrangian \rf{hihgderLag01} describes the spin-1 field related to
non-unitary representation of the conformal algebra $so(d,2)$. The field
$\phi^a$ has conformal dimension independent of space-time dimension,
$\Delta_{\phi^a} = 1$.

Let us now discuss on-shell D.o.F of the conformal theory under
consideration. For this purpose it is convenient to use fields transforming
in irreps of the $so(d-2)$ algebra. Namely, we decompose on-shell D.o.F into
irreps of the $so(d-2)$ algebra. One can prove (for details, see Appendix A)
that on-shell D.o.F are described by $k+1$ vector fields $\phi_{k'}^i$ and
$k$ scalar fields $\phi_{k'}$:
\beq \label{DOFspi01ad01}
&&\phi_{k'}^i \qquad \ \ k'\in [k]_2\,;
\nonumber\\[-12pt]
&&
\\[-12pt]
&& \phi_{k'}, \qquad k' \in [k-1]_2\,,
\nonumber
\eeq
where vector indices of the $so(d-2)$ algebra take values $i=1,2,\dots,d-2$
and we use the notation as in \rf{sumnot02}.
The fields $\phi_{k'}^i$ and $\phi_{k'}$ transform in the respective vector
and scalar representations of the $so(d-2)$ algebra. We note that the scalar
on-shell D.o.F \rf{DOFspi01ad01} appear only in $d\geq 6$ (i.e., $k\geq 1$).
Total number of on-shell D.o.F given in \rf{DOFspi01ad01}
is equal to
\be \label{DOFspi01} \nbf = \frac{1}{2}d(d-3)\,. \ee
Namely, we note that $\nbf$ \rf{DOFspi01} is a sum of the respective on-shell
D.o.F for the vector fields $\nbf(\phi_{k'}^i)$ and on-shell D.o.F for the
scalar fields $\nbf(\phi_{k'})$,
\beq
&& \nbf = \sum_{k'\in [k]_2} \nbf (\phi_{k'}^i) + \sum_{k'\in [k-1]_2}
\nbf(\phi_{k'}) \,,
\nonumber\\
&& \nbf(\phi_{k'}^i) = d-2, \qquad  \nbf(\phi_{k'}) = 1\,.
\eeq

\subsection{ Ordinary-derivative formulation
of spin-1 conformal field}\label{sec-04-sub02}

{\bf Field content}. To discuss ordinary-derivative gauge invariant
formulation of spin-1 conformal field in flat space of dimension $d\geq 4$ we
use $k+1$ vector fields $\phi_{k'}^a$ and $k$ scalar fields $\phi_{k'}$:
\newpage
\beq \label{covDOFspi01ad01}
&&\phi_{k'}^a\,, \qquad \ \ \ k' \in [k]_2\,;
\nonumber\\[-16pt]
&& \hspace{6cm}  k \equiv \frac{d-4}{2}\,,
\\[-16pt]
&& \phi_{k'}, \qquad\quad k'\in [k-1]_2\,;
\nonumber
\eeq
where we use the notation as in \rf{sumnot02}.
The fields $\phi_{k'}^a$ and $\phi_{k'}$ transform in the respective vector
and scalar irreps of the Lorentz algebra $so(d-1,1)$ and have the conformal dimensions
\be \label{delspi1def01} \Delta_{\phi_{k'}^a} = \frac{d-2}{2} + k'\,,\qquad
\qquad \Delta_{\phi_{k'}} = \frac{d-2}{2} + k'\,.\ee
Note that scalar fields $\phi_{k'}$ \rf{covDOFspi01ad01} appear only when $ d
\geq 6$ (i.e., $k\geq 1$).

Comparison of light-cone gauge fields \rf{DOFspi01ad01} and Lorentz fields
\rf{covDOFspi01ad01} demonstrates the general rule we use to obtain the field
content of gauge invariant ordinary-derivative formulation. Namely, all that
is required is to replace on-shell light-cone gauge fields of the $so(d-2)$
algebra by the
respective fields of the Lorentz algebra $so(d-1,1)$.%
\footnote{ Such a rule can be used when there is one-to-one mapping between
spin labels of the $so(d-2)$ algebra and those of the Lorentz algebra
$so(d-1,1)$.}

In order to obtain gauge invariant description in an easy--to--use form we
use oscillators $\alpha^a$, $\zeta$, $\upsilon^\oplussm$,
$\upsilon^\ominussm$ \rf{manold-31102011-01}. Namely, we collect fields
\rf{covDOFspi01ad01} into the ket-vector $\phik$ defined by
\beq
&&  \label{phispin1def01} \phik =  |\phi_1 \rangle  + \zeta |\phi_0\rangle
\,,
\\
\label{phispin1def02} && |\phi_1\rangle \equiv \sum_{k'\in [k]_2}  \frac{1}{
(\frac{k+k'}{2})!} \alpha^a
(\upsilon^\oplussm)^{^{\frac{k+k'}{2}}}(\upsilon^\ominussm)^{^{\frac{k-k'}{2}}}
 \, \phi_{k'}^a  |0\rangle\,,
\\
\label{phispin1def03} && |\phi_0\rangle \equiv \sum_{k'\in [k-1]_2} \frac{1}{
(\frac{k-1+k'}{2})!} (\upsilon^\oplussm)^{^{\frac{k-1+k'}{2}}}
(\upsilon^\ominussm)^{^{\frac{k-1-k'}{2}}}   \, \phi_{k'}^{\vphantom{5pt}}
|0\rangle\,.
\eeq
From \rf{phispin1def01}-\rf{phispin1def03}, we see that the ket-vectors
$\phik$, $|\phi_1\rangle$, and $|\phi_0\rangle$ satisfy the constraints
\beq \label{oldman12012011-01}
&& (N_\alpha + N_\zeta)\phik =  \phik\,,\qquad\quad
(N_\zeta + N_\upsilon)\phik = k \phik\,,
\\
&& \label{oldman12012011-02} N_\upsilon\phikst{1} = k
\phikst{1}\,,\hspace{2cm} N_\upsilon\phikst{0} = (k-1) \phikst{0}\,.  \eeq
Constraints \rf{oldman12012011-01} tell us that the ket-vector $\phik$ is
degree-1 homogeneous polynomial in the oscillators $\alpha^a$, $\zeta$,  and
degree-$k$ homogeneous polynomial in the oscillators $\zeta$,
$\upsilon^\oplussm$, $\upsilon^\ominussm$. Constraints \rf{oldman12012011-02}
tell us that the ket-vectors $\phikst{1}$ and $\phikst{0}$ are the respective
degree-$k$ and $k-1$ homogeneous polynomials in the oscillators
$\upsilon^\oplussm$, $\upsilon^\ominussm$.

Having described the field content, we are ready to discuss Lagrangian in the
framework of ordinary-derivative approach. We would like to discuss two
representations for the Lagrangian. We discuss these representations in turn.

{\bf 1st representation for Lagrangian}. Lagrangian we found takes the form
\beq
\label{Lagspi1gen01} \LL & = & \frac{1}{2} \phibr E \phik\,,
\\
&& E  \equiv  E_\smtwo + E_\smone + E_\smzero \,,
\\
\label{Mawsecordope01} && E_\smtwo = \Box - \alpar\albpar\,,
\\
&& E_\smone =  \eb_1\alpar + e_1\albpar\,,
\\
&& E_\smzero = m_1\,,
\\
\label{man19122010-01} && e_1 = \zeta  \bar\upsilon^\ominussm\,,
\qquad
\eb_1 = -\upsilon^\ominussm   \bar\zeta\,,
\qquad
m_1 = \upsilon^\ominussm  \bar\upsilon^\ominussm (N_\zeta-1)\,,
\eeq
where the notation we use can be found in
\rf{manold-31102011-02}-\rf{manold-31102011-06}. We note that operator
$E_\smtwo$ appearing in \rf{Mawsecordope01} is the standard second-order Maxwell
operator rewritten in terms of the oscillators.

{\bf 2nd representation for Lagrangian}. For the reader convenience, we discuss this
representation because it allows us to introduce Lorentz like gauge condition
for the spin-1 conformal field. This is to say that Lagrangian
\rf{Lagspi1gen01} can be represented as
\beq
\LL & = &  \half \phibr (\Box - m^2 )\phik + \half
\langle \Cb\phi|\Cb \phi\rangle \,,
\nonumber\\
\label{olfman19012011-01} && \Cb \equiv  \bar\alpha\partial  - \eb_1 \,,
\qquad \quad
m^2 \equiv \upsilon^\ominussm \bar\upsilon^\ominussm \,,
\eeq
where $\eb_1$ is defined in \rf{man19122010-01}. We now note that it is
operator $\Cb$ \rf{olfman19012011-01} that defines Lorentz like gauge
condition for the spin-1 conformal field,
\be \label{oldman-01112011-02}    \Cb\phik = 0\,. \ee
Interesting feature of the 2nd representation is that mass-like operator
$m^2$ \rf{olfman19012011-01} takes the same form as for the scalar field (see
\rf{scafielm2def01x}).

{\bf Gauge transformations}. We now discuss gauge symmetries of the conformal
theory under consideration. To this end we introduce the gauge transformation
parameters
\be \label{listgaugtraparspi101}
\xi_{k'-1}^{\vphantom{5pt}}
\hspace{1.3cm} k' \in [k]_2\,,
\ee
where $k$ is given in \rf{covDOFspi01ad01}. The gauge transformation
parameters $\xi_{k'}$ are scalar fields of the Lorentz algebra $so(d-1,1)$.
As usually, we collect $\xi_{k'}$ in the ket-vector $\xik$ defined by
\be\label{ketvecepsdef01}
\xik \equiv \sum_{k'\in [k]_2} \frac{1}{
(\frac{k+k'}{2})!}(\upsilon^\oplussm)^{^{\frac{k+k'}{2}}}
(\upsilon^\ominussm)^{^{\frac{k-k'}{2}}}    \, \xi_{k'-1}^{\vphantom{5pt}}
|0\rangle\,.
\ee
The ket-vector of gauge transformation parameters $\xik$
satisfies the algebraic constraint
\be
N_\upsilon\xik= k \xik\,,
\ee
which tells us that $\xik$ is a degree-$k$ homogeneous polynomial in the
oscillators $\upsilon^\oplussm$, $\upsilon^\ominussm$.

In terms of the ket-vectors $\phik$ and $\xik$, gauge transformations we
found take the form
\be \label{gautraspi1def01}
\delta \phik =  G \xik  \,,\qquad \quad G \equiv  \alpar - e_1\,,
\ee
where $e_1$ is given in \rf{man19122010-01}.

{\bf Component form of Lagrangian and gauge transformations}. The component
form of Lagrangian is obtained by plugging ket-vector \rf{phispin1def01} into
\rf{Lagspi1gen01}. Doing so, we obtain the component form of the 1st
representation for Lagrangian
\beq \label{18052011-oldman01}
&&  \LL = \sum_{k' \in [k]_2} \LL_{k'}\,,
\\
&& \LL_{k'} =  \half \phi_{-k'}^a E_\Max \phi_{k'}^a + \half
\phi_{-k'+1}^{}\Box\phi_{k'-1}^{}
+ \phi_{-k'+1} \partial^a \phi_{k'}^a  - \half \phi_{-k'}^a\phi_{k'+2}^a \,,
\\
\label{manold-01112011-03} && \hspace{1cm} E_\Max\phi^a = \Box \phi^a  -
\partial^a \partial^b\phi^b \,.
\eeq
We now see explicitly that two-derivative contributions to Lagrangian
\rf{18052011-oldman01} are the standard Maxwell kinetic terms for the vector fields
$\phi_{k'}^a$ and the standard Klein-Gordon kinetic terms for the scalar
fields $\phi_{k'}$. Note however that, in addition to the two-derivative
contributions, the Lagrangian involves one-derivative contributions and
derivative-independent mass-like contributions.

The second representation for Lagrangian given in \rf{olfman19012011-01} can
also be rewritten in the component form. Namely, Lagrangian
\rf{olfman19012011-01} can be presented as in \rf{18052011-oldman01} with
$\LL_{k'}$ given by
\beq
\LL_{k'} & = &  \half \phi_{-k'}^a \Box \phi_{k'}^a + \half
\phi_{-k'+1}^{}\Box\phi_{k'-1}^{}
\nonumber\\
& + & \half C_{-k'+1} C_{k'+1} -  \frac{1}{2}\phi_{-k'}^a \phi_{k'+2}^a -
\half \phi_{-k'+1}\phi_{k'+1}\,,
\\
\label{oldman-01112011-01} &&  C_{k'+1}  = \partial^a \phi_{k'}^a  +  \phi_{k'+1}\,.
\eeq
We note that it is quantities $C_{k'+1}$ \rf{oldman-01112011-01} that define
Lorentz like gauge condition for the spin-1 conformal field, $C_{k'+1}=0$,
$k'\in [k]_2$ (see \rf{oldman-01112011-02}).

Lagrangian \rf{Lagspi1gen01} can be represented as in \rf{18052011-oldman01}
with manifestly gauge invariant $\LL_{k'}$ given by
\beq
\label{olfman19012011-04a1} && \LL_{k'} =
-\frac{1}{4}F^{ab}(\phi_{-k'})F^{ab}(\phi_{k'}) - \half F_{-k'}^a
F_{k'+2}^a\,,
\\
&& \qquad \quad F^{ab}(\phi) \equiv \partial^a \phi^b - \partial^b\phi^a\,,
\qquad
\label{olfman19012011-04} F_{k'}^a \equiv \phi_{k'}^a +
\partial^a\phi_{k'-1}\,.
\eeq
Using component form of gauge transformations given below in
\rf{gautraspi1def02},\rf{gautraspi1def03}, we see that field strengths
\rf{olfman19012011-04} are gauge invariant.

Plugging \rf{phispin1def01},\rf{ketvecepsdef01} into \rf{gautraspi1def01}, we
obtain the corresponding component form of the gauge transformations,
\beq
\label{gautraspi1def02} && \delta \phi_{k'}^a  = \partial^a \xi_{k'-1}\,,
\\
\label{gautraspi1def03}&& \delta \phi_{k'} = - \xi_{k'}\,.
\eeq

Two remarks are in order.

\noindent {\bf i}) From \rf{gautraspi1def03}, we see that the scalar fields
$\phi_{k'}$ transform as Stueckelberg fields, i.e., the scalar fields are the
Stueckelberg fields in the framework of ordinary-derivative approach.

\noindent {\bf ii}) The vector field $\phi_{-k}^a$ has the same conformal
dimension as the vector field $\phi$ entering the higher-derivative approach
\rf{hihgderLag01}. Therefore we can use the identification $\phi_{-k}^a =
\phi^a$. The remaining vector fields $\phi_{k'}^a$, $k'=-k+2,-k+4,\ldots
k-2,k$, are auxiliary fields in our approach. Gauging away all scalar fields,
$\phi_{k'}=0$, and using equations of motion for auxiliary fields obtained
from Lagrangian \rf{18052011-oldman01}, $\phi_{k'}^a = E_\Max \phi_{k'-2}^a$,
we can express all auxiliary vector fields in terms of
$\phi_{-k}^a\equiv\phi^a$,
\be \label{06082011-01} \phi_{k'}^a = \Box^{^{\frac{k+k'-2}{2}}}
E_\Max\phi^a\,,\qquad k'=-k+2,-k+4,\ldots, k-2,k\,. \ee
Plugging \rf{06082011-01} into \rf{18052011-oldman01}, we obtain
higher-derivative Lagrangian \rf{hihgderLag01}, i.e., our approach and
higher-derivative approach are equivalent.

{\bf Conformal symmetries}. To complete our formulation we provide
realization of the conformal algebra symmetries on space of the ket-vector
$\phik$. All that is required is to fix operators $M^{ab}$, $\Delta$ and
$R^a$ for the case of spin-1 conformal field and then use relations given in
\rf{conalggenlis01}-\rf{conalggenlis04}. For the case of spin-1 field, the
realization of the spin operator of the Lorentz algebra and conformal
dimension operator on space of $\phik$ \rf{phispin1def01} is given by
\beq
&&  M^{ab} = \alpha^a \bar\alpha^b - \alpha^b \bar\alpha^a \,,
\nonumber\\[-12pt]
&&
\\[-12pt]
&& \Delta  =  \frac{d-2}{2}+\Delta'\,,
\qquad\quad  \Delta' \equiv  N_{\upsilon^\oplussm} - N_{\upsilon^\ominussm}
\,,
\nonumber
\eeq
where the realization of the conformal dimension operator $\Delta$ on space
of $\phik$ can be read from \rf{delspi1def01}. The realization of the operator
$R^a$ on space of $\phik$ is given by
\beq
\label{Rgsmspi101} R^a & = &r^a + R_\smG^a\,,
\\
&& r^a = \rb_{0,1} \alpha^a + r_{0,1} \bar\alpha^a + r_{1,1}
\partial^a\,,
\\
\label{Rgsmspi104} && R_\smG^a = G r_\smG^a\,,
\qquad \ \ \ r_\smG^a = r_{\smG,1}\bar\alpha^a\,,
\\
\label{Rgsmspi105} && \rb_{0,1} = 2\upsilon^\oplussm\bar\zeta\,,
\qquad
\qquad r_{0,1} = - 2\zeta \bar\upsilon^\oplussm\,,
\qquad \quad r_{1,1} = -2\upsilon^\oplussm \bar\upsilon^\oplussm \,,
\\
\label{Rgsmspi108} && r_{\smG,1} = \upsilon^\oplussm \rwt_{\smG,1}
\bar\upsilon^\oplussm \,, \qquad \rwt_{\smG,1} = \rwt_{\smG,1}(\Delta')\,,
\eeq
where $G$ appearing \rf{Rgsmspi104} is defined in \rf{gautraspi1def01} and
$\rwt_{\smG,1}$ \rf{Rgsmspi108} is an arbitrary function of $\Delta'$.

The following remarks are in order.

\noindent {\bf i}) From \rf{Rgsmspi101}-\rf{Rgsmspi108}, we see that $r^a$
part of the operator $R^a$ is fixed uniquely, while $R_\smG^a$ part, in view
of arbitrary $\rwt_{\smG,1}$, is sill to be arbitrary. Reason of the
arbitrariness in $R_\smG^a$ is obvious. Global transformations of gauge
fields are defined up to gauge transformations governed by $R_\smG^a$.

\noindent {\bf ii}) We check that $R^a$ transformations with $R_\smG^a=0$
satisfy the commutator $[K^a,K^b]=0$. In terms of the component fields, $R^a$
transformations with $R_\smG^a=0$ take the form
\beq
\label{manold-10112011-02} && \delta_{R^a} \phi_{k'}^b =  -(k+k')\eta^{ab}
\phi_{k'-1} - \half (k+k')(k+2-k')\partial^a \phi_{k'-2}^b\,,
\\
\label{manold-10112011-03} && \delta_{R^a} \phi_{k'} =  (k+1-k')
\phi_{k'-1}^a - \half (k-1+k')(k+1-k')\partial^a \phi_{k'-2}\,.
\eeq

\noindent {\bf iii}) To find all $R_\smG^a$ which satisfy the commutator
$[K^a,K^b]=0$, we note the relation
\be \label{22052011-oldman01}
[K^a,K^b] = G r_\smG^{ab}\,,
\qquad\quad
r_\smG^{ab} \equiv r^a r_\smG^b + r_\smG^a r^b + r_\smG^a G r_\smG^b - (a
\leftrightarrow b)\,.
\ee
We see that the commutator $[K^a,K^b]$ is proportional to operator of gauge
transformations $G$ \rf{gautraspi1def01}, as it should be in gauge theory.
From \rf{22052011-oldman01}, we see that the requirement $[K^a,K^b]=0$
amounts to the equation $r_\smG^{ab}=0$. Solution to this equation is given
by
\be \label{rwtg1sol01} \rwt_{\smG,1} = \frac{4}{\Delta' + c_0}\,,\ee
where $c_0$ is real-valued constant. Note that the simple
representation for $R^a$ corresponding to $R_\smG^a=0$ is achieved by taking
$c_0=\infty$.

{\bf Spin-1 conformal field in 6d}. To illustrate our approach we consider
spin-1 field in $6d$. This is the simplest conformal theory of spin-1 field
involving Stueckelberg field. For the case of spin-1 field in $6d$, our field
content involves two vector fields and one scalar field (see
\rf{covDOFspi01ad01}),
\beq \label{11082011-01}
&& \phi_\smminone^a \qquad\  \ \phi_{_1}^a
\nonumber\\
&& \hspace{0.8cm} \phi_{_0}
\eeq
For these fields, the Lagrangian takes the form (see
\rf{18052011-oldman01},\rf{olfman19012011-04a1})
\be \label{11082011-02}
\LL =  - \half F^{ab}(\phi_{-1}) F^{ab}(\phi_1) - \half (\phi_1^a +
\partial^a\phi_0)^2\,.
\ee
Gauge symmetries are described by the two gauge transformation parameters
$\xi_{_{-2}}$, $\xi_{_0}$ (see \rf{listgaugtraparspi101}) and appropriate
gauge transformations are given by (see
\rf{gautraspi1def02},\rf{gautraspi1def03})
\be \label{11082011-03}
\delta\phi_{_{-1}}^a  =\partial^a\xi_{_{-2}}\,, \qquad\quad \delta\phi_{_1}^a
=\partial^a\xi_{_0}\,, \qquad\quad \delta\phi_{_0} = - \xi_{_0}\,.
\ee
$R^a$ transformations taken in the minimal scheme, $R_\smG^a=0$, are given by
(see \rf{manold-10112011-02}, \rf{manold-10112011-03})
\beq
&& \delta_{_{R^a}}\phi_{-1}^b = 0 \,,
\\
&& \delta_{_{R^a}}\phi_1^b = -2\eta^{ab} \phi_0 -2\partial^a \phi_{-1}^b\,,
\\
&& \delta_{_{R^a}} \phi_0 = 2\phi_{-1}^a \,.
\eeq
From \rf{11082011-03}, we see that the scalar field $\phi_{_0}$ is Stueckelberg field.
Gauging away this field, $\phi_{_0} = 0$, we see that our Lagrangian
\rf{11082011-02} takes the simplified form
\be \label{11082011-04}
\LL =  - \half F^{ab}(\phi_{-1}) F^{ab}(\phi_1) - \half \phi_1^a \phi_1^a\,.
\ee
Now, using equations of motion for the auxiliary vector field $\phi_{_1}^a$
obtained from Lagrangian \rf{11082011-04}, we can express $\phi_{_1}^a$ in
terms of $\phi_{_{-1}}^a$,
\be \label{11082011-05}  \phi_{_1}^a  = \partial^b F^{ba}(\phi_{_{-1}})\,.
\ee
Plugging $\phi_{_1}^a$ \rf{11082011-05} into Lagrangian \rf{11082011-04}, we
obtain the higher-derivative Lagrangian given in \rf{hihgderLag01} when $d=6$
(i.e., $k=1$). Thus, our ordinary-derivative approach is equivalent to the
standard one and our field $\phi_{_{-1}}^a$ is identified with the conformal
vector field $\phi^a$ in Sec. \ref{sec-04-sub01}.

We note that Lagrangian, gauge transformations and operator $R^a$ are fixed
by requiring that:

\noindent i) Lagrangian does not involve higher than second order terms in
derivatives;

\noindent ii) gauge transformations and operator $R^a$ do not involve higher
than first order terms in derivatives;

\noindent iii) Lagrangian is invariant under gauge transformations and
conformal algebra transformations.

These requirements fix Lagrangian and gauge transformations uniquely. The
operator $R^a$ is fixed uniquely up to the gauge transformation operator (as
it should be for gauge fields). For the derivation of Lagrangian, gauge
transformations and operator $R^a$, see Appendix B.

\newsection{ Spin-2 conformal field}\label{sec-05}

We proceed our study of ordinary-derivative formulation of conformal field
theory with discussion of spin-2 conformal field. In the literature, such
field is often referred to as conformal Weyl graviton. As before, to make
contact with studies in earlier literature we start with the presentation of
the standard higher-derivative approach to the spin-2 conformal field. In due
course, we review our result concerning the counting of on-shell D.o.F for
the spin-2 conformal field.

\subsection{ Higher-derivative formulation of  spin-2 conformal
field}\label{sec-05-sub01}

To discuss higher-derivative formulation of spin-2 conformal field one uses
rank-2 $so(d-1,1)$ Lorentz algebra tensor field $\phi^{ab}$ having the
conformal dimension $\Delta_{\phi^{ab}}=0$. The field $\phi^{ab}$ is
symmetric, $\phi^{ab}=\phi^{ba}$, and traceful $\phi^{aa} \ne 0$.
Higher-derivative Lagrangian for the field $\phi^{ab}$ is given by
\be \label{Lconfie2lag01}  \LL= \frac{1}{f^2}C_\lin^{abce} \Box^{k-1}
C_\lin^{abce}\,,\qquad  f^2 \equiv 4\frac{d-3}{d-2}\,, \qquad
k\equiv \frac{d-2}{2}\,,\ee
where $C_\lin^{abce}$ is the linearized Weyl tensor. Using representation of
the Weyl tensor in terms of curvatures and the Gauss-Bonnet relation
\beq
C^{abce} &=& R^{abce} - \frac{1}{d-2}(\eta^{ac}R^{be} - \eta^{bc}R^{ae} +
\eta^{be}R^{ac} - \eta^{ae}R^{bc})
\nonumber\\
& + & \frac{1}{(d-1)(d-2)}(\eta^{ac} \eta^{be} - \eta^{ae} \eta^{bc}) R\,,
\\
&&\hspace{-2cm}  R_\lin^{abce} \Box R_\lin^{abce} - 4 R_\lin^{ab} \Box
R_\lin^{ab} + R_\lin\Box R_\lin = 0 \quad (\hbox{ up to total derivative})\,,
\eeq
we obtain the representation for Lagrangian \rf{Lconfie2lag01} in terms of
linearized Ricci curvatures,
\be \label{2m04122010-02}
\LL = R_\lin^{ab} \Box^{k-1} R_\lin^{ab} - \frac{d}{4(d-1)}
R_\lin\Box^{k-1}R_\lin \,, \ee
which is also useful for certain purposes. Using the explicit representation
of the linearized Ricci curvatures in terms of the gauge field $\phi^{ab}$,
\beq
&& \hspace{1cm} R_\lin^{ab} = \frac{1}{2}\left(-\Box\phi^{ab} + \partial^a
\partial^c\phi^{bc} +\partial^b
\partial^c\phi^{ac} - \partial^a\partial^b \phi^{cc} \right)\,,
\nonumber\\[-10pt]
\label{2m04122010-06} &&
\\[-10pt]
&& \hspace{1cm} R_\lin  = \partial^a\partial^b \phi^{ab} - \Box \phi^{aa}\,,
\nonumber
\eeq
we get other well-known form of the Lagrangian,
\beq
\label{Lconfie2lag01n01} && \LL = \frac{1}{4} \phi^{ab} \Box^{k+1} P^{ab\,ce}
\phi^{ce} \,,
\\
&& P^{ab\,ce} \equiv \half( \pi^{ac} \pi^{be} + \pi^{ae} \pi^{bc})
-\frac{1}{d-1} \pi^{ab}\pi^{ce} \,,
\qquad \quad
\pi^{ab} \equiv \eta^{ab}   - \frac{\partial^a \partial^b}{\Box} \,,
\eeq
where $k$ is given in \rf{Lconfie2lag01}. Lagrangian \rf{Lconfie2lag01n01}
can also be represented as
\beq \label{07072009-01}
\LL & = & \frac{1}{4} \phi^{ab}\Box^{k+1} \phi^{ab} -\frac{1}{4(d-1)}
\phi^{aa}\Box^{k+1} \phi^{bb}
\nonumber\\
& + & \half \partial^b\phi^{ab}\Box^k \partial^c\phi^{ac} + \frac{1}{2(d-1)}
\phi^{aa} \Box^k \partial^b\partial^c\phi^{bc}
+ \frac{d-2}{4(d-1)} \partial^a\partial^b\phi^{ab} \Box^{k-1}
\partial^c\partial^e\phi^{ce}\,.\qquad
\eeq
Lagrangian \rf{Lconfie2lag01} is invariant under linearized diffeomorphism
and Weyl gauge transformations
\be
\label{secspin2con10} \delta \phi^{ab} = \partial^a \xi^b +
\partial^b \xi^a  + \eta^{ab} \xi\,,
\ee
where $\xi^a$ and $\xi$ are the respective diffeomorphism and Weyl gauge
transformation parameters.

We now discuss on-shell D.o.F of the conformal theory under consideration. As
before, to discuss on-shell D.o.F we use fields transforming in irreps of
$so(d-2)$ algebra. One can prove (for details, see Appendix C)  that on-shell
D.o.F are described by $k+1$ rank-2 traceless symmetric tensor fields
$\phi_{k'}^{ij}$, $k$ vector fields $\phi_{k'}^i$, and $k-1$ scalar fields
$\phi_{k'}$ of the $so(d-2)$ algebra:
\beq
\label{spin2DoF01} &&\phi_{k'}^{ij}\,, \qquad k' \in [k]_2\,;
\\
\label{spin2DoF02} &&\phi_{k'}^i\,, \qquad k' \in [k-1]_2\,;
\\
\label{spin2DoF03} && \phi_{k'}\,, \qquad k' \in  [k-2]_2\,,
\eeq
where $i,j=1,\ldots,d-2$ and we use the notation as in \rf{sumnot02}. Note
that scalar on-shell D.o.F \rf{spin2DoF03} appear in spin-2 conformal field
theory only in $d\geq 6$ (i.e., $k\geq 2$). Total number of on-shell D.o.F
shown in \rf{spin2DoF01}-\rf{spin2DoF03} is given by
\beq \label{spin2DoF04}
&& \nbf = \frac{1}{4}d(d-3)(d+2) \,.
\eeq
We note that $\nbf$ \rf{spin2DoF04} is a sum of D.o.F for the $so(d-2)$
algebra
fields given in \rf{spin2DoF01}-\rf{spin2DoF03}:%
\footnote{ Total D.o.F given in \rf{spin2DoF04} was found in
Ref.\cite{Fradkin:1985am}. For the case of $d=4$ spin-2 conformal field
theory, decomposition of $\nbf$ \rf{spin2DoF05} into irreps of the $so(d-2)$
algebra was carried out in Ref.\cite{Lee:1982cp}. In Appendix C, we use the
light-cone approach to generalize result of the latter reference to case of
arbitrary $d\geq 4$.}
\beq \label{spin2DoF05}
&& \nbf = \sum_{k'\in [k]_2} \nbf (\phi_{k'}^{ij}) + \sum_{k'\in [k-1]_2}
\nbf(\phi_{k'}^i) + \sum_{k'\in [k-2]_2} \nbf(\phi_{k'})\,,
\\
&& \nbf(\phi_{k'}^{ij}) = \frac{d(d-3)}{2}, \quad\qquad  \nbf(\phi_{k'}^i) =
d-2,  \quad\qquad \nbf(\phi_{k'}) = 1\,.
\eeq
For various space-time dimension $d$, the $\nbf$ and decomposition
\rf{spin2DoF05} are as follows:
\be \label{nuvardim}
\begin{array}{lll}
d=4  & \nbf = 6   & 2 \times \ {\bf 2}_2 +  1\times {\bf 2}_1 + 0 \times {\bf
1}_0
\\
d=6  & \nbf = 36 & 3\times \ {\bf 9}_2 + 2\times {\bf 4}_1 + 1\times {\bf
1}_0
\\
d=8    & \nbf = 100  & 4\times \!{\bf 20}_2 + 3\times {\bf 6}_1 + 2\times
{\bf 1}_0
\\
d=10  \qquad\qquad & \nbf = 210 \qquad & 5\times \!{\bf 35}_2 + 4\times {\bf
8}_1 + 3\times {\bf 1}_0
\end{array}
\ee
In \rf{nuvardim} in expressions like $X \times {\bf Y}_s$, the ${\bf Y}$
stands for dimension of spin-$s$ irreps of $so(d-2)$ algebra, while $X$
stands for multiplicity of the spin-$s$ irreps of $so(d-2)$ algebra.

\subsection{ Ordinary-derivative
formulation of spin-2 conformal field}\label{sec-05-sub02}

{\bf Field content}. To discuss ordinary-derivative and gauge invariant
formulation of spin-2 conformal field in flat space of dimension $d\geq 4$ we
use $k+1$ rank-2 symmetric traceful tensor fields $\phi_{k'}^{ab}$,
$\phi_{k'}^{ab}=\phi_{k'}^{ba}$, $\phi_{k'}^{aa}\ne 0$, $k$ vector fields
$\phi_{k'}^a$, and $k-1$ scalar fields $\phi_{k'}$ of the Lorentz algebra
$so(d-1,1)$:
\beq
\label{covspin2DoF01} &&\phi_{k'}^{ab}\,, \qquad k' \in [k]_2\,;
\\
\label{covspin2DoF02} &&\phi_{k'}^a\,, \qquad k'= [k-1]_2\,;
\\
\label{covspin2DoF03} && \phi_{k'}\,, \qquad k'= [k-2]_2\,;
\\
\label{20082011-01} && \hspace{2cm} k \equiv \frac{d-2}{2}\,,
\eeq
where we use the notation as in \rf{sumnot02}.  Note that scalar fields
\rf{covspin2DoF03} appear only in $d\geq 6$ (i.e., $k\geq 2$). Fields in
\rf{covspin2DoF01}-\rf{covspin2DoF03} have the conformal dimensions
\be \label{Pdelspi2def01} \Delta_{\phi_{k'}^{ab}} = \frac{d-2}{2} + k'\,,
\qquad \Delta_{\phi_{k'}^a} = \frac{d-2}{2} + k'\,,
\qquad \Delta_{\phi_{k'}} = \frac{d-2}{2} + k'\,.\ee

As before, we use oscillators $\alpha^a$, $\zeta$, $\upsilon^\oplussm$,
$\upsilon^\ominussm$ \rf{manold-31102011-01} to collect fields
\rf{covspin2DoF01}-\rf{covspin2DoF03} into the ket-vector $\phik$ defined by
\beq
&&  \label{phispin2def01} \phik =  |\phi_{_2} \rangle  + \zeta |\phi_{_1}
\rangle + \frac{\zeta^2}{\sqrt{2}} |\phi_{_0}\rangle \,,
\\
\label{phispin2def01x1} && |\phi_2\rangle \equiv \sum_{k'\in [k]_2}
\frac{1}{2 (\frac{k+k'}{2})! } \alpha^a\alpha^b
(\upsilon^\oplussm)^{^{\frac{k+k'}{2}}}
(\upsilon^\ominussm)^{^{\frac{k-k'}{2}}} \, \phi_{k'}^{ab} |0\rangle\,,
\\
\label{phispin2def01x2} && |\phi_1\rangle \equiv \sum_{k'\in [k-1]_2} \frac{1}{
 (\frac{k-1+k'}{2})! } \alpha^a
(\upsilon^\oplussm)^{^{\frac{k-1+k'}{2}}}
(\upsilon^\ominussm)^{^{\frac{k-1-k'}{2}}}  \phi_{k'}^a |0\rangle\,,
\\
\label{phispin2def04} && |\phi_0\rangle \equiv \sum_{k'\in [k-2]_2} \frac{1}{
 (\frac{k-2+k'}{2})! }(\upsilon^\oplussm)^{^{\frac{k-2+k'}{2}}}
(\upsilon^\ominussm)^{^{\frac{k-2-k'}{2}}}  \, \phi_{k'}^{\vphantom{5pt}}
|0\rangle\,.
\eeq
From \rf{phispin2def01}-\rf{phispin2def04}, we see that the ket-vector
$\phik$ is degree-2 homogeneous polynomial in the oscillators $\alpha^a$,
$\zeta$ and degree-$k$ homogeneous polynomial in the oscillators $\zeta$,
$\upsilon^\oplussm$, $\upsilon^\ominussm$. Also, note that the ket-vectors
$\phikst{2}$, $\phikst{1}$, and $\phikst{0}$ are the respective degree-$k$,
$k-1$, and $k-2$ homogeneous polynomials in the oscillators
$\upsilon^\oplussm$, $\upsilon^\ominussm$. In other words, the ket-vectors
satisfy the relations
\beq \label{21072011-11}
&& (N_\alpha + N_\zeta)\phik =  2\phik\,,
\qquad
(N_\zeta + N_\upsilon)\phik = k \phik\,,
\\
\label{21072011-13} && N_\upsilon\phikst{2} = k \phikst{2}\,,\hspace{1.9cm}
N_\upsilon\phikst{1} = (k-1) \phikst{1}\,,\qquad N_\upsilon\phikst{0} = (k-2)
\phikst{0}\,. \qquad\quad \eeq
We now discuss two representations for gauge invariant Lagrangian in turn.

{\bf 1st representation for Lagrangian}. Lagrangian we found takes the form
\beq
\label{spi2lag01} \LL & = & \frac{1}{2} \phibr E \phik\,,
\\
E  & = & E_\smtwo + E_\smone + E_\smzero\,,
\\
\label{EHsecordope01} && E_\smtwo \equiv \Box - \alpar \albpar +
\frac{1}{2}(\alpar)^2\bar\alpha^2 + \frac{1}{2} \alpha^2 (\albpar)^2
- \frac{1}{2}\alpha^2 \Box \bar\alpha^2\,,
\\
&& E_\smone \equiv  \eb_1(\alpar  -\alpha^2 \albpar) + e_1 ( \albpar - \alpar
\bar\alpha^2)\,,
\\
&& E_\smzero \equiv m_1 + \alpha^2\bar\alpha^2m_2 + \mb_3 \alpha^2 + m_3
\bar\alpha^2\,,
\\
\label{man19122010-02} && e_1 =  \zeta \Bigl(\frac{d-N_\zeta}{d-2N_\zeta}\Bigr)^{1/2}
\bar\upsilon^\ominussm\,,
\qquad \quad
\eb_1 = - \upsilon^\ominussm \Bigl(\frac{d-N_\zeta}{d-2N_\zeta}\Bigr)^{1/2}
\bar\zeta\,,
\\
&& m_1 = \upsilon^\ominussm  \bar\upsilon^\ominussm
\frac{d+2-N_\zeta}{d+2-2N_\zeta}(N_\zeta-1)\,,
\qquad\quad
m_2 = \half \upsilon^\ominussm  \bar\upsilon^\ominussm\,,
\\
&& m_3 = \half \Bigl(\frac{d-1}{d-2}\Bigr)^{1/2} \zeta^2
\bar\upsilon^\ominussm \bar\upsilon^\ominussm\,,
\hspace{2.4cm}
\mb_3 = \half \Bigl(\frac{d-1}{d-2}\Bigr)^{1/2} \upsilon^\ominussm
\upsilon^\ominussm \bar\zeta^2\,.\qquad
\eeq
For the notation, see \rf{manold-31102011-02}-\rf{manold-31102011-06}. We
note that $E_\smtwo$ \rf{EHsecordope01} is the standard second-order
Einstein-Hilbert operator rewritten in terms of the oscillators.

{\bf 2nd representation for Lagrangian}. This representation allows us to
introduce de Donder like gauge condition for the spin-2 conformal field.
Namely, Lagrangian \rf{spi2lag01} can be rewritten as
\beq
\LL & = & \half \phibr (1-\frac{1}{4}\alpha^2\bar\alpha^2)(\Box - m^2 )\phik +
\half \langle \Cb\phi|\Cb \phik \,,
\nonumber\\
\label{olfman19012011-06} && \Cb \equiv \bar\alpha\partial - \half
\alpha\partial \bar\alpha^2 - \eb_1 + \half e_1\bar\alpha^2 \,,
\qquad\quad
m^2 \equiv \upsilon^\ominussm \bar\upsilon^\ominussm \,,
\eeq
where $e_1$, $\eb_1$ are given in \rf{man19122010-02}. We now note that it is
operator $\Cb$ \rf{olfman19012011-06} that defines de Donder like gauge
condition for the spin-2 conformal field,
\be \label{manold-31102011-11} \Cb\phik = 0\,. \ee

{\bf Gauge transformations}. To discuss gauge symmetries of Lagrangian
\rf{spi2lag01} we introduce the following gauge transformation parameters:
\beq \label{listgauparspi201}
&& \xi_{k'-1}^a\,, \hspace{1.2cm} k' \in [k]_2\,;
\nonumber\\[-13pt]
&&
\\[-13pt]
&& \xi_{k'-1}^{\vphantom{5pt}} \hspace{1.4cm} k' \in
[k-1]_2\,,
\nonumber
\eeq
where $k$ is given in \rf{20082011-01}. The gauge transformation parameters
$\xi_{k'}^a$ and $\xi_{k'}$ are the respective vector and scalar fields of
the Lorentz algebra $so(d-1,1)$. As usually, we collect gauge transformation
parameters \rf{listgauparspi201} in the ket-vector
$\xik$ defined by
\beq \label{epsspi2def01}
&& \xik = |\xi_1\rangle + \zeta |\xi_{_0}\rangle\,,
\\
\label{epsspi2def01x1} && |\xi_1\rangle \equiv \sum_{k'\in
[k]_2}\frac{1}{(\frac{k+k'}{2})! } \alpha^a
(\upsilon^\oplussm)^{^{\frac{k+k'}{2}}}
(\upsilon^\ominussm)^{^{\frac{k-k'}{2}}} \, \xi_{k'-1}^a |0\rangle\,,
\\
\label{epsspi2def01x2} && |\xi_0\rangle \equiv \sum_{k'\in [k-1]_2} \frac{1}{
(\frac{k-1+k'}{2})! } (\upsilon^\oplussm)^{^{\frac{k-1+k'}{2}}}
(\upsilon^\ominussm)^{^{\frac{k-1-k'}{2}}} \, \xi_{k'-1}^{\vphantom{5pt}}
|0\rangle\,.
\eeq
The ket-vectors $\xik$, $|\xi_1\rangle$, $|\xi_0\rangle$ satisfy the
algebraic constraints
\beq
&&  \label{oldman12012011-03}  (N_\alpha + N_\zeta)\xik =  \xik\,,
\quad\qquad
(N_\zeta + N_\upsilon)\xik= k \xik\,,
\\
&& \label{oldman12012011-04} N_\upsilon |\xi_1\rangle =
k|\xi_1\rangle\,,\hspace{2cm} N_\upsilon|\xi_0\rangle = ( k - 1)
|\xi_0\rangle\,.
\eeq
Constraints \rf{oldman12012011-03} tell us that $\xik$ is a degree-1
homogeneous polynomial in the oscillators $\alpha^a$, $\zeta$ and degree-$k$
homogeneous polynomial in the oscillators $\zeta$, $\upsilon^\oplussm$,
$\upsilon^\ominussm$. Constraints \rf{oldman12012011-04} imply that
$|\xi_1\rangle$ and $|\xi_0\rangle$ are the respective degree-$k$ and $k-1$
homogeneous polynomials in $\upsilon^\oplussm$, $\upsilon^\ominussm$.

We now note that gauge transformations can entirely be written in terms of
ket-vectors $\phik$ \rf{phispin2def01} and $\xik$ \rf{epsspi2def01}. This is
to say that gauge transformations we found take the form
\beq \label{gautraspi2def01}
\delta \phik  & = &  G \xik  \,, \qquad \quad G \equiv   \alpar - e_1
-\frac{1}{d-2} \alpha^2 \eb_1\,,
\eeq
where $e_1$ and $\eb_1$ are defined in \rf{man19122010-02}.

{\bf Component form of Lagrangian and gauge transformations}. The component
form of Lagrangian can easily be obtained by plugging ket-vector
\rf{phispin2def01} into \rf{spi2lag01}. Doing so, we obtain the component
form of the 1st representation for Lagrangian \rf{spi2lag01},
\beq \label{olfman19012011-08}
\LL & = & \sum_{k' \in [k]_2} \LL_{k'} \,,
\\
\label{olfman19012011-08a1} \LL_{k'} & = & \frac{1}{4} \phi_{-k'}^{ab}
E_\EH\phi_{k'}^{ab} + \half \phi_{-k'-1}^a E_\Max\phi_{k'+1}^a + \half
\phi_{-k'}^{}\Box\phi_{k'}^{}
\nonumber\\
& + &  \phi_{-k'+1}^a (\partial^b \phi_{k'}^{ba}  - \partial^a
\phi_{k'}^{bb} - u \partial^a \phi_{k'}^{})
\nonumber\\
& - & \frac{1}{4} \phi_{-k'}^{ab} \phi_{k'+2}^{ab}  +  \frac{1}{4}
\phi_{-k'}^{aa} \phi_{k'+2}^{bb}
+ \frac{u}{2}\phi_{-k'}^{aa} \phi_{k'+2}^{} +
\frac{d}{2(d-2)}\phi_{-k'}\phi_{k'+2} \,,
\eeq
\be \label{olfman19012011-09} u \equiv \Bigl(2\frac{d-1}{d-2}\Bigr)^{1/2}\,,
\ee
where the Maxwell operator is given in \rf{manold-01112011-03}, while the
Einstein-Hilbert  operator $E_\EH$ is given by
\be
\label{25072011-01} E_\EH\phi^{ab} = \Box \phi^{ab} -\partial^a
\partial^c\phi^{bc} -
\partial^b \partial^c\phi^{ac}  + \partial^a \partial^b \phi^{cc}
+ \eta^{ab}(\partial^c\partial^e\phi^{ce} - \Box \phi^{cc}) \,.
\ee
From \rf{olfman19012011-08a1}, we see that two-derivative contributions to
Lagrangian \rf{olfman19012011-08} are the Einstein-Hilbert kinetic terms for
the rank-2 tensor fields $\phi_{k'}^{ab}$, the Maxwell kinetic terms for the
vector fields $\phi_{k'}^a$, and the Klein-Gordon kinetic terms for the scalar
fields $\phi_{k'}$. Besides the two-derivative contributions, the Lagrangian
involves one-derivative contributions and derivative-independent
contributions.

In order to obtain component form of the 2nd representation for Lagrangian
\rf{olfman19012011-06} we should plug ket-vector \rf{phispin2def01} in
\rf{olfman19012011-06}. Doing so, we get Lagrangian \rf{olfman19012011-08}
with $\LL_{k'}$ given by
\beq
\LL_{k'} & = & \frac{1}{4} \phi_{-k'}^{ab} \Box \phi_{k'}^{ab} - \frac{1}{8}
\phi_{-k'}^{aa} \Box \phi_{k'}^{bb} + \half \phi_{-k'}^a \Box \phi_{k'}^a +
\half \phi_{-k'}^{}\Box\phi_{k'}^{}
\nonumber\\
& + &  \half C_{-k'+1}^a C_{k'+1}^a + \half C_{-k'+1} C_{k'+1} +
\nonumber\\
& - & \frac{1}{4} \phi_{-k'}^{ab} \phi_{k'+2}^{ab} + \frac{1}{8}
\phi_{-k'}^{aa} \phi_{k'+2}^{bb}  -  \frac{1}{2}\phi_{-k'}^a \phi_{k'+2}^a -
\half \phi_{-k'}\phi_{k'+2}\,,
\\
\label{olfman19012011-10} && C_{k'+1}^a  \equiv \partial^b \phi_{k'}^{ab} -
\half
\partial^a\phi_{k'}^{bb} + \phi_{k'+1}^a\,,
\\
\label{olfman19012011-11} && C_{k'+1}  \equiv \partial^a \phi_{k'}^a + \half
\phi_{k'+1}^{bb} + u \phi_{k'+1}\,.
\eeq
We note that it is quantities given in \rf{olfman19012011-10},
\rf{olfman19012011-11} that define de Donder like gauge condition for the
spin-2 conformal field, $C_{k'+1}^a=0$, $k'\in [k]_2$, $C_{k'+1}=0$, $k'\in
[k-1]_2$ (see \rf{manold-31102011-11}).

{\bf Pauli-Fierz representation}. As a side remark, we note Pauli-Fierz
like representation for Lagrangian \rf{olfman19012011-08}\,,
\beq
\LL & = & \sum_{k' \in [k]_2} \LL_{k'} \,,
\\
\LL_{k'} & = &  \frac{1}{4} \Phi_{-k'}^{ab} E_\EH\Phi_{k'}^{ab} -
\frac{1}{4}\Phi_{-k'}^{ab}\Phi_{k'+2}^{ab} + \frac{1}{4} \Phi_{-k'}^{aa}
\Phi_{k'+2}^{bb}\,,
\nonumber\\
\label{manold-31102011-09} \Phi_{k'}^{ab} & \equiv &  \phi_{k'}^{ab} +
\partial^a \phi_{k'-1}^b +
\partial^b \phi_{k'-1}^a
+ \frac{2}{u} \partial^a \partial^b \phi_{k'-2} + \frac{2}{(d-2)u} \eta^{ab}
\phi_{k'}^{}\,,
\eeq
where we introduce conformal Pauli-Fierz fields $\Phi_{k'}^{ab}$
\rf{manold-31102011-09}. It is easy to check that the conformal Pauli-Fierz
fields are invariant under gauge transformations given below in
\rf{gautraspi2def02}-\rf{gautraspi2def04}.

The component form of gauge transformations can straightforwardly be obtained
by plugging ket-vectors \rf{phispin2def01},\rf{epsspi2def01} into
\rf{gautraspi2def01},
\beq
\label{gautraspi2def02} && \delta \phi_{k'}^{ab} = \partial^a \xi_{k'-1}^b +
\partial^b \xi_{k'-1}^a + \frac{2}{d-2}\eta^{ab}\xi_{k'}\,,
\\
\label{gautraspi2def03}&& \delta \phi_{k'}^a  = \partial^a \xi_{k'-1} -
\xi_{k'}^a\,,
\\
\label{gautraspi2def04} && \delta \phi_{k'} = - u \xi_{k'}\,.
\eeq

Two remarks are in order.

\noindent {\bf i}) From \rf{gautraspi2def03},\rf{gautraspi2def04}, we see
that the vector and scalar fields transform as Stueckelberg fields, i.e.,
these fields are the Stueckelberg fields in the framework of the
ordinary-derivative approach.

\noindent {\bf ii}) Because the tensor field $\phi_{-k}^{ab}$ has the same
conformal dimension as the tensor field $\phi^{ab}$ entering the
higher-derivative approach \rf{Lconfie2lag01} we can use the identification
$\phi_{-k}^{ab} = \phi^{ab}$. The remaining tensor fields $\phi_{k'}^{ab}$,
$k'=-k+2,-k+4,\ldots k-2,k$, are auxiliary fields. Gauging away all vector
and scalar fields, $\phi_{k'}^a=0$, $\phi_{k'}=0$, and using equations of
motion for auxiliary tensor fields obtained from \rf{olfman19012011-08},
$\phi_{k'}^{ab} - \eta^{ab}\phi_{k'}^{cc} = E_\EH\phi_{k'-2}^{ab}$, we
express all auxiliary tensor fields in terms of $\phi_{-k}^{ab}$,
\be
\label{07082011-02} \phi_{k'}^{ab} = \Box^{^{\frac{k+k'-2}{2}}} (-
2R_\lin^{ab} + \frac{1}{d-1}\eta^{ab} R_\lin) +
\frac{d-2}{d-1}\theta_{k'}\Box^{^{\frac{k+k'-4}{2}}}
\partial^a\partial^b R_\lin\,,
\ee
where $\theta_{-k+2} = 0$, $\theta_{k'}=1$ for $k'=-k+4,-k+6,\ldots,k-2,k$
and the linearized Ricci curvatures for $\phi^{ab} \equiv \phi_{-k}^{ab}$ are
given in \rf{2m04122010-06}. Plugging solution for auxiliary fields
\rf{07082011-02} into \rf{olfman19012011-08} and using the identification
$\phi_{-k}^{ab} = \phi^{ab}$, we obtain higher-derivative Lagrangian
\rf{Lconfie2lag01}, i.e., our approach and higher-derivative approach are
equivalent.

{\bf Conformal symmetries}. In order to realize the conformal algebra
symmetries on space of ket-vector $\phik$ \rf{phispin2def01} we should fix
operators $M^{ab}$, $\Delta$ and $R^a$ and use these operators in
\rf{conalggenlis01}-\rf{conalggenlis04}. For the case of spin-2 field, the
realization of spin operator of the Lorentz algebra and conformal dimension
operator on space of ket-vector $\phik$ \rf{phispin2def01} is given by
\beq
&& M^{ab} = \alpha^a \bar\alpha^b - \alpha^b \bar\alpha^a \,,
\nonumber\\[-12pt]
&&
\\[-12pt]
&& \Delta  =  \frac{d-2}{2}+\Delta'\,,
\qquad\quad \Delta' \equiv  N_{\upsilon^\oplussm} - N_{\upsilon^\ominussm} \,,
\nonumber
\eeq
where the realization of the conformal dimension operator $\Delta$ on space
of $\phik$ can be read from \rf{Pdelspi2def01}. The realization of the operator
$R^a$ on space of $\phik$ is given by
\beq
\label{Rgsmspi201} && \hspace{-1cm} R^a  = r^a + R_\smG^a\,,
\\
\label{Rgsmspi201x1} && r^a = r_{0,1} \bar\alpha^a + \rb_{0,1} \bigl(
\alpha^a - \frac{1}{d-2}\alpha^2\bar\alpha^a \bigr) + r_{1,1}
\partial^a\,,
\\
\label{Rgsmspi204} && R_\smG^a  =   G r_\smG^a\,,
\qquad\quad r_\smG^a \equiv r_{\smG,1} \bar\alpha^a  + r_{\smG,2} \alpha^a +
r_{\smG,3} \alpha^a \bar\alpha^2\,,\qquad
\\
\label{Rgsmspi206}
&& r_{0,1} = 2 \zeta \Bigl(\frac{d-N_\zeta}{d-2N_\zeta}\Bigr)^{1/2}
\bar\upsilon^\oplussm\,,
\qquad \quad
\rb_{0,1} = - 2 \upsilon^\oplussm
\Bigl(\frac{d-N_\zeta}{d-2N_\zeta}\Bigr)^{1/2} \bar\zeta\,,
\\
\label{Rgsmspi208}  && r_{1,1} = -2\upsilon^\oplussm \bar\upsilon^\oplussm
\,,
\\
\label{Rgsmspi209}  && r_{\smG,1} =  \upsilon^\oplussm  \rwt_{\smG,1}
\bar\upsilon^\oplussm \,,
\hspace{1.3cm} r_{\smG,2} =  \upsilon^\oplussm \upsilon^\oplussm
\rwt_{\smG,2} \bar\zeta^2 \,,
\qquad r_{\smG,3} =  \upsilon^\oplussm \rwt_{\smG,3} \bar\upsilon^\oplussm
\,,
\\
\label{Rgsmspi211} && \rwt_{\smG,1} =\rwt_{\smG,1} (N_\zeta,\Delta')\,,
\qquad \rwt_{\smG,2} =\rwt_{\smG,2} (\Delta')\,,\hspace{1.3cm} \rwt_{\smG,3}
=\rwt_{\smG,3} (\Delta')\,,\qquad
\eeq
where $G$ in \rf{Rgsmspi204} is given in \rf{gautraspi2def01}, and
$\rwt_{\smG,1}$, $\rwt_{\smG,2}$, $\rwt_{\smG,3}$ \rf{Rgsmspi211} are
arbitrary functions of $N_\zeta$ and $\Delta'$.

The following remarks are in order.

\noindent {\bf i}) $r^a$ part of the operator $R^a$ is fixed uniquely, while
$R_\smG^a$ part, in view of arbitrary $\rwt_{\smG,1}$, $\rwt_{\smG,2}$,
$\rwt_{\smG,3}$, is still to be arbitrary. Reason of the arbitrariness in
$R_\smG^a$ is obvious. Global transformations of gauge fields are defined up
to gauge transformations governed by $R_\smG^a$.

\noindent {\bf ii}) We check that $R^a$ transformations with $R_\smG^a=0$
satisfy the commutator $[K^a,K^b]=0$. In terms of the component fields, $R^a$
transformations with $R_\smG^a=0$ take the form
\beq
&& \delta_{_{R^a}} \phi_{k'}^{bc} =  -(k+k')(\eta^{ab} \phi_{k'-1}^c +
\eta^{ac} \phi_{k'-1}^b - \frac{2}{d-2}\eta^{bc} \phi_{k'-1}^a)
\nonumber\\
&& \hspace{1.3cm} - \ \half (k+k')(k+2-k')\partial^a \phi_{k'-2}^{bc}\,,
\\
&& \delta_{_{R^a}} \phi_{k'}^b =  (k+1-k')\phi_{k'-1}^{ab} - u (k-1+k')
\eta^{ab} \phi_{k'-1}
\nonumber\\
&& \hspace{1.3cm} - \ \half (k-1+k')(k+1-k')\partial^a \phi_{k'-2}^b\,,
\\
&& \delta_{_{R^a}} \phi_{k'} =  (k-k') u \phi_{k'-1}^a - \half
(k-2+k')(k-k')\partial^a \phi_{k'-2}\,.
\eeq

\noindent {\bf iii}) To find all $R_\smG^a$ which satisfy the commutator
$[K^a,K^b]=0$ we note the relation $[K^a,K^b] =Gr_\smG^{ab}$, where $G$,
$r_\smG^{ab}$, $r^a$, $r_\smG^a$ are given in \rf{gautraspi2def01},
\rf{22052011-oldman01}, \rf{Rgsmspi201x1}, \rf{Rgsmspi204} respectively. In
other words, $[K^a,K^b]$ is proportional to operator of gauge transformations
$G$ \rf{gautraspi2def01}, as it should be in gauge theory. We see that the
requirement $[K^a,K^b]=0$ amounts to the equation $r_\smG^{ab}=0$. There are
the following two solutions to this equation:

{\bf 1st solution}:
\beq
\label{rwtg1solspi201} && \rwt_{\smG,1}(0,\Delta') = \frac{4}{\Delta' +
c_0}\,,
\qquad \qquad\qquad
\rwt_{\smG,1}(1,\Delta') = \frac{4}{\Delta' + c_1}\,,
\nonumber\\
&& \rwt_{\smG,2} = \frac{c_2}{\Delta' +  1 + c_0}\,, \qquad\qquad\qquad\quad
\rwt_{\smG,3}= 0\,,
\nonumber\\
&& \hspace{1cm}  c_2 = \frac{2\sqrt{2}}{u k (k - c_1)}((k+1) c_1 - k c_0);
\eeq

{\bf 2nd solution}:
\beq
\label{rwtg1solspi202} && \rwt_{\smG,1}(0,\Delta') = 0\,,
\qquad\qquad \qquad
\rwt_{\smG,1}(1,\Delta') = \frac{4}{\Delta' + c_3}\,,
\nonumber\\
\label{rwtg1solspi202n02} && \rwt_{\smG,2} = \frac{c_5}{\Delta' + 1 + c_4}\,,
\qquad\qquad \rwt_{\smG,3} = - \frac{2}{\Delta'  + c_4}\,, \qquad
\nonumber\\
&& \hspace{1cm}  c_5 = \frac{2\sqrt{2}}{u k (c_3-k)}(c_3 + k c_4 + k(k+2)
)\,,
\eeq
where $k$ and $u$ in \rf{rwtg1solspi201},\rf{rwtg1solspi202n02} are given in
\rf{20082011-01} and \rf{olfman19012011-09}. Quantities $c_0$, $c_1$, $c_3$,
$c_4$ are real-valued constants. Note that the simplest representative
$R_\smG^a=0$ is achieved by taking $c_0= c_1=\infty$ and $c_3 = c_4=\infty$.

We note that the Lagrangian, gauge transformations, and operator $R^a$ are
determined by imposing the same requirements we used  for the case of spin-1
conformal field (see at the end of Section \ref{sec-04-sub02}). Details of
the derivation may be found in Appendix D.

{\bf Spin-2 conformal field in 4d}. To illustrate our approach we finish
discussion in this section by considering spin-2 conformal field in $4d$.
This is the simplest conformal theory of spin-2 field involving Stueckelberg
field. For the case of spin-2 conformal field in $4d$, our field content
involves two rank-2 tensor fields and one vector field (see
\rf{covspin2DoF01},\rf{covspin2DoF02}),
\beq \label{25072011-02}
&& \phi_\smminone^{ab} \qquad\ \ \ \phi_{_1}^{ab}
\nonumber\\
&& \hspace{0.9cm} \phi_{_0}^a
\eeq
The corresponding Lagrangian \rf{olfman19012011-08} takes the form
\be \label{d4lagcon01}
\LL = \half \phi_{_1}^{ab} E_\EH\phi_\smminone^{ab} + \half \phi_{_0}^a
E_\Max \phi_{_0}^a
+  \phi_{_0}^a (\partial^b \phi_{_1}^{ab} - \partial^a \phi_{_1}^{bb} )
- \frac{1}{4} \phi_{_1}^{ab} \phi_{_1}^{ab} +  \frac{1}{4} \phi_{_1}^{aa}
\phi_{_1}^{bb}\,,
\ee
where the operators $E_\EH$, $E_\Max$ are given in \rf{25072011-01} and
\rf{manold-01112011-03}. Gauge symmetries are described by the following
gauge transformation parameters (see
\rf{listgauparspi201}):
\beq
&& \xi_\smmintwo^a \qquad\quad \xi_{_0}^a
\nonumber\\
&& \hspace{0.8cm} \xi_\smminone
\eeq
and appropriate gauge transformations are given by (see
\rf{gautraspi2def02},\rf{gautraspi2def03})
\beq
&& \delta\phi_\smminone^{ab} =\partial^a\xi_\smmintwo^b + \partial^b
\xi_\smmintwo^a + \eta^{ab}\xi_\smminone\,,
\nonumber\\
\label{25072011-03} && \delta\phi_{_1}^{ab} =\partial^a\xi_{_0}^b +
\partial^b \xi_{_0}^a\,,
\\
&& \delta\phi_{_0}^a = \partial^a\xi_\smminone  - \xi_{_0}^a\,.
\nonumber
\eeq
$R^a$ transformations taken in the minimal scheme, $R_\smG^a=0$, are given by
\beq
&& \delta_{R^a}\phi_\smminone^{bc} = 0\,,
\nonumber\\
&& \delta_{R^a} \phi_{_1}^{bc} = -2(\eta^{ab}\phi_{_0}^c +
\eta^{ac}\phi_{_0}^b) + 2 \eta^{bc}\phi_{_0}^a
-2\partial^a\phi_\smminone^{bc}\,,
\\
&& \delta_{R^a}\phi_{_0}^b = 2\phi_\smminone^{ab}\,.
\nonumber
\eeq
From \rf{25072011-03}, we see that the vector field $\phi_{_0}^a$ transforms
as Stueckelberg field under the gauge transformations, i.e., this field can
be gauged away by using $\xi_{_{0}}^a$ gauge symmetries. Gauging away the
vector field, $\phi_{_0}^a = 0$, we see that our Lagrangian \rf{d4lagcon01}
takes the simplified form
\be \label{2man29112010-06}
\LL  = -\phi_{_1}^{ab} G_\lin^{ab} - \frac{1}{4}\phi_{_1}^{ab} \phi_{_1}^{ab}
+ \frac{1}{4} \phi_{_1}^{aa}\phi_{_1}^{bb}\,.
\ee
Here we use  the well-known relations for the linearized Einstein tensor
$G_\lin^{ab}$ and operator $E_\EH$ \rf{25072011-01},
\be \label{2m17122010-01}
G_\lin^{ab} = -\half E_\EH\phi_\smminone^{ab} \,, \qquad G_\lin^{ab} \equiv
R_\lin^{ab}(\phi_\smminone) -\half \eta^{ab} R_\lin(\phi_\smminone)\,,
\ee
where the linearized Ricci curvatures for the field $\phi_\smminone^{ab}$ in
\rf{2m17122010-01} are obtained by substituting the field
$\phi_\smminone^{ab}$ in the respective expressions in \rf{2m04122010-06}.
Using equations of motion for the rank-2 tensor field $\phi_{_1}^{ab}$
obtained from Lagrangian \rf{2man29112010-06}, $\phi_{_1}^{ab} -
\eta^{ab}\phi_{_1}^{cc} = - 2G_\lin^{ab}$, we obtain the following  solution
for $\phi_{_1}^{ab}$:
\be \label{2man29112010-06a2} \bar\phi_{_1}^{ab} = - 2 R_\lin^{ab} +
\frac{1}{3}\eta^{ab} R_\lin\,. \ee
Plugging solution $\bar\phi_{_1}^{ab}$ \rf{2man29112010-06a2} into Lagrangian
\rf{2man29112010-06}, we obtain the higher-derivative Lagrangian given in
\rf{2m04122010-02}. Thus, we see that our ordinary-derivative approach is
equivalent to the standard one and our field $\phi_\smminone^{ab}$ is
identified with the excitation of conformal graviton field $\phi^{ab}$ in
Sec. \ref{sec-05-sub01}.

\newsection{ Ordinary-derivative Lagrangian of interacting
$4d$ conformal gravity}\label{sec-06}

We begin our discussion of interacting theory of $4d$ conformal gravity with
the description of a field content. Field content of the interacting theory
is simply obtained by promoting the Minkowski space free fields
\rf{25072011-02} to the fields in curved space-time described by metric
tensor field $g_{\mu\nu}$. As usual, this metric tensor field is considered
to be the conformal graviton field. As we have already said, the field
$\phi_\smminone^{ab}$ describes excitation of the conformal graviton, i.e.,
in the interacting theory, the field $\phi_\smminone^{ab}$ is related to the
metric tensor field $g_{\mu\nu}$. In place of the free field
$\phi_{_1}^{ab}$, we will use symmetric rank-2 tensor field
$\varphi_{\mu\nu}$. Also, following commonly used nomenclature, we use
notation $b_\mu$ in place of the field $\phi_{_0}^a$. To summarize, the field
content we use to develop the ordinary-derivative approach to the interacting
$4d$ conformal gravity is
given by%
\beq \label{2m04122010-05}
& g_{\mu\nu} \qquad \varphi_{\mu\nu} &
\nonumber\\
& b_\mu &
\eeq
Ordinary-derivative Lagrangian we found takes the following form:
\beq \label{oldman19012011-12}
e^{-1} g_\W^2 \LL  & = &  - \varphi_{\mu\nu} \Gwh^{(\mu\nu)} -
\frac{1}{4}F_{\mu\nu}F^{\mu\nu} - \frac{1}{4} \varphi_{\mu\nu}
\varphi^{\mu\nu} + \frac{1}{4} \varphi_\mu^\mu \varphi_\nu^\nu\,,
\\
&& \Gwh^{(\mu\nu)} \equiv R^{\mu\nu} + \half (D^\mu b^\nu + D^\nu b^\mu +
b^\mu b^\nu)
- \half g^{\mu\nu} ( R + 2Db - \half b^2 )\,,
\\
\label{oldman19012011-12a2} && F_{\mu\nu}\ \equiv \ \partial_\mu b_\nu -
\partial_\nu b_\mu\,,
\eeq
where $Db \equiv D_\mu b^\mu$, $D_\mu b^\nu =\partial_\mu b^\nu +
\Gamma_{\mu\rho}^\nu b^\rho$, $b^2 \equiv b_\mu b^\mu$, $e\equiv \sqrt{-\det
g_{\mu\nu}}$ and $g_\W$ stands for dimensionless coupling constant. Our
curvature conventions are $R^\lambda{}_{\mu\nu\rho} =
\partial_\nu\Gamma_{\mu\rho}^\lambda-\ldots$,
$R_{\mu\nu}=R^\lambda{}_{\mu\lambda\nu}$, $R=R_\mu^\mu$. The derivation of
Lagrangian \rf{oldman19012011-12} may be found at the end of this Section.

The Lagrangian \rf{oldman19012011-12} is invariant under the  standard
diffeomorphism transformations. Also, the Lagrangian \rf{oldman19012011-12}
is invariant under the conformal boost and Weyl gauge transformations given
by
\beq
\label{25072011-07} && \delta g_{\mu\nu}= - 2\sigma g_{\mu\nu}\,,
\\
\label{25072011-08} && \delta \varphi_{\mu\nu}  = D_\mu \xi_\nu^{\rm K}  +
D_\nu \xi_\mu^{\rm K} + b_\mu \xi_\nu^{\rm K} + b_\nu \xi_\mu^{\rm K} -
g_{\mu\nu} b^\rho \xi_\rho^{\rm K} \,,
\\
\label{25072011-09} && \delta b_\mu = \partial_\mu \xi^{\rm D} - \xi_\mu^{\rm
K}\,,\qquad \xi^{\rm D} \equiv - 2\sigma\,,
\eeq
where $\sigma$ and $\xi_\mu^{\rm K}$ are parameters of the respective Weyl
and conformal boost gauge transformations. From \rf{25072011-07} and
\rf{25072011-09}, we see that Weyl dimensions of $g_{\mu\nu}$ and
$\varphi_{\mu\nu}$ are given by $\Delta_{_{\rm W}}(g_{\mu\nu})=-2$,
$\Delta_{_{\rm W}}(\varphi_{\mu\nu})=0$. From \rf{25072011-09}, we see that
the vector field $b_\mu$ transforms as Stueckelberg field.

Before we proceed, let us
 note how the standard higher-derivative Weyl Lagrangian
is obtained from our Lagrangian \rf{oldman19012011-12}. Using Lagrangian
\rf{oldman19012011-12}, we consider equations of motion for the field
$\varphi_{\mu\nu}$, $\partial_{\varphi_{\mu\nu}}\LL=0$, which allow us to
solve the field $\varphi_{\mu\nu}$ in terms of the remaining fields,
\be \label{sec0145a1}  \varphi_{\mu\nu} = - 2\Gwh_{(\mu\nu)}
+\frac{2}{3}g_{\mu\nu} \Gwh_{(\sigma\lambda)} g^{\sigma\lambda}\,.
\ee
Plugging \rf{sec0145a1} into Lagrangian \rf{oldman19012011-12}, we obtain the
standard higher-derivative Weyl Lagrangian, which can be represented in two
equivalent (up to total derivative) well-known forms
\be
\label{sec0147} \LL = \frac{e}{g_\W^2} (R_{\mu\nu}R^{\mu\nu} - \frac{1}{3}
R^2)\,,
\qquad\quad
\LL =  \frac{e}{2g_\W^2} C_{\mu\nu\sigma\lambda}C^{\mu\nu\sigma\lambda}\,,
\ee
where $C^{\mu\nu\sigma\lambda}$ is the Weyl tensor. We note that overall sign
of Lagrangian we use in \rf{oldman19012011-12} leads to the commonly used
overall sign of Weyl Lagrangian \rf{sec0147}.

{\bf Pure Einstein AdS gravity from Weyl gravity}. As is well known,
classical equations  of the Weyl gravity  admit Einstein spaces as special
solutions, so in that sense the Einstein gravity  can be viewed  as a
particular solution of the  Weyl conformal gravity theory that
breaks the conformal gauge symmetry  of the  latter.%
\footnote{ Recent interesting discussion of a relation between  the  Einstein
and Weyl gravities utilizing AdS background may be found in
Ref.\cite{Maldacena:2011mk}.}
Here, using our ordinary-derivative approach, we would like to discuss the
relation between the  Einstein and Weyl gravities via breaking conformal
gauge symmetries. Let us first discuss the case of pure Einstein gravity. To
this end we use $\xi_\mu^{\rm K}$ gauge symmetry \rf{25072011-09} to gauge
away the vector field, $b_\mu = 0$, and note that equations of motion for the
metric field $g_{\mu\nu}$ and the auxiliary tensor field $\varphi_{\mu\nu}$
obtained from Lagrangian \rf{oldman19012011-12} have the following solution:
\beq
\label{25072011-05a0} && \bar{g}_{\mu\nu} = \hbox{ metric of (A)dS space}\,,
\\
\label{25072011-05} && \bar\varphi_{\mu\nu} = - 2\rho \bar{g}_{\mu\nu}\,,
\eeq
where our convention for the curvature of (A)dS space is as follows
\be \bar{R}_{\mu\nu\sigma\lambda} = \rho(
\bar{g}_{\mu\sigma}\bar{g}_{\nu\lambda} - \bar{g}_{\mu\lambda}
\bar{g}_{\nu\sigma})\,, \qquad
\rho = \left\{\begin{array}{cl}
-\frac{1}{r^2} & \hbox{for AdS space}\,,
\\[5pt]
+\frac{1}{r^2} & \hbox{for dS space}\,,
\end{array}\right.
\ee
and $r$ is a radius of (A)dS space. Motivated by the relation
\rf{25072011-05} we suggest the following way for breaking the conformal
gauge symmetries. Namely, we propose to eliminate the auxiliary tensor field
$\varphi_{\mu\nu}$
from the  Lagrangian \rf{oldman19012011-12} by using the relation%
\footnote{ To our knowledge, for the first time, solution \rf{25072011-05a0},
\rf{25072011-05} and the use of constraint \rf{25072011-06} for the
derivation of Einstein AdS gravity from Weyl conformal gravity were discussed
in Ref.\cite{metdubna} (see pages 51-54 in Ref.\cite{metdubna}).}
\be \label{25072011-06}  \varphi_{\mu\nu} = - 2\rho g_{\mu\nu}\,, \ee
where $g_{\mu\nu}$ in r.h.s \rf{25072011-06} is arbitrary metric tensor.
Obviously, the relation \rf{25072011-06} does not respect
\rf{25072011-07}-\rf{25072011-09}, i.e  the use of this relation breaks the
conformal gauge symmetries. Plugging \rf{25072011-06} into Lagrangian
\rf{oldman19012011-12} and assuming the gauge condition $b_\mu=0$, we find
\be \label{25072011-10} e^{-1}\LL = -\frac{2\rho}{g_\W^2} (R - 6 \rho) \,.\ee
Comparing this with the Einstein-Hilbert Lagrangian for (A)dS gravity
\be \label{25072011-11} e^{-1}\LL_\EH = \frac{1}{2\kappa^2} (R - 6 \rho)
\,,\ee
we find the relation
\be \label{25072011-12} g_\W^2 = - 4\rho \kappa^2 \,,\ee
which implies that
\be \label{25072011-14} \rho < 0 \,,\ee
i.e., we are left with Einstein AdS gravity. As a side  remark, we note that
it is the commonly used overall sign of the Weyl gravity \rf{sec0147} that
leads to Einstein AdS  gravity.%
\footnote{ It will be interesting to understand the possibility of getting
the massless higher-spin $AdS_4$ fields theory \cite{Vasiliev:1990en} from
the conformal higher-spin fields theory via breaking conformal symmetries.
Recent interesting discussion of conformal symmetries of the massless
higher-spin $AdS_4$ fields theory may be found in
Ref.\cite{Vasiliev:2007yc}.}

{\bf Einstein AdS gravity with massive vector field from Weyl gravity}. Our
approach allows us to obtain Einstein AdS gravity interacting with massive
vector field from Weyl gravity. To this end we note that the vector field
$b_\mu$ enters ordinary-derivative Lagrangian \rf{oldman19012011-12} with
correct sign of the kinetic term. It seems therefore reasonable to keep this
vector field in the procedure of breaking conformal gauge symmetries. Namely,
as before, we use constraint \rf{25072011-06} to break conformal gauge
symmetries and plug this constraint in ordinary-derivative Lagrangian
\rf{oldman19012011-12}. Doing so, we get
\be \label{25072011-15} e^{-1}\LL = -\frac{2\rho}{g_\W^2} (R - 6 \rho) +
\frac{1}{g_\W^2}(-\frac{1}{4}F_{\mu\nu}F^{\mu\nu} + 3\rho b_\mu b^\mu)\,.\ee
As before, we see that we have to impose the relations
\rf{25072011-12},\rf{25072011-14}, i.e., we are left with Einstein AdS
gravity interacting with massive vector field $b_\mu$. As seen from
\rf{25072011-15}, a mass parameter of the vector field is given by
\be \label{26072011-01} m^2 = - 6 \rho\,.\ee
We see that restriction \rf{25072011-14} leads not only to the correct sign
of kinetic terms for the graviton field but also to the correct mass term,
$m^2 > 0$, for the vector field $b_\mu$. For the reader convenience we note
that lowest value of the energy operator for vector field with mass parameter
as in \rf{26072011-01} is given by $ E_0(b_\mu) = \frac{4}{r}$.%
\footnote{ Interrelation between $E_0$ and mass parameter for arbitrary spin
AdS field is given in formula (5.74) in Ref.\cite{Metsaev:2003cu}.}
Recall that, for the graviton field in $AdS_4$ background, energy lowest
value is given by $E_0(g_{\mu\nu}) = \frac{3}{r}$.

{\bf Derivation of ordinary-derivative Lagrangian}. We now discuss details of
the derivation of Lagrangian \rf{oldman19012011-12}. The Lagrangian could be
derived by using method in Ref.\cite{Metsaev:2010kp}. We note however that
derivation of interacting $4d$ conformal gravity can be streamlined by using
the gauge invariant formulation of $4d$ conformal gravity given in
Ref.\cite{Kaku:1977pa}. Field content in Ref.\cite{Kaku:1977pa} involves
vielbein $e_\mu^a$, compensator $b_\mu$ and conformal boost gauge field
$f_\mu{}^a$. Gauge invariant Lagrangian for these fields  takes the form
\beq
\label{27072011-01}  \LL & = &  -\frac{1}{8g_\W^2}
\RR_{\mu\nu}{}^{ab}\RR_{\sigma\lambda}{}^{ce}\epsilon^{abce}\epsilon^{\mu\nu\sigma\lambda}\,,
\\
\label{27072011-02} && \RR_{\mu\nu}{}^{ab} = R_{\mu\nu}{}^{ab} + e_\mu^a
B_\nu{}^b - e_\nu^a B_\mu{}^b + e_\nu^b B_\mu{}^a - e_\mu^b B_\nu{}^a\,,
\qquad B_\mu{}^a = B_\mu{}^\nu e_\nu^a\,,\qquad
\\
\label{23072011-10} && B_\nu{}^\mu \equiv qD_\nu b^\mu + q^2 b_\nu b^\mu -
\half q^2 b^2 \delta_\nu^\mu + q f_\nu{}^\mu \,, \qquad \quad q =\half\,,
\eeq
where we introduce the target space fields $g_{\mu\nu}=e_\mu^a e_\nu^a$,
$b_\mu = b^a e_\mu^a$, $f_\mu{}^\nu = f_\mu{}^a e^{\nu a}$. From now on, we
use field content involving metric tensor $g_{\mu\nu}$, vector field $b_\mu$
and auxiliary tensor field $f_\mu{}^\nu$. Note that the curvature
$R_{\mu\nu}{}^{ab}$ in r.h.s \rf{27072011-02} is related to the Riemann
curvature in a standard way $R_{\mu\nu\sigma\lambda} = e_\sigma^a e_\lambda^b
R_{\mu\nu}{}^{ab}$. Also, note that field $f_{\mu\nu}=f_\mu{}^\rho
g_{\rho\nu}$ is not symmetric $f_{\mu\nu}\ne f_{\nu\mu}$. Ignoring the
Gauss-Bonnet contribution, Lagrangian \rf{27072011-01} can be represented as
\beq \label{23072011-09}
e^{-1} g_\W^2 \LL & = & - 4 R_\mu^\nu B_\nu{}^\mu + 2 RB -4 B_\mu{}^\nu
B_\nu{}^\mu + 4 B^2\,, \qquad B \equiv B_\mu{}^\mu\,.
\eeq
Lagrangian \rf{23072011-09} is invariant under the gauge transformations
\beq
&& \delta g_{\mu\nu}= 2q \xi^{\rm D} g_{\mu\nu}\,,
\\
&& \delta f_\mu{}^\nu  = D_\mu \xi^{{\rm K}\, \nu}  + q ( b_\mu\xi^{{\rm
K}\,\nu} + b^\nu \xi_\mu^{\rm K} - \xi_\lambda^{\rm K} b^\lambda
\delta_\mu^\nu) - 2q\xi^{\rm D} f_\mu{}^\nu \,,
\\
&& \delta b_\mu = \partial_\mu \xi^{\rm D} - \xi_\mu^{\rm K}\,.
\eeq
Before deriving of ordinary-derivative Lagrangian we recall how the standard
higher-derivative Lagrangian is obtained from Lagrangian \rf{23072011-09}. To
this end we consider equations of motion for the field $f_\mu{}^\nu$,
$\partial_{f_\mu{}^\nu}\LL=0$, which allow us to solve the field
$f_\mu{}^\nu$ in terms of the remaining fields,
\be \label{sec0145}  B_\mu{}^\nu = -\half(R_\mu^\nu -\frac{1}{6}
\delta_\mu^\nu R)\,.
\ee
Plugging this solution into Lagrangian \rf{23072011-09}, we obtain the
standard higher-derivative Weyl Lagrangian given in \rf{sec0147}.

We now turn to the derivation of the ordinary-derivative Lagrangian. To this
end we introduce the decomposition of the tensor field $f_{\mu\nu} \equiv
f_\mu{}^\rho g_{\rho\nu}$ into symmetric and antisymmetric tensor fields,
\be\label{23072011-11} f_{\mu\nu}  =  \half \varphi_{\mu\nu} + \half
(f_{\mu\nu} - f_{\nu\mu})\,, \qquad \quad\varphi_{\mu\nu} \equiv f_{\mu\nu}+
f_{\nu\mu}\,.\ee
Plugging \rf{23072011-11} into \rf{23072011-10}, we obtain the appropriate
decomposition for field $B_{\mu\nu} \equiv B_\mu{}^\rho g_{\rho\nu}$,
\beq
\label{23072011-12} &&  B_{\mu\nu} = \half B_{\mu\nu}^\sym + \half
B_{\mu\nu}^\as\,,
\\
\label{23072011-15} && B_{\mu\nu}^\sym = q \varphi_{\mu\nu} + Q_{\mu\nu}
\,,\qquad B_{\mu\nu}^\as \equiv q( f_{\mu\nu} - f_{\nu\mu}) + q F_{\mu\nu}\,,
\\
\label{23072011-16} && Q_{\mu\nu} \equiv q (D_\mu b_\nu +  D_\nu b_\mu ) +
2q^2 b_\mu b_\nu - q^2 b^2 g_{\mu\nu}\,,\qquad Q_\mu^\mu = 2 q D b - 2 q^2
b^2 \,,\qquad
\eeq
where $F_{\mu\nu}$ is defined in \rf{oldman19012011-12a2}.  Plugging
$B_{\mu\nu}$ \rf{23072011-12} into Lagrangian \rf{23072011-09}, we  obtain
\be \label{23072011-08} e^{-1} g_\W^2 \LL = -2 B_{\mu\nu}^\sym R^{\mu\nu} +
B^\sym R - B_{\mu\nu}^\sym B^{\sym\,\mu\nu} + B^\sym B^\sym + B_{\mu\nu}^\as
B^{\as\, \mu\nu}\,,\ee
where $B^\sym \equiv B_{\mu\nu}^\sym g^{\mu\nu}$. Equations of motion for
$B_{\mu\nu}^\as$ give $B_{\mu\nu}^\as = 0$. Using this in \rf{23072011-08},
we get
\be \label{23072011-14} e^{-1} g_\W^2 \LL = -2 B_{\mu\nu}^\sym R^{\mu\nu} +
B^\sym R - B_{\mu\nu}^\sym B^{\sym\,\mu\nu} + B^\sym B^\sym\,. \ee
Plugging $B_{\mu\nu}^\sym$ \rf{23072011-15},\rf{23072011-16} into
\rf{23072011-14}, we get ordinary-derivative Lagrangian
\rf{oldman19012011-12}. Note that the relation $B_{\mu\nu}^\as=0$ and the 2nd
relation in \rf{23072011-15} imply the relation $f_{\mu\nu}-f_{\nu\mu}=
-F_{\mu\nu}$.

\newsection{Spin-$\frac{1}{2}$ conformal fermionic Dirac field}\label{sec-07}

As a warm up we start with fermionic spin-$\frac{1}{2}$ Dirac field. To make
contact with studies in literature we begin with presentation of the
standard higher-derivative formulation for the fermionic field.

\subsection{ Higher-derivative
formulation of spin-$\frac{1}{2}$ conformal fermionic
field}\label{sec-07-sub01}

In the framework of the standard approach, spin-$\frac{1}{2}$ conformal
complex-valued non-chiral Dirac field $\psi$ propagating in flat space of
arbitrary dimension $d$ is described by the Lagrangian,%
\footnote{ Chiral fermionic field can be introduced by allowing the field
$\psi$ to be positive-chirality (or negative-chirality).}
\be\label{ferstandlag01} {\rm i}\LL = \bar\psi \Box^k \parline\psi\,,\qquad
\bar\psi=\psi^\dagger\gamma^0\,, \qquad  k \hbox{ arbitrary positive
integer}\,. \ee
Here and below the spinor indices are implicit. For $k=0$ ($k\geq 1$),
Lagrangian \rf{ferstandlag01} describes field associated with unitary
(non-unitary) representation of the conformal algebra $so(d,2)$. The field
$\psi$ has the conformal dimension
\be \label{fercondim0001} \Delta_\psi = \frac{d-1}{2} - k\,.\ee

\subsection{ Ordinary-derivative formulation of spin-$\frac{1}{2}$
conformal fermionic field}\label{sec-07-sub02}

{\bf Field content}. In the framework of ordinary-derivative approach, a
dynamical system that on-shell equivalent to the single non-chiral Dirac
field $\psi$ with Lagrangian \rf{ferstandlag01} and conformal dimension
\rf{fercondim0001} is described by $2k+1$ non-chiral spin-$\half$ Dirac fields,%
\footnote{ Chiral fermionic fields can be introduced by allowing the fields
in the1st line in \rf{ferscafiecol01} to be positive-chirality (or
negative-chirality), while the fields in the 2nd line in \rf{ferscafiecol01}
to be negative-chirality (or positive-chirality).}%
\beq
&& \label{ferscafiecol01}
\psi_{k'}\,,\qquad k' \in [k]_2\,,
\nonumber\\[-14pt]
&&  \hspace{6cm} k \hbox{ arbitrary positive integer}\,.
\\[-14pt]
&& \psi_{k'}\,,\qquad k' \in [k-1]_2\,,
\nonumber
\eeq
The conformal dimensions of fields $\psi_{k'}$ \rf{ferscafiecol01} are given
by:
\be \label{delfer1-2def01} \Delta_{\psi_{k'}} = \frac{d-1}{2} + k'\,.\ee

In order to obtain the Lagrangian description of fields \rf{ferscafiecol01}
in an easy--to--use form we use oscillators $\upsilon^\oplussm$,
$\upsilon^\ominussm$ \rf{manold-31102011-01} and collect fields
\rf{ferscafiecol01} into the ket-vector $\psik$ defined by
\be \label{psidef0001} \psik  = \left(
\begin{array}{l}
|\psi_\u \rangle
\\[5pt]
|\psi_\d\rangle \end{array} \right)\,,
\ee

\vspace{-0.3cm}
\beq \label{Ppsi0def01}
|\psi_\u\rangle  & \equiv &  \sum_{k'\in [k]_2} \frac{1}{(\frac{k+k'}{2})!}}
(\upsilon^\oplussm)^{^{\frac{k+k'}{2}}}
(\upsilon^\ominussm)^{^{\frac{k-k'}{2}}} \, \psi_{k' |0\rangle\,,
\\
\label{Ppsi0def02} |\psi_\d\rangle & \equiv  & \sum_{k'\in [k-1]_2} \frac{1}{
(\frac{k-1+k'}{2})!} (\upsilon^\oplussm)^{^{\frac{k-1+k'}{2}}}
(\upsilon^\ominussm)^{^{\frac{k-1-k'}{2}}} \, \psi_{k'}|0\rangle\,,
\eeq
where a bra-vector $\psibr$ is defined according the rule $\psibr =
(\psik)^\dagger\gamma^0$. Ket-vector $\psik$ \rf{psidef0001} satisfies the
following algebraic constraints:
\be
\label{algconfer01} (N_\upsilon - k )\pi_+ \psik = 0 \,,
\qquad \quad
(N_\upsilon  - k+1)\pi_- \psik =0 \,,
\ee
where matrices $\pi_\pm$ are defined in \rf{manold-03112011-01}. Constraints
\rf{algconfer01} tell us that ket-vectors $|\psi_\u\rangle$ \rf{Ppsi0def01}
and $|\psi_\d\rangle$ \rf{Ppsi0def02} are the respective degree-$k$ and $k-1$
homogeneous polynomials in the oscillators $\upsilon^\oplussm$,
$\upsilon^\ominussm$.

{\bf Lagrangian}. Lagrangian can entirely be presented in terms of the
ket-vector $\psik$,
\be
\label{oldman-29012011-01} {\rm i}\LL = \psibr E \psik\,,
\qquad\quad
E \equiv \parline + e_1^\smGamma \,,
\qquad \quad e_1^\smGamma \equiv  \bar\upsilon^\ominussm \sigma_- +
\upsilon^\ominussm \sigma_+ \,.
\ee
For the reader convenience we note that, in terms of component fields
\rf{ferscafiecol01}, Lagrangian \rf{oldman-29012011-01} takes the form
\be \label{20082011-oldman-01} {\rm i}\LL = \sum_{k'\in [k]_1}
\bar\psi_{-k'}
\parline \psi_{k'} + \bar\psi_{-k'} \psi_{k'+1}\,.  \ee
To illustrate \rf{20082011-oldman-01}, we note that, for $k=1,2$, Lagrangian
\rf{20082011-oldman-01} takes the form
\beq
{\rm i}\LL_{k=1} & = & \bar\psi_{_{-1}}\parline \psi_{_1} +\bar\psi_{_1}
\parline \psi_{_{-1}} + \bar\psi_{_0}\parline \psi_{_0}
\nonumber\\
& + &  \bar\psi_{_0}\psi_{_1} + \bar\psi_{_1}\psi_{_0}\,,
\\
{\rm i}\LL_{k=2} & =  & \bar\psi_{_{-2}}\parline \psi_{_2} +\bar\psi_{_2}
\parline \psi_{_{-2}} + \bar\psi_{_{-1}}\parline \psi_{_1} + \bar\psi_{_1}\parline
\psi_{_{-1}} + \bar\psi_{_0}\parline \psi_{_0}
\nonumber\\
& +  & \bar\psi_{_{-1}}\psi_{_2} + \bar\psi_{_2}\psi_{_{-1}} +
\bar\psi_{_0}\psi_{_1} + \bar\psi_{_1}\psi_{_0} \,.
\eeq

We now make comment on the interrelation of ordinary-derivative Lagrangian
\rf{20082011-oldman-01} and higher-derivative Lagrangian \rf{ferstandlag01}.
Noticing that the field $\psi_{-k}$ has the same conformal dimension as the
field $\psi$ entering the higher-derivative approach \rf{fercondim0001}, we
use the identification $\psi_{-k} = \psi$. The remaining fields $\psi_{k'}$,
$k'=-k+1,-k+2,\ldots k-1,k$, are auxiliary fields. Using equations of motion
for the auxiliary fields obtained from \rf{20082011-oldman-01},
$\parline\psi_{k'} + \psi_{k'+1}=0$, we can express all auxiliary fields in
terms of $\psi_{-k}\equiv\psi$,
\be \label{manold-02112011-03} \psi_{k'} = (-\parline)^{k+k'}\psi\,,\qquad
k'=-k+1,-k+2,\ldots, k-1,k\,. \ee
Plugging \rf{manold-02112011-03} into \rf{20082011-oldman-01}, we obtain
higher-derivative Lagrangian \rf{ferstandlag01}, i.e., our approach and
higher-derivative approach are equivalent.

\bigskip
{\bf Conformal symmetries}. To complete the ordinary-derivative description
of the spin-$\frac{1}{2}$ field we provide realization of the conformal
symmetries on space of $\psik$ \rf{psidef0001}. All that is required is to
find the operators $M^{ab}$, $\Delta$ and $R^a$ for the case of
spin-$\frac{1}{2}$ conformal fermionic field and plug these operators in
\rf{conalggenlis01}-\rf{conalggenlis04}. The realization of the spin operator
of the Lorentz algebra and conformal dimension operator  $\Delta$ on space of
$\psik$ \rf{psidef0001} is given by
\beq
&& M^{ab} = \half \gamma^{ab}\,, \qquad \Delta  =  \frac{d-1}{2}+\Delta'\,,
\qquad \Delta' \equiv  N_{\upsilon^\oplussm} - N_{\upsilon^\ominussm}\,,
\eeq
where the realization of the conformal dimension operator $\Delta$ on space
of $\psik$ can be read from \rf{delfer1-2def01}. Requiring the Lagrangian to
be invariant under conformal boost transformations, we find the realization
of the operator $R^a$ on space of $\psik$,
\beq
\label{Rgsmspi1-201} R^a & = & r^a +  R_\smE^a\,,
\\
&& r^a = r_{0,1}^\smGamma \gamma^a + r_{1,1} \partial^a\,,
\\
\label{Rgsmspi1-204} && R_\smE^a  =   r_{\smE,1}\gamma^a   E\,,
\\
\label{Rgsmspi1-205} && \qquad \ \ r_{0,1}^\smGamma = \bar\upsilon^\oplussm
\sigma_- - \upsilon^\oplussm \sigma_+ \,, \qquad  r_{1,1} =
-2\upsilon^\oplussm \bar\upsilon^\oplussm \,,
\\
\label{Rgsmspi1-207} && \qquad \ \ r_{\smE,1} = \upsilon^\oplussm
\rwt_{\smE,1 - } \bar\upsilon^\oplussm \pi_- +  \upsilon^\oplussm
\rwt_{\smE,1 + } \bar\upsilon^\oplussm \pi_+\,,
\\
\label{Rgsmspi1-208} && \qquad \ \ \rwt_{\smE,1 \pm  }  = \rwt_{\smE,1 \pm
}(\Delta')\,, \qquad \ \ \rwt_{\smE,1 \pm  }^\dagger   = \rwt_{\smE,1 \pm
}\,,
\eeq
where $E$ appearing in \rf{Rgsmspi1-204} is given in \rf{oldman-29012011-01}.
Quantities $\rwt_{\smE,1 \pm }$ \rf{Rgsmspi1-208}, subject to hermicity
condition \rf{Rgsmspi1-208}, are arbitrary functions of the operator
$\Delta'$.

A few remarks are in order.

\noindent {\bf i}) $r^a$ part of the operator $R^a$ is fixed uniquely, while
$R_\smE^a$ part, is still to be arbitrary. Reason of the arbitrariness in
$R_\smE^a$ is obvious. Global transformations of fermionic fields are defined
up to the contribution $\tau E\psik$, where $\tau$ is an arbitrary operator
satisfying the hermitian conjugation condition $\tau^\dagger = \gamma^0
\tau\gamma^0$. This condition, in view of \rf{Rgsmspi1-208}, is respected by
the operator $r_{\smE,1} \gamma^a $ entering  $R_\smE^a$ \rf{Rgsmspi1-204}.

\noindent {\bf ii)} We check that $R^a$ transformations with $R_\smE^a=0$
satisfy the commutator $[K^a,K^b]=0$. In terms of the component fields, $R^a$
transformations with $R_\smE^a=0$ take the form%
\beq
&&\hspace{-1.3cm}  \delta_{_{R^a}} \psi_{k'} =  -\frac{k+k'}{2}\gamma^a
\psi_{k'-1} - \ \half (k+k')(k+2-k')\partial^a \psi_{k'-2}\,, \hspace{2cm}
k'\in [k]_2\,,
\\
&& \hspace{-1.3cm} \delta_{_{R^a}} \psi_{k'} =  \frac{k+1-k'}{2}\gamma^a
\psi_{k'-1}  - \ \half (k-1+k')(k+1-k')\partial^a \psi_{k'-2}^b\,,\qquad
k'\in [k-1]_2\,.
\eeq

\noindent {\bf iii}) To find all $R_\smE^a$ which satisfy the commutator
$[K^a,K^b]=0$ we note the relation $[K^a,K^b] = r_\smE^{ab} E$, where
$r_\smE^{ab}$ is given below in \rf{20082011-oldman-02}. The requirement
$[K^a,K^b]=0$ amounts to the equation $r_\smE^{ab}=0$. Solution to this
equation is given by:
\beq
&& \rwt_{\smE,1+} = c_{1+} \,, \qquad    \hbox{ for } \ \ k =1;
\qquad\qquad
\rwt_{\smE,1\pm } = 0\,, \quad  \hbox{ for } \  \ k \geq 2\,,
\eeq
where $c_{1+}$ is a real-valued constant. Note that, for $k =1$,
$\rwt_{\smE,1-} \equiv 0$.

\newsection{ Spin-$\frac{3}{2}$ conformal fermionic Dirac field}\label{sec-08}

We now extend our discussion of fermionic fields to the case of
spin-$\frac{3}{2}$ conformal fermionic field. As before to make contact with
studies in earlier literature we start with the presentation of the standard
higher-derivative formulation for the fermionic spin-$\frac{3}{2}$ Dirac
field. In due course, we present two new representations for
higher-derivative Lagrangian and review our result concerning the counting of
on-shell D.o.F for spin-$\frac{3}{2}$ conformal field.

\subsection{ Higher-derivative formulation of spin-$\frac{3}{2}$ conformal
fermionic Dirac field}\label{sec-08-sub01}

In the framework of the standard approach, spin-$\frac{3}{2}$ conformal
fermionic Dirac field propagating in flat space of arbitrary dimension $d\geq
4$ is described by the Lagrangian
\be \label{3-2counDoFt03} {\rm i} \LL =  \bar\psi^a \Box^k P_{3/2}^{ab}
\parline \psi^b \,,\qquad k \equiv \frac{d-2}{2}\,, \ee
where $\psi^a$ stands for non-chiral vector-spinor complex-valued
Dirac field.%
\footnote{ Chiral spin-$\frac{3}{2}$ field can be introduced by allowing the
field $\psi^a$ to be positive-chirality (or negative-chirality).}
The spinor indices of the field $\psi^a$ are implicit. The field $\psi^a$ has
the conformal dimension $\Delta_{\psi^a}=1/2$. The operator $P_{3/2}^{ab}$
and its basic properties are as follows
\cite{Fradkin:1985am}:%
\footnote{ In Ref.\cite{Fradkin:1985am}, operator $P_{3/2}^{ab}$ was given
for $d=4$. Generalization to arbitrary $d$ is straightforward. All that is
required is to respect relations \rf{3-2counDoFt04} and the hermicity
condition $({\rm i}\gamma^0 P_{3/2}^{ab}\parline)^\dagger = {\rm i}\gamma^0
P_{3/2}^{ba}\parline$.}
\beq
 \label{3-2counDoFt04} && P_{3/2}^{ab} \equiv \pi^{ab} - \frac{1}{d-1}
\pi^{ac}\pi^{be}\gamma^{c} \gamma^e\,,
\qquad \quad
\pi^{ab} \equiv \eta^{ab} - \frac{\partial^a\partial^b}{\Box} \,,
\nonumber\\[-12pt]
&&
\\[-12pt]
\label{3-2counDoFt06} && \partial^a P_{3/2} ^{ab}=0\,,\qquad
 P_{3/2} ^{ab} \partial^b = 0 \,,
\qquad
\gamma^a P_{3/2} ^{ab}=0\,,\qquad P_{3/2} ^{ab} \gamma^b = 0 \,.
\nonumber
\eeq
Lagrangian \rf{3-2counDoFt03} is invariant under the gauge transformations
\be \label{3-2gaugsym01}
\delta \psi^a = \partial^a \xi +  \gamma^a \lambda\,,
\ee
where $\xi$ and $\lambda$ are gauge transformations parameters.

For Lagrangian \rf{3-2counDoFt03}, we find two new representations, which, to
our knowledge, have not been discussed in earlier literature,
\beq
\label{10082011-01} && \hspace{-1cm} {\rm i}\LL = \bar{F}^a \gamma^{abc}
\Box^{k-1}\partial^b F^c\,,
\qquad \quad {\rm i}\LL = \bar{W}^a \Box^{k-2} \gamma^{abc}
\partial^b W^c\,,
\\
\label{10082011-03} && \hspace{1cm} F^{ab} \equiv \partial^a\psi^b
-\partial^b \psi^a \,,
\\
\label{10082011-04} &&  \hspace{1cm} F^a \equiv \gamma^b F^{ab} -
\frac{1}{2(d-1)} \gamma^a \gamma^{bc} F^{bc}\,,
\\
\label{10082011-05} &&  \hspace{1cm} W^a \equiv \gamma^b \parline F^{ab} -
\frac{1}{2(d-1)} \gamma^a \gamma^{bc} \parline F^{bc}\,.
\eeq
We note that field strengths $F^a$ \rf{10082011-04}, $W^a$ \rf{10082011-05}
are manifestly  invariant under $\xi$-gauge transformation \rf{3-2gaugsym01},
while, under $\lambda$-gauge transformation, these field strengths transform
as
\be \label{10082011-05a1} \delta F^a = (d-2)\partial^a \lambda\,,\qquad
\qquad  \delta W^a = (2-d)\partial^a \parline \lambda\,. \ee
Using \rf{10082011-05a1}, we see that Lagrangians \rf{10082011-01} are
invariant under the $\lambda$-gauge transformation.

We now discuss on-shell D.o.F of the conformal theory under consideration.
For this purpose, we use fields transforming in spinor representations of the
$so(d-2)$ algebra and classify on-shell D.o.F by spin labels of $so(d-2)$. We
prove (see Appendix E) that on-shell D.o.F are described by $2k+1$ non-chiral
vector-spinor fields $\psi_{k'}^i$ and $2k-1$ non-chiral
spinor fields $\psi_{k'}$:%
\footnote{ If the generic conformal field $\psi$ has positive (or negative)
chirality, then fields in \rf{ferspin2DoF01},\rf{ferspin2DoF03} should have
positive (or negative) chirality, while fields in
\rf{ferspin2DoF02},\rf{ferspin2DoF04} should have negative (or positive)
chirality.}%

\beq
\label{ferspin2DoF01} && \psi_{k'}^i \,,
\hspace{2cm} k' \in [k]_2\,;
\\
\label{ferspin2DoF02} && \psi_{k'}^i\,, \hspace{2cm} k' \in [k-1]_2\,;
\\
\label{ferspin2DoF03} && \psi_{k'}\,, \hspace{2cm} k' \in [k-1]_2\,;
\\
\label{ferspin2DoF04} && \psi_{k'}\,,    \hspace{2cm} k' \in [k-2]_2\,,
\eeq
where we use the notation as in \rf{sumnot02}. The {\it complex-valued}
fermionic fields $\psi_{k'}^i$ and $\psi_{k'}$ transform in the respective
non-chiral vector-spinor and non-chiral spinor
representations of the $so(d-2)$ algebra.%
\footnote{ These fields satisfy the standard constraints $\gamma^i
\psi_{k'}^i=0$, $\Pi^{\minushat}\psi_{k'}^i=0$, $\Pi^{\minushat}\psi_{k'}=0$
(for details, see Appendix E).}
We note that on-shell D.o.F given in \rf{ferspin2DoF04} appear only for
$d\geq 6$ (i.e., $k\geq 2$).

Total number of {\it complex-valued} on-shell D.o.F shown in
\rf{ferspin2DoF01}-\rf{ferspin2DoF04} is given by
\beq \label{ferspin2DoF04n1}
&& \nbf = 2^{^{\frac{d-2}{2}}} d(d-3) \,.
\eeq
We note that $\nbf$ \rf{ferspin2DoF04n1} is a sum of D.o.F for fields given in
\rf{ferspin2DoF01}-\rf{ferspin2DoF04}:%
\footnote{ Total number of D.o.F $\nbf$ \rf{ferspin2DoF04n1} for
spin-$\frac{3}{2}$ conformal fermionic field  in $d=4$ was found in
Ref.\cite{Fradkin:1985am}. Decomposition of $\nbf$ \rf{ferspin2DoF05} into
irreps of the $so(d-2)$ algebra for the case of $d=4$ spin-$\frac{3}{2}$
conformal field was carried out in Ref.\cite{Lee:1982cp}. In Appendix E, we
use light-cone approach to generalize these results to the case of arbitrary
$d\geq 4$.}
\beq \label{ferspin2DoF05}
&& \nbf = \sum_{k'\in [k]_2} \nbf (\psi_{k'}^i) + \sum_{k'\in [k-1]_2} \nbf
(\psi_{k'}^i) + \sum_{k'\in [k-1]_2} \nbf(\psi_{k'}) + \sum_{k'\in [k-2]_2}
\nbf(\psi_{k'})\,,
\\
\label{decferspin2DoF01} && \hspace{1cm} \nbf(\psi_{k'}^i) =
2^{\frac{d-2}{2}}(d-3) \,,
\hspace{2cm} k' \in [k]_2\,;
\nonumber\\
&& \hspace{1cm}  \nbf(\psi_{k'}^i) = 2^{\frac{d-2}{2}}(d-3) \,, \hspace{2cm}
k' \in [k-1]_2\,;
\nonumber\\
&& \hspace{1cm}  \nbf(\psi_{k'}) = 2^{\frac{d-2}{2}}\,, \hspace{3.2cm} k' \in
[k-1]_2;
\nonumber\\
\label{decferspin2DoF04} && \hspace{1cm} \nbf(\psi_{k'}) = 2^{\frac{d-2}{2}}
\,, \hspace{3.2cm} k' \in [k-2]_2\,.
\eeq
%

\subsection{ Ordinary-derivative formulation of spin-$\frac{3}{2}$ conformal
field}\label{sec-08-sub02}

{\bf Field content}. To discuss ordinary-derivative and gauge invariant
formulation of spin-$\frac{3}{2}$ conformal non-chiral Dirac field in flat
space of dimension $d\geq 4$ we use $2k+1$ non-chiral vector-spinor Dirac
fields $\psi_{k'}^a$ and $2k-1$ non-chiral spinor Dirac fields $\psi_{k'}$:
\beq
\label{3-2psicol01} && \psi_{k'}^a \,,
\hspace{2cm} k' \in [k]_2;
\\
\label{3-2psicol02} && \psi_{k'}^a\,, \hspace{2cm} k' \in [k-1]_2\,;
\\
\label{3-2psicol03} && \psi_{k'}\,, \hspace{2cm} k' \in [k-1]_2\,;
\\
\label{3-2psicol04} && \psi_{k'}\,,    \hspace{2cm} k' \in [k-2]_2\,;
\\
&& \hspace{1.2cm} \label{oldman-26012011-01} k \equiv \frac{d-2}{2}\,,
\eeq
where the spinor indices of the fermionic fields $\psi_{k'}^a$ and
$\psi_{k'}$ are implicit. The fields $\psi_{k'}^a$ and $\psi_{k'}$ are the
respective non-chiral vector-spinor fields and
non-chiral spinor fields of the Lorentz algebra $so(d-1,1)$.%
\footnote{ Chiral fermionic fields can be introduced by allowing fields in
\rf{3-2psicol01}, \rf{3-2psicol03} to be positive-chirality (or
negative-chirality), while fields in \rf{3-2psicol02},\rf{3-2psicol04} to be
negative-chirality (or positive-chirality).}
We note that fields in \rf{3-2psicol04} appear only for $d\geq 6$ (i.e.,
$k\geq 2$). Also, we note that fields in \rf{3-2psicol01}-\rf{3-2psicol04}
have the conformal dimensions
\be \label{delspi3-2def01} \Delta_{\psi_{k'}^a} = \frac{d-1}{2} + k'\,,\qquad
\qquad \Delta_{\psi_{k'}} = \frac{d-1}{2} + k'\,.\ee

In order to obtain the gauge invariant description in an easy--to--use form
we use oscillators $\alpha^a$, $\zeta$, $\upsilon^\oplussm$,
$\upsilon^\ominussm$ \rf{manold-31102011-01} and collect fields
\rf{3-2psicol01}-\rf{3-2psicol04} into the ket-vector $\psik$ defined by
\be \label{3-2psidef0001}
\psik = \psikst{1} + \zeta \psikst{0}\,, \ee
where the ket-vectors $\psikst{1}$, $\psikst{0}$ are defined by%
\beq
\label{3-2psidef0001nn01} && \psikst{1}  = \left(
\begin{array}{l}
|\psi_{\u 1}\rangle
\\[6pt]
|\psi_{\d 1}\rangle \end{array} \right),
\qquad
\psikst{0}  = \left(
\begin{array}{l}
|\psi_{\u 0}\rangle
\\[6pt]
|\psi_{\d 0}\rangle \end{array} \right)\,,
\\
\label{psi10def01}
&& |\psi_{\u 1}\rangle  \equiv \sum_{k'\in [k]_2} \frac{1}{
(\frac{k+k'}{2})!} \alpha^a (\upsilon^\oplussm)^{^{\frac{k+k'}{2}}}
(\upsilon^\ominussm)^{^{\frac{k-k'}{2}}} \, \psi_{ k'}^a |0\rangle\,,
\\
\label{psi10def02} && |\psi_{\d 1}\rangle \equiv \sum_{k'\in [k-1]_2}
\frac{1}{ (\frac{k-1+k'}{2})!} \alpha^a
(\upsilon^\oplussm)^{^{\frac{k-1+k'}{2}}}
(\upsilon^\ominussm)^{^{\frac{k-1-k'}{2}}} \, \psi_{k'}^a |0\rangle\,,
\\
\label{psi10def03} && |\psi_{\u 0}\rangle  \equiv \sum_{k'\in [k-1]_2}
\frac{1}{ (\frac{k-1+k'}{2})!} (\upsilon^\oplussm)^{^{\frac{k-1+k'}{2}}}
(\upsilon^\ominussm)^{^{\frac{k-1-k'}{2}}}  \, \psi_{ k'}|0\rangle\,,
\\
\label{psi10def04} && |\psi_{\d 0}\rangle \equiv \sum_{k'\in [k-2]_2}
\frac{1}{ (\frac{k-2+k'}{2})!} (\upsilon^\oplussm)^{^{\frac{k-2+k'}{2}}}
(\upsilon^\ominussm)^{^{\frac{k-2-k'}{2}}} \, \psi_{k'} |0\rangle\,,
\eeq
and $k$ is given in \rf{oldman-26012011-01}. We see that ket-vector
\rf{3-2psidef0001} satisfies the algebraic constraints:
\beq \label{3-2Nalzet01}
&& (N_\alpha + N_\zeta - 1 ) \psik = 0 \,,
\\
\label{NzetNups01} && (N_\zeta + N_\upsilon - k )\pi_+ \psik = 0 \,,
\\
\label{NzetNups02} && (N_\zeta + N_\upsilon  - k+1)\pi_- \psik =0 \,.
\eeq
From algebraic constraints \rf{3-2Nalzet01}-\rf{NzetNups02}, we learn that
\\
a) ket-vector $\psik$ \rf{3-2psidef0001} is degree-1 homogeneous
polynomial in the oscillators $\alpha^a$, $\zeta$;
\\
b) ket-vector $|\psi_{\u 1}\rangle$ \rf{psi10def01} is degree-$k$
homogeneous polynomial in $\upsilon^\oplussm$,
$\upsilon^\ominussm$;\\
c) ket-vectors $|\psi_{\d 1}\rangle$ \rf{psi10def02}, $|\psi_{\u
0}\rangle$ \rf{psi10def03} are degree $k-1$ homogeneous polynomials in
$\upsilon^\oplussm$, $\upsilon^\ominussm$;\\
e) ket-vector $|\psi_{\d 0}\rangle$ \rf{psi10def04} is degree $k-2$
homogeneous polynomial in $\upsilon^\oplussm$, $\upsilon^\ominussm$.

{\bf Lagrangian}. Lagrangian we found takes the form
\beq \label{oldman-31012011-01}
{\rm i}\LL &  = &  \psibr E \psik\,,
\\
\label{spi3-2eqmotope01}
&& E  = E_\smone + E_\smzero\,,
\\
\label{RarSchEHsecordope01}
&& E_\smone  \equiv
\parline
- \alpha\partial\gamma\bar\alpha
- \gamma\alpha\bar\alpha\partial
+ \gamma\alpha
\parline\gamma\bar\alpha
\,,
\\
&& E_\smzero =  (1-\gamma\alpha\gamma\bar\alpha)e_1^\smGamma + \gamma\alpha
\eb_1 + e_1 \gaalb\,,
\\
\label{oldman-26012011-02}
&& \hspace{1cm} e_1^\smGamma = \frac{d}{d - 2N_\zeta }
(\bar\upsilon^\ominussm \sigma_- + \upsilon^\ominussm\sigma_+)\,,
\\
\label{oldman-26012011-03} &&  \hspace{1cm} e_1 = u_\f \zeta
\bar\upsilon^\ominussm \,,
\qquad\quad
\eb_1 = - u_\f \upsilon^\ominussm \bar\zeta\,, \qquad u_\f \equiv
\Bigl(\frac{d-1}{d-2}\Bigr)^{1/2}\,,
\eeq
where $E_\smone$ \rf{RarSchEHsecordope01} is the Rarita-Schwinger operator
rewritten in terms of the oscillators.

{\bf Gauge transformations}. We now discuss gauge symmetries of the
Lagrangian. To this end we introduce the gauge transformation parameters,
\beq
\label{oldman-21082011-01} && \xi_{k'-1}\,,
\hspace{1.3cm} k' \in [k]_2\,;
\nonumber\\[-12pt]
&&
\\[-12pt]
&& \xi_{k'-1}
\hspace{1.5cm} k' \in [k-1]_2\,.
\nonumber
\eeq
The gauge transformations parameters $\xi_{k'}$ are non-chiral spinor fields
of the Lorentz algebra $so(d-1,1)$. We collect the gauge
transformation parameters in the ket-vector $\xik$ defined by
\beq \label{oldman-31012011-02}
&& \xik  = \left(
\begin{array}{l}
|\xi_\u\rangle
\\[6pt]
|\xi_\d\rangle \end{array} \right)\,,
\\
\label{epszeroferdef01}
|\xi_\u\rangle  & \equiv &  \sum_{k'\in [k]_2} \frac{1}{ (\frac{k+k'}{2})!}
(\upsilon^\oplussm)^{^{\frac{k+k'}{2}}}
(\upsilon^\ominussm)^{^{\frac{k-k'}{2}}} \, \xi_{ k'-1} |0\rangle\,,
\\
\label{epszeroferdef02} |\xi_\d\rangle & \equiv  & \sum_{k'\in [k-1]_2}
\frac{1}{ (\frac{k-1+k'}{2})!} (\upsilon^\oplussm)^{^{\frac{k-1+k'}{2}}}
(\upsilon^\ominussm)^{^{\frac{k-1-k'}{2}}} \, \xi_{k'-1}|0\rangle\,.
\eeq
Ket-vector of gauge transformation parameters $\xik$ \rf{oldman-31012011-02}
satisfies the algebraic constraints
\be
\label{epsNzetNups01} (N_\upsilon - k )\pi_+ \xik = 0 \,,
\qquad \quad (N_\upsilon  - k+1)\pi_- \xik =0 \,,
\ee
which tell us that ket-vector $|\xi_\u\rangle$ \rf{epszeroferdef01} is
degree-$k$ homogeneous polynomial in $\upsilon^\oplussm$,
$\upsilon^\ominussm$, while ket-vector $|\xi_\d\rangle$
\rf{epszeroferdef02} is degree $k-1$ homogeneous polynomial in
$\upsilon^\oplussm$, $\upsilon^\ominussm$.

Gauge transformations take the form
\be \label{spi3-2gautradef01}
\delta \psik = G\xik\,, \qquad \quad G \equiv \alpha\partial -e_1 +
\frac{1}{d-2}\gaal e_1^\smGamma \,, \ee
where $e_1^\smGamma$ and $e_1$ are given in \rf{oldman-26012011-02},
\rf{oldman-26012011-03}.

{\bf Component form of Lagrangian and gauge transformations}. Component form
of Lagrangian \rf{oldman-31012011-01} is obtained by plugging ket-vector
\rf{3-2psidef0001} into \rf{oldman-31012011-01}. Doing so, we obtain the
component form of the Lagrangian,
\beq
&& \hspace{-1.8cm} \LL = \sum_{k' \in [k]_1} \LL_{k'} \,,
\nonumber\\
\label{oldman-31012011-03} {\rm i}\LL_{k'} & = &
\bar\psi_{-k'}^a\gamma^{abc}\partial^b \psi_{k'}^c + \bar\psi_{-k'}
\parline \psi_{k'}
\nonumber\\
& - &   \bar\psi_{-k'}^a\gamma^{ab} \psi_{k'+1}^b  + u_\f
(\bar\psi_{-k'}\gamma^a\psi_{k'+1}^a - \bar\psi_{-k'}^a \gamma^a \psi_{k'+1})
+ \frac{d}{d-2}\bar\psi_{-k'} \psi_{k'+1}\,,
\eeq
where $u_\f$ is given in \rf{oldman-26012011-03}. We see that one-derivative
contributions to Lagrangian \rf{oldman-31012011-03} are the standard
Rarita-Schwinger kinetic terms for the spin-$\frac{3}{2}$ fields
$\psi_{k'}^a$ and the standard Dirac kinetic terms for the spin-$\half$
fields $\psi_{k'}$. Besides the one-derivative contributions, the Lagrangian
involves derivative-independent contributions. Appearance of the
derivative-independent contributions is a characteristic feature of the
ordinary-derivative approach to conformal fermionic fields.

Component form of the gauge transformations is obtained by plugging
ket-vectors \rf{3-2psidef0001}, \rf{oldman-31012011-02} into
\rf{spi3-2gautradef01},
\beq
\label{23072011-05a1} && \delta \psi_{k'}^a = \partial^a\xi_{k'-1} +
\frac{1}{d-2}\gamma^a \xi_{k'}\,,
\\
\label{23072011-05a2} && \delta \psi_{k'} = - u_\f\xi_{k'}\,,
\eeq
where $u_\f$ is given in \rf{oldman-26012011-03}. Two remarks are in order.

\noindent {\bf i}) From \rf{23072011-05a2}, we see that the fields
$\psi_{k'}$ transform as Stueckelberg fields.

\noindent {\bf ii}) The field $\psi_{-k}^a$ has the same conformal dimension
as the field $\psi^a$ entering the higher-derivative approach
\rf{3-2counDoFt03}, i.e., we can use the identification $\psi_{-k}^a =
\psi^a$. The remaining fields $\psi_{k'}^a$, $k'=-k+1,-k+2,\ldots k-1,k$, are
auxiliary fields. Gauging away all spin-$\half$ fields, $\psi_{k'}=0$, we
make sure that equations of motion for the auxiliary fields obtained from
\rf{oldman-31012011-03} take the form
\be \gamma^{abc}\partial^b\psi_{k'}^c - \gamma^{ab} \psi_{k'+1}^b = 0 \,. \ee
These equations tell us that we can express all auxiliary vector-spinor
fields in terms of $\psi_{-k}^a\equiv \psi^a$,
\beq  \label{09082011-02}
&& \hspace{-1cm} \psi_{k'}^a  =  \Box^{^{\frac{k+k'-1}{2}}} F^a +
\frac{d-2}{2(d-1)}\theta_{k'} \Box^{^{\frac{k+k'-3}{2}}}
\partial^a \gamma^{bc} \parline F^{bc}\,,
\qquad
\hbox{ for } \ k'=[k-1]_2\,,
\\
&& \hspace{-1cm} \psi_{k'}^a  =  -\Box^{^{\frac{k+k'-2}{2}}} W^a -
\frac{d-2}{2(d-1)} \Box^{^{\frac{k+k'-2}{2}}}
\partial^a \gamma^{bc} F^{bc}\,,
\nonumber\\
\label{09082011-03} &&  \hspace{3cm} \hbox{ for } \
k'=-k+2,-k+4,\ldots,k-4,k-2,k\,,
\\
&&  \theta_{-k+1}=0\,,\qquad \theta_{k'}=1\,, \qquad k'=-k+3,-k+5,\ldots,
k-3,k-1\,,
\eeq
where field strengths $F^{ab}$, $F^a$, $W^a$ are defined as in
\rf{10082011-03}-\rf{10082011-05}. Plugging solution for auxiliary fields
\rf{09082011-02},\rf{09082011-03} into \rf{oldman-31012011-03}, we obtain
higher-derivative Lagrangian \rf{10082011-01}, i.e., our approach and the
higher-derivative approach are equivalent.

{\bf Conformal symmetries}. To complete ordinary-derivative description of
the spin-$\frac{3}{2}$ field we provide realization of the conformal algebra
symmetries on space of ket-vector $\psik$ \rf{3-2psidef0001}. All that is
required is to fix operators $M^{ab}$, $\Delta$ and $R^a$ for the case of
spin-$\frac{3}{2}$ conformal fermionic field and then use these operators in
\rf{conalggenlis01}-\rf{conalggenlis04}. The realization of the spin operator
of the Lorentz algebra and conformal dimension operator $\Delta$ on space of
$\psik$ takes the form

\beq
&& M^{ab} = \alpha^a\bar\alpha^b - \alpha^b \bar\alpha^a  + \half
\gamma^{ab}\,,
\\
\label{manold-31102011-12} && \Delta  =  \frac{d-1}{2}+\Delta'\,, \qquad
\quad \Delta' \equiv N_{\upsilon^\oplussm} - N_{\upsilon^\ominussm} \,,
\eeq
where the realization of the conformal dimension operator $\Delta$ on space
of $\psik$ given in \rf{manold-31102011-12} can be read from
\rf{delspi3-2def01}. The realization of the operator $R^a$ on space of
$\psik$ is given by
\beq \label{spi3-2Edef01}
&& R^a =  r^a + R_\smE^a + R_\smG^a \,,
\\
\label{spi3-2Edef02} && r^a  =  r_{0,1}^\smGamma (\gamma^a -\frac{2}{d-2}
\gaal \bar\alpha^a) +  r_{0,1}\bar\alpha^a + \rb_{0,1} (\alpha^a -
\frac{1}{d-2}\gaal \gamma^a) + r_{1,1}\partial^a\,,
\\
\label{spi3-2Edef04} && R_\smG^a  =  G r_\smG^a \,,
\qquad\quad
R_\smE^a  =   r_\smE^a E \,,
\\
\label{spi3-2Edef05nn1} && r_{0,1}^\smGamma = \frac{d}{d-2N_\zeta} (
\bar\upsilon^\oplussm \sigma_- - \upsilon^\oplussm \sigma_+ ) \,,
\\
\label{spi3-2Edef06} && r_{0,1}  =  2 u_\f \zeta \bar\upsilon^\oplussm \,,
\qquad
\rb_{0,1} = - 2 u_\f \upsilon^\oplussm \bar\zeta\,,
\qquad
r_{1,1} = -2\upsilon^\oplussm \bar\upsilon^\oplussm \,,
\eeq

\vspace{-0.8cm}
\beq
\label{spi3-2Edef09} && r_\smG^a  =    r_{\smG,1} \bar\alpha^a  + r_{\smG,2}
\gamma^a  + r_{\smG,3} \gamma^a \gaalb \,,
\\
\label{spi3-2Edef10} && r_\smE^a  =  r_{\smE,1} \gamma^a + r_{\smE,2}
\alpha^a \gaalb + r_{\smE,3} \gaal\bar\alpha^a + r_{\smE,4} \gaal \gamma^a
\gaalb
\nonumber\\
&& \hspace{0.7cm} + \   r_{\smE,5} \alpha^a +  r_{\smE,6}\bar\alpha^a +
r_{\smE,7} \gaal \gamma^a  + r_{\smE,8}\gamma^a \gaalb +   r_{\smE,9}
\gamma^a \,,
\\
\label{spi3-2Edef11} && r_{\smG,n} = \upsilon^\oplussm
\rwt_{\smG,n-}\bar\upsilon^\oplussm \pi_-  +  \upsilon^\oplussm
\rwt_{\smG,n+} \bar\upsilon^\oplussm \pi_+ \,,\hspace{2.5cm} n=1,3\,;
\nonumber\\
\label{spi3-2Edef12} && r_{\smG,n} = \upsilon^\oplussm
\rwt_{\smG,n-}\bar\upsilon^\oplussm \bar\zeta\sigma_-  + \upsilon^\oplussm
\upsilon^\oplussm \rwt_{\smG,n+} \bar\zeta \sigma_+ \,,\hspace{2cm} n=2\,;
\\
\label{spi3-2Edef13} && r_{\smE,n} = (\upsilon^\oplussm
\rwt_{\smE,n-}\bar\upsilon^\oplussm \pi_-  + \upsilon^\oplussm \rwt_{\smE,n+}
\bar\upsilon^\oplussm \pi_+)(1-N_\zeta) \,,\qquad n=1\,;
\nonumber\\
\label{spi3-2Edef14} && r_{\smE,n} = \upsilon^\oplussm
\rwt_{\smE,n-}\bar\upsilon^\oplussm \pi_-  + \upsilon^\oplussm \rwt_{\smE,n+}
\bar\upsilon^\oplussm \pi_+ \,,\hspace{2cm} n=2,3,4\,;
\nonumber\\
\label{spi3-2Edef15} && r_{\smE,n} = \upsilon^\oplussm \rwt_{\smE,n-}
\bar\upsilon^\oplussm \bar\zeta \sigma_- + \upsilon^\oplussm
\upsilon^\oplussm \rwt_{\smE,n+}  \bar\zeta \sigma_+ \,,\hspace{2cm} n=5,7\,;
\nonumber\\
\label{spi3-2Edef16} && r_{\smE,n} = \zeta \rwt_{\smE,n-}
\bar\upsilon^\oplussm \bar\upsilon^\oplussm \sigma_- + \zeta
\upsilon^\oplussm \rwt_{\smE,n+} \bar\upsilon^\oplussm \sigma_+ \,,
\hspace{2cm} n=6,8\,;
\nonumber\\
\label{spi3-2Edef17} && r_{\smE,n} = (\upsilon^\oplussm
\rwt_{\smE,n-}\bar\upsilon^\oplussm \pi_-  + \upsilon^\oplussm \rwt_{\smE,n+}
\bar\upsilon^\oplussm \pi_+)N_\zeta \,, \hspace{2cm} n=9\,;
\\
\label{spi3-2Edef18} && \qquad \rwt_{\smG,n\pm}=
\rwt_{\smG,n\pm}(\Delta')\,,\qquad \qquad n=1,2,3\,.
\\
\label{spi3-2Edef19} && \qquad \rwt_{\smE,n\pm}= \rwt_{\smE,n\pm}(\Delta')\,,
\qquad \qquad n=1,\ldots,9\,;
\\
\label{herconrulcon01}
&& \qquad \rwt_{\smE,n\pm}^\dagger =  \rwt_{\smE,n\pm} \qquad a =1,4,9 \,;
\nonumber\\
\label{herconrulcon02}
&& \qquad \rwt_{\smE,2\pm}^\dagger = \rwt_{\smE,3\pm}\,, \qquad
\rwt_{\smE,5\pm }^\dagger =  - r_{\smE,6\mp}\,,\qquad
\rwt_{\smE,7\pm}^\dagger = - \rwt_{\smE,8\mp}\,,
\eeq
where $u_\f$ \rf{spi3-2Edef06} is given in \rf{oldman-26012011-03}. Operators
$G$ and $E$ \rf{spi3-2Edef04} are given in \rf{spi3-2gautradef01} and
\rf{spi3-2eqmotope01} respectively. Quantities $\rwt_{\smG,n \pm }$,
$\rwt_{\smE,n\pm }$ \rf{spi3-2Edef18},\rf{spi3-2Edef19} are arbitrary
functions of the operator $\Delta'$. The quantities $\rwt_{\smE,n\pm }$ are
subject to hermitian conjugation rules in
\rf{herconrulcon02}. Details of the derivation of the
Lagrangian, gauge transformations and operator $R^a$ may be found in Appendix
F.

The following remarks are in order.

\noindent {\bf i}) $r^a$ part of the operator $R^a$ is fixed uniquely, while
$R_\smG^a$ and $R_\smE^a$ parts, in view of arbitrary $\rwt_{\smG,n\pm }$,
$\rwt_{\smE,n\pm }$, are still to be arbitrary. The arbitrariness in
$R_\smG^a$ is related to the fact that global transformations of gauge fields
are defined up to gauge transformations. Reason of the arbitrariness in
$R_\smE^a$ is the same as for the case of spin-$\frac{1}{2}$ conformal field
(see discussion below Eq.\rf{Rgsmspi1-208}).

\noindent {\bf ii)} We check that $R^a$ transformations with $R_\smG^a=0$,
$R_\smE^a=0$ satisfy the commutator $[K^a,K^b]=0$. In terms of the component
fields, $R^a$ transformations with $R_\smG^a=0$,  $R_\smE^a=0$ take the form
\beq
&&\hspace{-1.3cm}  \delta_{R^a} \psi_{k'}^b =  -\frac{k+k'}{2}(\gamma^a
\psi_{k'-1}^b - \frac{1}{k}\gamma^b \psi_{k'-1}^a) -(k+k') u_\f (\eta^{ab} -
\frac{1}{2k}\gamma^b \gamma^a)\psi_{k'-1}
\nonumber\\
&& - \ \half (k+k')(k+2-k')\partial^a \psi_{k'-2}^b\,, \hspace{2cm} k'\in
[k]_2\,,
\\
&& \hspace{-1.3cm} \delta_{R^a} \psi_{k'}^b =  \frac{k+1-k'}{2}(\gamma^a
\psi_{k'-1}^b - \frac{1}{k}\gamma^b \psi_{k'-1}^a) -(k-1+k') u_\f (\eta^{ab}
- \frac{1}{2k}\gamma^b \gamma^a)\psi_{k'-1}
\nonumber\\
&&  - \ \half (k-1+k')(k+1-k')\partial^a \psi_{k'-2}^b\,,\qquad\quad k'\in
[k-1]_2\,,
\\
&& \hspace{-1.3cm} \delta_{R^a} \psi_{k'} =  -
\frac{k+1}{2k}(k-1+k')\gamma^a\psi_{k'-1} + u_\f (k+1-k') \psi_{k'-1}^a
\nonumber\\
&&  - \ \half (k-1+k')(k+1-k')\partial^a \psi_{k'-2}\,, \qquad \quad k'\in
[k-1]_2\,,
\\
&& \hspace{-1.3cm} \delta_{R^a} \psi_{k'} =
\frac{k+1}{2k}(k-k')\gamma^a\psi_{k'-1} + u_\f (k-k') \psi_{k'-1}^a
\nonumber\\
&& - \ \half (k-2+k')(k-k')\partial^a \psi_{k'-2}\,,\qquad \qquad \ \ k'\in
[k-2]_2\,.
\eeq

\noindent {\bf iii}) To find all  $R_\smG^a$, $R_\smE^a$ which satisfy the
commutator $[K^a,K^b]=0$, we note the relations
\beq
\label{28072011-01} && [K^a,K^b] =  W^{ab}\,,
\\
\label{28072011-02} && \hspace{2cm} W^{ab} \equiv G r_\smG^{ab} + r_\smE^{ab}
E +   G r_\smGE^{ab} E\,,
\\
&& \hspace{2cm} r_\smG^{ab} \equiv r^a r_\smG^b + r_\smG^a r^b + r_\smG^a G
r_\smG^b - (a \leftrightarrow b)\,,
\\
\label{20082011-oldman-02} && \hspace{2cm} r_\smE^{ab} \equiv r^a r_\smE^b +
r_\smE^b r^{a\hat{\dagger}} + r_\smE^a E r_\smE^b - (a \leftrightarrow b)\,,
\qquad r^{a\hat{\dagger}} \equiv -\gamma^0 r^{a\dagger}\gamma^0\,, \qquad
\\
&& \hspace{2cm} r_\smGE^{ab} \equiv r_\smG^a r_\smE^b - (a \leftrightarrow
b)\,.
\eeq
From \rf{28072011-01}, we see that the requirement $[K^a,K^b]=0$ amounts to
the equations
\be \label{01082011-01} W^{ab} = 0 \,.\ee
We now present solutions of Eqs.\rf{01082011-01} for $d =4$, $d=6$, and
$d\geq 8$ in turn.

\noindent {\bf Case $d =4$} ($k = 1$). For this case, solution to
Eqs.\rf{01082011-01} is given by
\beq
&& \rwt_{\smE,n+} = c_{\smE,n+}, \qquad \ \  n=1,2,3,4;
\nonumber\\
&& \rwt_{\smE,n-} \equiv 0 \,, \qquad\qquad n=1,2,3,4;
\nonumber\\
&& \rwt_{\smE,n\pm} \equiv   0\,,\qquad\qquad n= 5,6,7,8,9;
\nonumber\\
&& \rwt_{\smG,n\pm} = c_{\smG,n\pm}, \qquad \ \ n=1,3;
\nonumber\\
&& \rwt_{\smG,n\pm} \equiv  0\,,\qquad n= 2\,,
\eeq
where $c_{\smE,1+}$, $c_{\smE,4+}$ are real-valued constants, while
$c_{\smE,2+}= c_{\smE,3+}^*$,  and $c_{\smG,n\pm}$ are complex-valued
constants.

\noindent {\bf Case $d =6$} ($k = 2$).  For this case, solution to
Eqs.\rf{01082011-01} is given by
\beq
\label{01082011-02} && \rwt_{\smE,1\pm} = 0\,,
\qquad
\rwt_{\smE,2+} =0 \,,\qquad \rwt_{\smE,2-} = 2Y (1+ 5e^{-\i \varphi})c\,,
\nonumber\\
&& \rwt_{\smE,3+}=0\,,\qquad \rwt_{\smE,3-} = 2\Yb(1+ 5e^{\i  \varphi})c\,,
\nonumber\\
&& \rwt_{\smE,4+} = c\,,\qquad \rwt_{\smE,4-} = Y\Yb \Bigl(88(1-\cos\varphi)
c + 5 \cos\varphi + 2\Bigr) c\,,
\nonumber\\
&& \rwt_{\smE,5+}=0\,,\qquad \rwt_{\smE,5-} =  \frac{1}{u_\f}
\rwt_{\smE,2-}\,,
\qquad
\rwt_{\smE,6+} = - \frac{1}{u_\f} \rwt_{\smE,3-}\,,\qquad \rwt_{\smE,6-}=0\,,
\nonumber\\
&& \rwt_{\smE,7+} = 4u_\f e^{\i  \varphi } c\,,\hspace{3cm} \rwt_{\smE,7-} =
-\frac{1}{2u_\f} \rwt_{\smE,3-} + \frac{1}{u_\f}\rwt_{\smE,4-}\,,
\nonumber\\
&& \rwt_{\smE,8+}  = \frac{1}{2u_\f} \rwt_{\smE,2-} -
\frac{1}{u_\f}\rwt_{\smE,4-}\,,\qquad \rwt_{\smE,8-} = -4u_\f e^{-\i  \varphi
} c\,,
\nonumber\\
&& \rwt_{\smE,9+} = \frac{1}{2u_\f^2}(\rwt_{\smE,2-} + \rwt_{\smE,3-}) -
\frac{1}{u_\f^2} \rwt_{\smE,4-}\,,\qquad  \rwt_{\smE,9-}\equiv 0 \,,
\nonumber\\
&& Y^{-1} \equiv 24(1 - e^{-\i  \varphi})c -1\,, \qquad \Yb^{-1} \equiv 24(1
- e^{\i  \varphi})c -1\,,
\eeq
where $u_\f \equiv \sqrt{5}/2$. Parameters $c$ and $\varphi$ appearing in
\rf{01082011-02} are real-valued constants, i.e., we see that solution for
$\rwt_{\smE,n\pm}$ is described by two real-valued constants. Expressions for
$\rwt_{\smG,n\pm}$ depend on $c$. For $c\ne 0$, we obtain
\beq
&& \rwt_{\smG,1+} = \frac{4}{\Delta'+c_{1+}}\,,
\hspace{2.4cm}
\rwt_{\smG,1-} = \frac{2(3+e^{\i  \varphi})}{c_{1+} + e^{\i  \varphi}}\,,
\nonumber\\
&& \rwt_{\smG,2+} = \frac{e^{\i  \varphi}}{u_\f( c_{1+} + 1 )}\,,
\hspace{2cm}
\rwt_{\smG,2-} = \frac{5+ 3e^{\i  \varphi}}{2u_\f(c_{1+} +e^{\i  \varphi})}\,,
\qquad\quad
\rwt_{\smG,3 \pm } = 0 \,,\qquad
\nonumber\\
&& c_{2+} \equiv \frac{1}{u_\f} e^{\i  \varphi}\,,
\hspace{2cm}c_{2-} \equiv \frac{5+3e^{\i  \varphi}}{u_\f(3+e^{\i  \varphi})}
\,,
\eeq
where $c_{1+}$ is complex-valued constant. For $c=0$, we obtain
\beq
&& \rwt_{\smG,1+} = \frac{4}{\Delta'+c_{1+}}\,,
\hspace{2cm}
\rwt_{\smG,1-} = \frac{4}{c_{1-}}\,,
\nonumber\\
&& \rwt_{\smG,2+} = \frac{c_{2+}}{1+ c_{1+}}\,,
\hspace{2.1cm}
\rwt_{\smG,2-} = \frac{c_{2-}}{c_{1-}}\,,
\qquad\quad
\rwt_{\smG,3 \pm } = 0 \,,\qquad
\nonumber\\
&& c_{2+}\equiv \frac{1}{u_\f (2-c_{1-})}(3c_{1-} -2 c_{1+})\,,
\qquad c_{2-} \equiv \frac{1}{u_\f (3-c_{1+})}(2c_{1-} -3c_{1+} + 5)\,,\qquad
\eeq
where $c_{1\pm}$ are complex-valued constants.

\noindent {\bf Case $d\geq 8$} ($k\geq 3$).  For this case, solution to
Eqs.\rf{01082011-01} is given by
\beq
\label{01082011-05} && \rwt_{\smE,n\pm} = 0 \,, \qquad n=1,2,\ldots, 9\,,
\nonumber\\
&& \rwt_{\smG,1\pm} = \frac{4}{\Delta'+c_{1\pm}}\,,
\nonumber\\
&& \rwt_{\smG,2+} = \frac{c_{2+}}{\Delta'+1+ c_{1+}}\,,
\qquad\quad
\rwt_{\smG,2-} = \frac{c_{2-}}{\Delta'+c_{1-}}\,,
\qquad\quad
\rwt_{\smG,3\pm } = 0 \,,\qquad
\nonumber\\
&& \qquad c_{2+} \equiv \frac{2}{u_\f k(k-c_{1-})}((k+1)c_{1-} -k c_{1+})\,,
\nonumber\\
\label{01082011-06} && \qquad c_{2-} \equiv \frac{2}{u_\f
k(k+1-c_{1+})}(kc_{1-} -(k+1)c_{1+} + 2k+1)\,,
\eeq
where $c_{1\pm}$ are complex valued constants.

{\bf Spin-$\frac{3}{2}$ conformal field in 4d}. To illustrate our approach we
consider spin-$\frac{3}{2}$ in $4d$, which is the simplest conformal theory
of spin-$\frac{3}{2}$ field involving Stueckelberg field. Our field content
involves three spin-$\frac{3}{2}$ fields and one spin-$\half$ field (see
\rf{3-2psicol01}-\rf{3-2psicol03}),
\beq
&& \psi_{_{-1}}^a \qquad  \psi_{_1}^a
\nonumber\\
&& \quad \ \ \ \psi_{_0}^a
\nonumber\\
&& \quad \ \ \ \psi_{_0}
\eeq
The corresponding Lagrangian \rf{oldman-31012011-03} takes the form
\beq
{\rm i}\LL & = & \bar\psi_{_{-1}}^a\gamma^{abc}\partial^b \psi_{_1}^c+
\bar\psi_{_0}^a\gamma^{abc}\partial^b \psi_{_0}^c +
\bar\psi_{_1}^a\gamma^{abc}\partial^b \psi_{_{-1}}^c + \bar\psi_{_0}\parline
\psi_{_0}
\nonumber\\
& - &  \bar\psi_{_1}^a \gamma^{ab} \psi_{_0}^b - \bar\psi_{_0}^a \gamma^{ab}
\psi_{_1}^b + u_\f (\bar\psi_{_0}\gamma^a\psi_{_1}^a -
\bar\psi_{_1}^a\gamma^a\psi_{_0})\,,
\eeq
where $u_\f = \sqrt{3/2}$. Gauge symmetries are described by three gauge
transformation parameters (see
\rf{oldman-21082011-01})
\beq
&& \xi_{_{-2}} \qquad \ \xi_{_0}
\nonumber\\
&& \quad \ \  \xi_{_{-1}}
\eeq
and appropriate gauge transformations are given by (see
\rf{23072011-05a1},\rf{23072011-05a2})
\beq
&&  \delta\psi_{_{-1}}^a = \partial^a\xi_{_{-2}} + \half
\gamma^a\xi_{_{-1}}\,,
\\
&&  \delta\psi_{_0}^a = \partial^a\xi_{_{-1}} + \half \gamma^a\xi_{_0}\,,
\\
&&  \delta\psi_{_1}^a = \partial^a\xi_{_0}\,,
\\
\label{manold-10112011-01} &&  \delta\psi_{_0} =  -u_\f \xi_{_0}\,.
\eeq
From \rf{manold-10112011-01}, we see that the field $\psi_{_0}$ transforms as
Stueckelberg field.

$R^a$ transformations taken in the minimal scheme, $R_\smG^a=0$,
$R_\smE^a=0$, are given by
\beq
&&  \delta_{R^a}\psi_{_{-1}}^b = 0\,,
\\
&&  \delta_{R^a}\psi_{_0}^b = \gamma^a \psi_{_{-1}}^b - \gamma^b
\psi_{_{-1}}^a\,,
\\
&&  \delta_{R^a}\psi_{_1}^b = \gamma^b \psi_{_0}^a - \gamma^a \psi_{_0}^b -
u_\f \gamma^a\gamma^b \psi_0 - 2\partial^a \psi_{_{-1}}^b\,,
\\
&&  \delta_{R^a}\psi_{_0} =  2u_\f \psi_{_{-1}}^a\,.
\eeq

\newsection{Conclusions}\label{conl-sec-01}

We have developed  the ordinary-derivative approach to conformal fields in
flat space of arbitrary dimension. In this paper, we applied this approach to
the study of low-spin fields. Because the approach we presented is based on
the use of oscillator realization of spin degrees of freedom of gauge fields
it allows straightforward generalization to higher-spin conformal fields.
Comparison of approach we developed with other approaches available in the
literature leads us to  the conclusion that our approach is very interesting
and attractive.

The results presented here should have a number of interesting applications
and generalizations, some of which are:

i) generalizations to supersymmetric conformal field theories and
applications to conformal supergravities in various dimensions
\cite{Bergshoeff:1980is}-\cite{Bergshoeff:1983sn}. The first step in this
direction is to understand how the supersymmetries are realized in the
framework of our approach. Note that, in this paper, we worked out
ordinary-derivative Lagrangians and realizations of conformal symmetries for
all fields that appear in supermultiplets of conformal supergravities and
involve higher-derivative contributions to Lagrangians of conformal
supergravity theories.

ii) extension of our approach to interacting (super)conformal low-spin and
higher-spin field theories \cite{Fradkin:1989md,Fradkin:1990ps}.%
\footnote{ In the framework higher-derivative formulation, uniqueness of
interacting spin-2 conformal field theory was discussed in
Ref.\cite{Boulanger:2001he}.}
Our approach to conformal theories is based on new realization of conformal
gauge symmetries via Stueckelberg fields. In our approach, use of
Stueckelberg fields is very similar to the one in gauge invariant formulation
of  massive fields. Stueckelberg fields provide interesting possibilities for
the study of interacting massive gauge fields (see e.g.
Refs.\cite{Zinoviev:2006im,Metsaev:2006ui}). Therefore we think that
application of our approach to the (super)conformal interacting fields should
lead to new interesting development.

iii) BRST approach turned out to be successful for the analysis of various
aspects of relativistic dynamics (see e.g. Ref.\cite{Siegel:1999ew}). Though
BRST approach was extended to higher-derivative theories (see e.g.
Ref.\cite{Nirov:1994rn}), it seems that this approach is conveniently adopted
for ordinary-derivative formulation. BRST approach was extensively developed
in recent time (see e.g.
Refs.\cite{Buchbinder:2001bs}-\cite{Buchbinder:2007vq}) and was applied to
the study of ordinary-derivative Lagrangian theories of massless and massive
fields in flat and AdS spaces. Because AdS theories and conformal theories
share many algebraic properties, application of the previously developed BRST
methods to the study of ordinary-derivative conformal field theories should
be relatively straightforward.

iv) There are other interesting approaches in the literature which could be
used to discuss the ordinary-derivative formulation of conformal theories.
This is to say that various formulations in terms of unconstrained fields in
flat space and AdS space may be found e.g. in
Refs.\cite{Francia:2002aa}-\cite{Francia:2005bu}.

v) Mixed-symmetry fields \cite{Aulakh:1986cb,Labastida:1986gy} have attracted
considerable interest in recent time (see e.g.
Refs.\cite{Burdik:2001hj}-\cite{Moshin:2007jt}). Higher-derivative
formulation of mixed-symmetry conformal fields was recently developed in
Ref.\cite{Vasiliev:2009ck}. We think that ordinary-derivative formulation of
mixed-symmetry fields might be useful for the study of various aspects of
conformal fields (see e.g. Ref.\cite{Metsaev:2008ba}). We note also that, in
$AdS_d$ space, massless mixed-symmetry fields, in contrast to massless fields
in Minkowski space whose physical degrees of freedom transform in irreps of
$o(d-2)$ algebra, reduce to a number of irreps of $so(d-2)$ algebra. In other
words, not every massless field in flat space admits a deformation to AdS
space with the same number of degrees of freedom \cite{Brink:2000ag}. It
would be interesting to understand this phenomenon from the point of view of
mixed-symmetry conformal fields.

vi) extension of our approach to light-cone gauge conformal fields and
application to analysis of interaction of conformal fields to composite
operators constructed out of the fields of supersymmetric YM
theory.%
\footnote{ Approach developed in Ref.\cite{Metsaev:2005ar} should streamline
such a analysis.}
We expect that a quantization of the Green-Schwarz $AdS$ superstring with a
Ramond - Ramond charge will be available only in the light-cone gauge
\cite{Metsaev:2000yf}-\cite{Metsaev:2000mv}. Therefore it seems that from the
stringy perspective of $AdS/CFT$ correspondence the light-cone approach to
conformal field theory is the fruitful direction to go.

We strongly believe that the approach developed in this paper will be
useful for better  understanding  conformal field theory.

\bigskip
{\bf Acknowledgments}. This work was supported by the INTAS project
03-51-6346, by the RFBR Grant No.05-02-17217, RFBR Grant for Leading
Scientific Schools, Grant No. 1578-2003-2, by Russian Science Support
Foundation, and by the Alexander von Humboldt Foundation Grant PHYS0167.

\setcounter{section}{0}\setcounter{subsection}{0}
\appendix{ Counting of on-shell D.o.F for spin-1 field }\label{app-01}

We analyze on-shell D.o.F for spin-1 conformal field that is described by
Lagrangian \rf{hihgderLag01}. To this end we use the framework of light-cone
gauge approach which turns out to be convenient
for our purpose.%
\footnote{ Discussion of alternative methods for counting on-shell D.o.F may
be found in Refs.\cite{Lee:1982cp,Buchbinder:1987vp}.}
Lagrangian \rf{hihgderLag01} leads to the following equations of motion:
\be \label{DoF01} \Box^{1+k} \phi^a - \Box^k  \partial^a \partial^b \phi^b
=0\,. \ee
Making use of the light-cone gauge,%
\footnote{ Light-cone coordinates in $\pm$ directions are defined as
$x^\pm=(x^{d-1} \pm x^0)/\sqrt{2}$ and $x^+$ is taken to be a light-cone
time. We adopt the conventions: $\partial^i =
\partial_i\equiv\partial/\partial x^i$, $\partial^\pm=\partial_\mp \equiv
\partial/\partial x^\mp$, $i,j=1,\ldots, d-2$. Lorentz algebra vector $X^a$
is decomposed as $X^a=(X^+,X^-,X^i)$. Note that $X^+=X_-$, $X^-=X_+$,
$X^i=X_i$.}
\be \label{DoF02} \phi^+ = 0\,, \ee
we obtain from `$+$' component of Eqs.\rf{DoF01}\,,
\be \label{DoF03} \partial^+ \Box^k \partial^b\phi^b = 0\,. \ee
As usually, kernel of the derivative $\partial^+$ is assumed to be trivial.
Therefore Eq.\rf{DoF03} amounts to the equation $\Box^k \partial^b\phi^b =
0$. Plugging the latter equation into \rf{DoF01}, we get $\Box^{1+k} \phi^a
=0$. To summarize, in the light-cone gauge, the generic Eqs.\rf{DoF01} amount
to the equations
\be
\label{boxk01} \Box^{1+k} \phi^a =0 \,,
\qquad\qquad
\Box^k \partial^a \phi^a =0\,.
\ee
To get ordinary-derivative form of Eqs.\rf{boxk01} we introduce $k+1$
vector fields and $k$ scalar fields,
\beq
&&  \label{DoF04} \phi_{k'}^a\,,\qquad k' \in [k]_2\,,
\\
&& \label{DoF05} \phi_{k'}\,,\qquad k' \in [k-1]_2\,,
\eeq
where we use the notation as in \rf{sumnot02} and the generic conformal
vector field $\phi^a$ is identified as $\phi_{-k}^a \equiv \phi^a$. We now
cast Eqs.\rf{boxk01} into the following ordinary-derivative form:
\beq \label{DoF07}
&& \Box \phi_{k'}^a - \phi_{k'+2}^a = 0\,,
\qquad k' \in [k]_2;
\\
\label{phi1phi0} && \partial^a \phi_{k'}^a - \phi_{k'+1} = 0\,,
\qquad k' \in [k]_2;
\\
\label{DoF08} && \Box \phi_{k'} - \phi_{k'+2} = 0\,,
\qquad k'= [k-1]_2\,,
\eeq
where we use the conventions $\phi_{k+2}^a\equiv 0$, $\phi_{k+1} \equiv 0$.
Note that light-cone gauge for generic field \rf{DoF02} and Eqs.\rf{DoF07}
lead to the light-cone gauge for vector fields $\phi_{k'}^a$ \rf{DoF04},
\be \label{DoF10} \phi_{k'}^+=0\,,\qquad k' \in [k]_2\,.\ee
Light-cone gauge \rf{DoF10} and constraints \rf{phi1phi0} allow us to express
$\phi_{k'}^-$ in terms of on-shell vector fields $\phi_{k'}^i$ and scalar
fields $\phi_{k'}$,

\be \phi_{k'}^- = -\frac{\partial^i}{\partial^+}\phi_{k'}^i +
\frac{1}{\partial^+}\phi_{k'+1}\,,\ee
i.e., on-shell, we have $k+1$ vector fields $\phi_{k'}^i$, $k'\in [k]_2$, and
$k$ scalar fields $\phi_{k'}$, $k'\in [k-1]_2$.

\appendix{ Spin-1 field: Derivation of Lagrangian, gauge transformations,
and operator $R^a$ } \label{appspin1}

We begin with the study of restrictions imposed by gauge symmetries. For the
spin-1 conformal field, general ordinary-derivative Lagrangian and gauge
transformations are given by
\beq
\label{20072011-04} \LL & = & \half \phibr E \phik \,,
\\
\label{20072011-05} && E = E_\smtwo + E_\smone + E_\smzero \,,
\\
\label{21072011-14} && E_\smtwo = \Box - \alpar\albpar\,,
\\
&& E_\smone =  \eb_1\alpar + e_1 \albpar\,,
\\
&& E_\smzero = m_1\,,
\\
\label{21072011-10} \delta \phik & = &  (G_\smone + G_\smzero) \xik \,,\qquad
G_\smone \equiv \alpar\,, \qquad G_\smzero \equiv b_1\,,
\\
\label{genl101} && e_1 = \zeta \ewt_1\bar\upsilon^\ominussm\,,
\hspace{2cm}
\eb_1 = \upsilon^\ominussm \ebwt_1 \bar\zeta\,,
\\
\label{genb101} && m_1 = \upsilon^\ominussm  \mwt_1 \bar\upsilon^\ominussm\,,
\hspace{1.5cm}
b_1 = \zeta \bwt_1 \bar\upsilon^\ominussm \,,
\\
\label{genb1012} && \ewt_1 = \ewt_1(\Delta')\,, \qquad\qquad\ \ \ \mwt_1 =
\mwt_1(N_\zeta,\Delta')\,,  \qquad \bwt_1 = \bwt_1(\Delta')\,,
\\
\label{genb1012a1} && \mwt_1^\dagger = \mwt_1\,, \qquad\qquad\qquad
\ewt_1^\dagger = -\ebwt_1\,.
\eeq
For the notation, see \rf{manold-31102011-02}-\rf{manold-31102011-06}.
We note that:\\
{\bf i)} The Maxwell operator $E_\smtwo$ \rf{21072011-14} is fixed by requiring
the $\phibr E_\smtwo\phik$ part of the Lagrangian to be invariant under
standard gradient $\alpar\xik$ part of gauge
transformations \rf{21072011-10};\\
{\bf ii)} dependence of operators $E$ \rf{20072011-05}, $G_\smone$,
$G_\smzero$ \rf{21072011-10} on the oscillators in \rf{genl101}-\rf{genb1012}
is fixed by requiring the ket-vectors $E\phik$, $(G_\smone + G_\smzero)\xik$
to satisfy constraints given in
\rf{oldman12012011-01},\rf{oldman12012011-02}; \\
{\bf iii)} quantities $\ewt_1$, $\mwt_1$, $\bwt_1$ in
\rf{genl101}-\rf{genb1012} depend $N_\zeta$ and $\Delta'$. Because $\phik$
\rf{phispin1def01} is polynomial of degree-1 in $\zeta$, the quantity
$\mwt_1$ should be determined only for $N_\zeta=0,1$.

Thus all that remains is to find dependence of the quantities $\ewt_1$,
$\mwt_1$, $\bwt_1$ on $N_\zeta$ and $\Delta'$. To this end we study
restrictions imposed by the gauge symmetries. Variation of Lagrangian
\rf{20072011-04} under gauge transformations \rf{21072011-10} takes the form
(up to total derivative)
\be \label{genb1014}
\delta \LL = \phibr  (e_1 + b_1)\Box
 + (m_1 + \eb_1b_1)\alpar + m_1b_1 \xik\,.
\ee
Requiring the variation to vanish gives the relations
\beq
\label{gauineq01} && (b_1 + e_1)\xik =0 \,,
\qquad
(m_1 +  \eb_1 b_1)\xik =0\,,
\qquad
m_1b_1 \xik = 0 \,.
\eeq
Using \rf{genl101},\rf{genb101}, we find that equations \rf{gauineq01} amount
to the equations
\beq
\label{genb1015} && \bwt_1 = -\ewt_1\,,
\qquad
\mwt_1(0,\Delta') = \ebwt_1 \ewt_1\,,
\qquad
\mwt_1(1,\Delta') \bwt_1^\smone  = 0\,,
\eeq
where $\bwt_1^\smone\equiv \bwt_1|_{\Delta'\rightarrow \Delta'+1}$. Equations
\rf{genb1015} are the restrictions imposed by the gauge symmetries. These
equations alone do now allow to determine $\ewt_1$, $\mwt_1$, $\bwt_1$
uniquely, i.e., the gauge symmetries alone are not enough to fix Lagrangian
and gauge transformations uniquely. Therefore we now consider restrictions
imposed by conformal boost symmetries.

{\bf Analysis of restrictions imposed by conformal boost symmetries}.
Requiring Lagrangian \rf{20072011-04} to be invariant (up to total
derivative) under conformal boost transformations given in \rf{20072011-03},
\rf{conalggenlis04}, we find the equations
\beq
\label{app3-15082009-04}  && R^{a \dagger}E + E R^a + E_\smzero^a +
E_\smone^a \approx 0\,,
\\
&& \hspace{1cm} E_\smone^a \equiv 2\Delta'\partial^a - (\Delta' +
\frac{d-4}{2})\alpha^a \albpar + ( - \Delta' + \frac{d-4}{2})\alpar
\bar\alpha^a\,,
\\
&& \hspace{1cm} E_\smzero^a \equiv \eb_1 (\Delta' + \frac{d-2}{2})\alpha^a -
( - \Delta' + \frac{d-2}{2})e_1 \bar\alpha^a\,.
\eeq
In \rf{app3-15082009-04} and below, to simplify our formulas, we adopt the
following convention. Let $A$ be some operator. We use the relation $A
\approx 0$ in place of $A\phik=0$, where $\phik$ is defined in
\rf{phispin1def01}.

The operator $R^a$ turns out to be degree-1 polynomial in the derivative.
Therefore it is convenient to represent the operator $R^a$ as the power series in
the derivative,
\be \label{kapkap00kpa01} R^a  =  R_\smzero^a + R_\smone^a\,,
\ee
where $R_\smn^a$ stands for the contribution involving $n$ derivatives. Using
power series expansions of operators $E$ \rf{20072011-05} and $R^a$
\rf{kapkap00kpa01}, we see that Eqs.\rf{app3-15082009-04} amount to the
following equations:
\beq
\label{E2kap1eq01}
&& E_\smtwo R_\smone^a + h.c. \approx  0 \,,
\\
\label{E2kap1eq02}
&& E_\smtwo R_\smzero^a +E_\smone R_\smone^a + h.c. \approx 0 \,,
\\
\label{E2kap1eq03}
&& ( E_\smone R_\smzero^a + E_\smzero R_\smone^a + h.c. ) + E_\smone^a
\approx 0 \,,
\\
\label{E2kap1eq04}
&& ( E_\smzero R_\smzero^a + h.c. )  + E_\smzero^a \approx 0 \,.
\eeq
Most general form of the operator $R^a$ that respects constraints in
\rf{oldman12012011-01},\rf{oldman12012011-02} is given by
\beq \label{manold-11112011-01}
R^a & =  & r_\smzero^a + r_\smone^a + R_\smG^a \,,
\\
&& r_\smzero^a = \rb_{0,1} \alpha^a + r_{0,1} \bar\alpha^a\,,
\\
&& r_\smone^a = r_{1,1} \partial^a + r_{1,5} \alpha^a \albpar\,,
\\
\label{manold07112011-01} && R_\smG^a =  G r_\smG^a\,,\qquad  G \equiv
G_\smone+ G_\smzero\,,
\\
&& r_\smG^a = r_{\smG,1}\bar\alpha^a\,,
\\
&& r_{0,1} = \zeta \rwt_{0,1} \bar\upsilon^\oplussm\,,
\qquad\quad \ \ \rb_{0,1} = \upsilon^\oplussm \rbwt_{0,1} \bar\zeta\,,
\\
&& r_{1,1} = \upsilon^\oplussm  \rwt_{1,1} \bar\upsilon^\oplussm\,,
\qquad\quad
r_{1,5} = \upsilon^\oplussm  \rwt_{1,5} \bar\upsilon^\oplussm\,,
\\
&& r_{\smG,1} = \upsilon^\oplussm  \rwt_{\smG,1} \bar\upsilon^\oplussm\,,
\eeq
\be \rbwt_{0,1} =\rbwt_{0,1}(\Delta')\,,\quad  \rwt_{0,1}
=\rwt_{0,1}(\Delta')\,,\quad  \rwt_{1,5} =\rwt_{1,5}(\Delta')\,,\quad
\rwt_{\smG,1} =\rwt_{\smG,1}(\Delta')\,, \ee
\be \rwt_{1,1} =\rwt_{1,1}(N_\zeta,\Delta')\,,
\ee
where $G_\smone$, $G_\smzero$ \rf{manold07112011-01} are given in
\rf{21072011-10}. Note that because $R^a$ \rf{manold-11112011-01} can be
presented as in \rf{kapkap00kpa01} with
\be
R_\smzero^a = r_\smzero^a + G_\smone r_\smG^a \,,
\qquad\quad
R_\smone^a = r_\smone^a + G_\smzero r_\smG^a\,,
\ee
and in view of the relation $E Gr_\smG^a\phik= 0$, all terms proportional to
$r_\smG^a$ cancel in Eqs.\rf{E2kap1eq01}-\rf{E2kap1eq04}. This implies that
we can replace $r_\smzero^a$ for $R_\smzero^a$ and $r_\smone^a$ for
$R_\smone^a$ in \rf{kapkap00kpa01} when we analyze
Eqs.\rf{E2kap1eq01}-\rf{E2kap1eq04}. We now analyze
Eqs.\rf{E2kap1eq01}-\rf{E2kap1eq04} in turn.

\noindent {\bf i)} Using the relation
\beq
E_\smtwo r_\smone^a & \approx &  r_{1,1}E_\smtwo\partial^a  + r_{1,5}\alpha^a
\Box\bar\alpha\partial - r_{1,5} \alpar\, \albpar\,  \partial^a\,,
\eeq
we find that requiring the $r_\smone^a$ to satisfy Eqs.\rf{E2kap1eq01} gives
\be r_{1,1}^\dagger = r_{1,1}\,,\qquad r_{1,5} = 0 \,. \ee

\noindent {\bf ii)} Using the relations
\beq
E_\smtwo r_\smzero^a & \approx &  \rb_{0,1}\alpha^a \Box + r_{0,1}
\Box\bar\alpha^a - \rb_{0,1} \alpar\, \partial^a\,,
\\
E_\smone r_\smone^a & \approx & \eb_1 r_{1,1}\alpar\, \partial^a  + e_1
r_{1,1}\albpar\,
\partial^a\,,
\eeq
we obtain that requiring the $r_\smzero^a$, $r_\smone^a$ to satisfy
Eqs.\rf{E2kap1eq02} amounts to the restrictions
\be \label{sec150026} \rb_{0,1} = [\eb_1,r_{1,1}]\,,  \qquad r_{0,1} \equiv -
\rb_{0,1}^\dagger\,.\ee

\noindent {\bf iii)}  Using the relations
\be
\label{sec150028} E_\smone r_\smzero^a  \approx   \eb_1 r_{0,1} \alpar
\bar\alpha^a + e_1 \rb_{0,1} \partial^a\,,
\qquad\quad
E_\smzero r_\smone^a \approx m_1 r_{1,1} \partial^a\,,
\ee
we make sure that Eqs.\rf{E2kap1eq03} amount to the equations
\beq
\label{sec150030} && -r_{0,1} \eb_1 + e_1 \rb_{0,1} + [m_1,r_{1,1}] +
2\Delta'\Bigr|_{N_\zeta =0,1 }  =0\,,
\\
\label{sec150031}&&  \eb_1 r_{0,1} - \Delta'  + \frac{d-4}{2}\Bigr|_{N_\zeta
=0 } =0 \,,
\\
\label{sec150032} &&  \rb_{0,1} e_1 + \Delta'  + \frac{d-4}{2}\Bigr|_{N_\zeta
=0 } =0 \,.
\eeq
In \rf{sec150030}-\rf{sec150032}, we use the shortcut $A|_{N_\zeta=0,1}=0$
for the respective two equations $A|\phi_1\rangle=0$, $A|\phi_0\rangle=0$,
where ket-vectors $|\phi_1\rangle$ and $|\phi_0\rangle$ are defined in
\rf{phispin1def02},\rf{phispin1def03}. Using \rf{sec150031}, we find
\be \label{sec150036}
\ebwt_1 \rwt_{0,1} = - 2 \,.
\ee
Eq.\rf{sec150036} implies $\ebwt_1 \ne 0$ for all $\Delta'$. Taking this into
account, using \rf{genb1012a1} and appropriate field redefinitions, we find
\be \label{sec150038} \ebwt_1 = - 1\,,\qquad  \ewt_1 = 1\,. \ee
Using \rf{sec150036}, we get the relation $\rwt_{0,1} = 2$. Using these
relations and \rf{genb1012},\rf{genb1015}, we find
\be  \label{sec150039n1} \mwt_1(0,\Delta') = -1\,,\qquad \mwt_1(1,\Delta') =
0\,,\qquad  \bwt_1 = -1\,. \ee
We proceed to examine Eqs.\rf{sec150030}. Using the results obtained, we find
that Eqs.\rf{sec150030} amount to
\be \rwt_{1,1} = - 2\,. \ee
With the results obtained, we make sure that Eq.\rf{sec150032} is satisfied
automatically.

\noindent {\bf iv)} Finally, we make sure that Eqs.\rf{E2kap1eq04} are
satisfied automatically.

\appendix{ Counting of on-shell D.o.F for spin-2 field }

We analyze on-shell D.o.F for spin-2 conformal field that is described by
Lagrangian \rf{Lconfie2lag01n01}. Lagrangian \rf{Lconfie2lag01n01} leads to
the equations of motion
\be \label{initeqofmot01}
\Box^{k+1} P^{ab\, ce} \phi^{ce} = 0\,,
\ee
which are invariant under linearized diffeomorphism gauge symmetries and Weyl
conformal gauge symmetry \rf{secspin2con10}. We analyze
Eqs.\rf{initeqofmot01} by using the light-cone gauge to fix the
diffeomorphism gauge symmetries and the traceless condition to fix the Weyl
gauge symmetry,
\beq \label{lcgaug01}
&& \phi^{+a} = 0 \,,
\qquad \phi^{aa}= 0 \,.
\eeq
Using \rf{lcgaug01}, we find that `$++$' component of equations
\rf{initeqofmot01},
\be \label{spin2DoF10}
\Box^{k+1} P^{++\, ce} \phi^{ce} = 0 \,,
\ee
amounts to the constraint
\be \label{spin2DoF11} \Box^{k-1} \partial^a\partial^a \phi^{ab} =0 \,.\ee
Plugging \rf{spin2DoF11} into \rf{initeqofmot01}, we obtain
\be \label{initeqofmot02}
\Box^{k+1}  \phi^{ab}  -\partial^a \Box^k \partial^c \phi^{bc}  -
\partial^b \Box^k \partial^c \phi^{ac} = 0 \,.
\ee
`$a+$' components of Eqs.\rf{initeqofmot02} give
\be \label{spin2DoF12} \Box^k \partial^b \phi^{ab} =  0 \,.\ee
To summarize, the generic equations \rf{initeqofmot01} amount to the
equations
\be
\label{spin2DoF13}  \Box^{k+1} \phi^{ab} =  0 \,,
\qquad
\Box^k \partial^b \phi^{ab} =0 \,,
\qquad
\Box^{k-1} \partial^a\partial^b \phi^{ab} =0 \,.
\ee
Introducing fields $\phi_{k'}^{ab}$, $k'\in [k]_2$, $\phi_{k'}^a$, $k'\in
[k-1]_2$, $\phi_{k'}$, $k'\in [k-2]_2$, and making use of the identification
$\phi_{-k}^{ab} \equiv \phi^{ab}$, we rewrite Eqs.\rf{spin2DoF13} into the
ordinary-derivative form,
\beq
\label{spin2DoF17}  && \Box \phi_{k'}^{ab} - \phi_{k'+2}^{ab} = 0\,,
\quad\qquad k'\in [k]_2;
\\
\label{spin2DoF18}  && \Box \phi_{k'}^a - \phi_{k'+2}^a = 0\,,
\quad\qquad k'\in [k-1]_2;
\\
\label{spin2DoF19}  && \Box \phi_{k'} - \phi_{k'+2} = 0\,,
\quad\qquad k'\in [k-2]_2;
\\
\label{spin2DoF20}  && \partial^b \phi_{k'}^{ab} - \phi_{k'+1}^a = 0\,,
\quad\qquad k'\in [k]_2;
\\
\label{spin2DoF21}  && \partial^a \phi_{k'}^a - \phi_{k'+1} = 0\,,
\quad\qquad k'\in [k-1]_2\,,
\eeq
where we assume the conventions $\phi_{k+2}^{ab}\equiv 0$,
$\phi_{k+1}^a\equiv 0$, $\phi_{k} \equiv 0$. Gauge conditions \rf{lcgaug01}
and Eqs.\rf{spin2DoF17},\rf{spin2DoF18} lead to
\beq
\label{spin2DoF23}  && \phi_{k'}^{+a} = 0\,,\quad \phi_{k'}^{ii}=0\,,
\quad\qquad \ k'\in [k]_2;
\\
\label{spin2DoF24}  && \phi_{k'}^+ = 0\,,
\hspace{3.5cm} k'\in [k-1]_2\,.
\eeq
Light-cone gauge \rf{spin2DoF23},\rf{spin2DoF24} and constraints
\rf{spin2DoF20},\rf{spin2DoF21} allow us to express non-dynamical fields
$\phi_{k'}^{-i}$, $\phi_{k'}^{--}$, and $\phi_{k'}^-$ in terms of dynamical
fields $\phi_{k'}^{ij}$, $\phi_{k'}^i$, and $\phi_{k'}$,
\beq
&& \phi_{k'}^{-i} = -\frac{\partial^j}{\partial^+}\phi_{k'}^{ij} +
\frac{1}{\partial^+}\phi_{k'+1}^i\,,
\\
&& \phi_{k'}^{--} = \frac{\partial^i
\partial^j}{\partial^+\partial^+} \phi_{k'}^{ij}
-2 \frac{\partial^i}{\partial^+\partial^+}\phi_{k'+1}^i +
\frac{1}{\partial^+\partial^+}\phi_{k'+2}\,,
\\
&& \phi_{k'}^- = -\frac{\partial^i}{\partial^+}\phi_{k'}^i +
\frac{1}{\partial^+}\phi_{k'+1}\,.
\eeq
To summarize, we are left with on-shell D.o.F given by $k+1$ traceless rank-2
tensor fields $\phi_{k'}^{ij}$, $k'\in [k]_2$, \ \ $k$ vector fields
$\phi_{k'}^i$, $k'\in [k-1]_2$, and $k-1$ scalar fields $\phi_{k'}$, $k'\in
[k-2]_2$, of the $so(d-2)$ algebra.

\appendix{ Spin-2 field: Derivation of Lagrangian, gauge transformations,
and operator $R^a$ }\label{appspin2}

We begin with the study of restrictions imposed by gauge symmetries. For the
spin-2 conformal field, general ordinary-derivative  Lagrangian and gauge
transformations are given by
\beq
\label{20072011-08} \LL &  = & \frac{1}{2} \phibr E \phik\,,
\\
\label{20072011-09} E  & = & E_\smtwo + E_\smone + E_\smzero\,,
\\
\label{21072011-05}  && E_\smtwo  \equiv  \Box - \alpar \albpar +
\frac{1}{2}(\alpar)^2\bar\alpha^2 + \frac{1}{2} \alpha^2 (\albpar)^2 -
\frac{1}{2}\alpha^2 \Box \bar\alpha^2\,,
\\
\label{21072011-05a1} && E_\smone  \equiv  \eb_1\alpar  + e_1 \albpar  +
\eb_2 \alpha^2 \albpar + e_2 \alpar \bar\alpha^2\,,
\\
&& E_\smzero  \equiv   m_1 + \alpha^2\bar\alpha^2m_2 + \mb_3 \alpha^2 + m_3
\bar\alpha^2\,,
\\
\label{21072011-01} \delta \phik & = &  (G_\smone+ G_\smzero) \xik\,,\qquad
G_\smone \equiv \alpar\,, \qquad G_\smzero \equiv  b_1 + \alpha^2 b_2\,,
\\
\label{21072011-06} && e_1 = \zeta \ewt_1 \bar\upsilon^\ominussm\,,
\qquad
\eb_1 =  \upsilon^\ominussm \ebwt_1 \bar\zeta\,,
\qquad
e_2 = \zeta \ewt_2 \bar\upsilon^\ominussm\,,
\qquad
\eb_2 =  \upsilon^\ominussm \ebwt_2 \bar\zeta\,,
\\
\label{21072011-07} && m_1 = \upsilon^\ominussm \mwt_1 \bar\upsilon^\ominussm
\,,\quad
m_2 = \upsilon^\ominussm \mwt_2 \bar\upsilon^\ominussm \,,\quad
\\
\label{21072011-08} && m_3 = \zeta^2 \mwt_3\bar\upsilon^\ominussm
\bar\upsilon^\ominussm\,, \quad
\mb_3 = \upsilon^\ominussm \upsilon^\ominussm \mbwt_3 \bar\zeta^2\,,
\\
\label{21072011-09} && b_1 = \zeta \bwt_1 \bar\upsilon^\ominussm\,,\qquad b_2
= \upsilon^\ominussm \bwt_2 \bar\zeta\,,
\\
\label{21072011-02} && \ewt_1=\ewt_1(N_\zeta,\Delta')\,,\quad
\ebwt_1=\ebwt_1(N_\zeta,\Delta')\,,\quad \ewt_2 =\ewt_2(\Delta')\,,\quad
\ebwt_2=\ebwt_2(\Delta')\,,
\\
\label{21072011-03} && \mwt_1=\mwt_1(N_\zeta,\Delta')\,,\quad
\mwt_2=\mwt_2(\Delta')\,,\quad \mwt_3 =\mwt_3(\Delta')\,,\quad
\mbwt_3=\mbwt_3(\Delta')\,,\qquad
\\
\label{21072011-04} && \bwt_1 =  \bwt_1(N_\zeta,\Delta')\,, \qquad \bwt_2 =
\bwt_2(\Delta')\,,
\\
\label{21072011-04x1} && \mwt_1^\dagger = \mwt_1\,, \quad \mwt_2^\dagger =
\mwt_2\,, \quad  \mwt_3^\dagger = \mbwt_3\,, \qquad \ewt_n^\dagger =
-\ebwt_n\,\qquad n=1,2\,.
\eeq
We note that:\\
{\bf i)} The Einstein-Hilbert operator $E_\smtwo$ \rf{21072011-05} is fixed by
requiring the $\phibr E_\smtwo\phik$ part of the Lagrangian to be invariant
under standard gradient $\alpar\xik$ part of gauge
transformations \rf{21072011-01};\\
{\bf ii)} dependence of operators $E$ \rf{20072011-09}, $G_\smone$,
$G_\smzero$ \rf{21072011-01} on the oscillators in
\rf{21072011-06}-\rf{21072011-04} is fixed by requiring the ket-vectors
$E\phik$, $(G_\smone+G_\smzero)\phik$ to satisfy constraints given in
\rf{21072011-11},\rf{21072011-13};\\
{\bf iii)} quantities $\ewt$, $\mwt$, $\bwt$ in
\rf{21072011-02}-\rf{21072011-04} depend $N_\zeta$ and $\Delta'$.  Note that
because $\phik$ \rf{phispin2def01} is polynomial of degree-2 in $\zeta$, the
quantity $\mwt_1$ should be determined only for $N_\zeta=0,1,2$, while the
quantities $\ewt_1$, $\ebwt_1$, $\bwt_1$ should be determined only for
$N_\zeta=0,1$.

Thus all that is required is to find the dependence of the quantities $\ewt$,
$\mwt$, $\bwt$ on $N_\zeta$ and $\Delta'$. To this end we study restrictions
imposed by the gauge symmetries. Variation of Lagrangian \rf{20072011-08}
under gauge transformations \rf{21072011-01} takes the form (up to total
derivative)
\be
\delta \label{oldman-14072011-01} \LL  =  \phibr (\VV_\smtwo + \VV_\smone +
\VV_\smzero)\xik\,,
\ee
where $\VV_{(n)}$ stands for contribution involving $n$ derivatives.
Obviously, gauge invariance of Lagrangian requires the equations
$\VV_{(n)}\xik = 0$, $n=2,1,0$. First, we consider the equation
$\VV_\smtwo\xik = 0$. To this end we compute $\VV_\smtwo$,
\beq
\VV_\smtwo
& = &  \Box X_{\smtwo,1}
+ \alpha\partial\bar\alpha\partial X_{\smtwo,2}
+ \alpha^2\Box X_{\smtwo,4}
+ (\alpha\partial)^2 X_{\smtwo,5}\,,
\nonumber\\
&& X_{\smtwo,1} \equiv  e_1 +  b_1\,,
\hspace{2cm}
X_{\smtwo,2} \equiv  e_1 + 2 e_2 - b_1\,,
\nonumber\\
&& X_{\smtwo,4} \equiv  \eb_2 - (d-2) b_2\,,
\qquad
X_{\smtwo,5} \equiv \eb_1 + (d-2) b_2\,,
\eeq
and note that solution to the equation $ \VV_\smtwo \xik = 0$ is given by
\beq
\label{21072011-17} && \bwt_1 = - \ewt_1\,, \qquad \bwt_2 =
-\frac{1}{d-2}\ebwt_1(0,\Delta') \,,
\\
\label{21072011-18} && \ewt_2 = - \ewt_1(0,\Delta')\,, \qquad  \ebwt_2 = -
\ebwt_1(0,\Delta')\,.
\eeq
Relations \rf{21072011-18} imply the following representation for $E_\smone$
\rf{21072011-05a1}:
\be \label{E1sol01}  E_\smone  \equiv  \eb_1(\alpar  -\alpha^2 \albpar) + e_1
(\albpar - \alpar \bar\alpha^2)\,.\ee
Second, we consider the equation $\VV_\smone\xik=0$. Using $E_\smone$
\rf{E1sol01}, we compute the gauge variation of Lagrangian \rf{20072011-08}
to obtain the following expression for $\VV_\smone$ \rf{oldman-14072011-01}:
\beq \label{E1sol01a1}
\VV_\smone
& = &  \alpha\partial X_{\smone,1}
+ \alpha^2\bar\alpha\partial X_{\smone,2}
+ \bar\alpha\partial X_{\smone,4}\,,
\nonumber\\
&& X_{\smone,1} \equiv m_1 + \eb_1 b_1 - 2 ( d - 1 )e_1 b_2\,,
\nonumber\\
&& X_{\smone,2} \equiv 2m_2 - \eb_1b_1\,,
\qquad
X_{\smone,4} \equiv 2m_3 + e_1 b_1 \,.
\eeq
Using \rf{21072011-17},\rf{E1sol01a1}, we make sure that the equation
$\VV_\smone \xik = 0$ amounts to the equations%
\beq \label{spin2m101}
\mwt_1 &  = &  (N_\zeta+1) \ebwt_1 \ewt_1 - 2\frac{d-1}{d - 2}\,N_\zeta \,
\ewt_1^{(-1,-1)} \ebwt_1^{(-1,1)}\,,\qquad \hbox{ for }  \ \ N_\zeta=0,1\,,
\\
\label{spin2m102} \mwt_2 &  = & - \half \ebwt_1(0,\Delta')
\ewt_1(0,\Delta')\,,
\\
\label{spin2m103} \mwt_3 & = & \frac{1}{2}
\ewt_1(1,\Delta')\ewt_1(0,\Delta'+1) \,,
\eeq
where, in \rf{spin2m101} and below, $\ewt_1^{(p,q)}$ stands for
$\ewt_1(N_\zeta+p,\Delta'+q)$. The same shortcut we use for $\mwt_1$. Also
note that Eqs.\rf{spin2m101}-\rf{spin2m103} should be considered on space of
ket-vector $\xik$ \rf{epsspi2def01}.

Finally, we consider the equation $\VV_\smzero\xik=0$. Expression for
$\VV_\smzero$ \rf{oldman-14072011-01} can be presented as
\beq
\VV_\smzero & = & X_{\smzero,1} +\alpha^2 X_{\smzero,2}\,,
\nonumber\\
&& X_{\smzero,1}  \equiv  m_1 b_1 + 2d m_3 b_2\,,
\qquad\quad
X_{\smzero,2}  \equiv    m_1 b_2 + 2d m_2 b_2 + \mb_3 b_1\,.\qquad
\eeq
Using \rf{21072011-17}, we find that the equation $\VV_\smzero \xik = 0$
amounts to the following equations:
\beq \label{spin2m104}
&& \mwt_1^{(1,0)} \ewt_1^{(0,1)} +  \frac{2d}{d - 2} N_\zeta \mwt_3^{(-1)}
\ebwt_1^{(-1,2)}\Bigr|_{N_\zeta=0,1} =0 \,,
\\
\label{spin2m105} &&   \frac{1}{d-2} \mwt_1^{(0,-1)} \ebwt_1^{(0,1)}  +
\frac{2d}{d-2} \mwt_2^{(-1)} \ebwt_1^{(0,1)}  + 2\mbwt_3
\ewt_1^{(1,0)}\Bigr|_{N_\zeta=0} =0 \,,
\eeq
where, in \rf{spin2m104} and below, we use the shortcuts $\mwt_2^{(p)}\equiv
\mwt_2|_{\Delta' \rightarrow \Delta'+p}$, $\mwt_3^{(p)}\equiv
\mwt_3|_{\Delta' \rightarrow \Delta'+p}$. In \rf{spin2m104},\rf{spin2m105},
we use shortcut $A|_{N_\zeta=0,1}=0$ for the respective two equations
$A|\xi_1\rangle=0$ and $A|\xi_0\rangle=0$, where the ket-vectors
$|\xi_1\rangle$ and $|\xi_0\rangle$ are defined in
\rf{epsspi2def01x1},\rf{epsspi2def01x2}.

Equations \rf{21072011-17},\rf{21072011-18}, \rf{spin2m101}-\rf{spin2m103},
and \rf{spin2m104}, \rf{spin2m105} are the restrictions imposed by the gauge
symmetries. These equations alone do not allow to determine quantities
$\ewt$, $\mwt$, $\bwt$ \rf{21072011-02}-\rf{21072011-04} uniquely. Therefore
we proceed with analysis of restrictions imposed by conformal boost
symmetries.

{\bf Analysis of restrictions imposed by conformal boost symmetries}.
Requiring Lagrangian \rf{20072011-08} to be invariant (up to total
derivative) under conformal boost transformations given in \rf{20072011-03},
\rf{conalggenlis04}, we find the equations
\beq
\label{app3-15082009-09}  && \hspace{-1.7cm} R^{a \dagger}E + E R^a +
E_\smzero^a + E_\smone^a \approx 0\,,
\\
E_\smone^a & \equiv & \Delta' (2-\alpha^2\bar\alpha^2) \partial^a
\nonumber\\
& - &  (\Delta' - N_\zeta + \frac{d-2}{2})\alpha^a \albpar + ( - \Delta'
-N_\zeta + \frac{d-2}{2})\alpar \bar\alpha^a
\nonumber\\
& + &  (\Delta' + \frac{d-2}{2})\alpar \alpha^a \bar\alpha^2 - ( - \Delta' +
\frac{d-2}{2})\alpha^2 \bar\alpha^a\albpar\,,
\\
E_\smzero^a & \equiv & \eb_1 (\Delta' -N_\zeta + \frac{d+2}{2})\alpha^a - ( -
\Delta' - N_\zeta + \frac{d+2}{2})e_1 \bar\alpha^a
\nonumber\\
& - &   (\Delta' + \frac{d-2}{2}) e_1 \alpha^a \bar\alpha^2 + \eb_1  ( -
\Delta' + \frac{d-2}{2}) \alpha^2 \bar\alpha^a\,.
\eeq
In \rf{app3-15082009-09} and below, we simplify our formulas as follows. Let
$A$ be some operator. We use the relation $A \approx 0$ in place of
$A\phik=0$, where $\phik$ is defined in \rf{phispin2def01}.

Because the operator $R^a$ turns out to be degree-1 polynomial in the
derivative we represent the operator $R^a$ as power series in the derivative,
\be \label{2kapkap00kpa01}
R^a =  R_\smzero^a+ R_\smone^a\,,
\ee
where $R_\smn^a$ stands for the contribution involving $n$ derivatives. Using
power series expansions of operators $E$ \rf{20072011-09} and $R^a$
\rf{2kapkap00kpa01}, we see that Eqs.\rf{app3-15082009-09} amount to the
following equations:
\beq
\label{2E2kap1eq01}
&& E_\smtwo R_\smone^a + h.c. \approx 0 \,,
\\
\label{2E2kap1eq02}
&& E_\smtwo R_\smzero^a +E_\smone R_\smone^a + h.c. \approx 0 \,,
\\
\label{2E2kap1eq03}
&& ( E_\smone R_\smzero^a + E_\smzero R_\smone^a + h.c. ) + E_\smone^a
\approx 0 \,,
\\
\label{2E2kap1eq04}
&& ( E_\smzero R_\smzero^a + h.c. )  + E_\smzero^a \approx 0 \,.
\eeq
Most general form of the operator $R^a$ that respects constraints
\rf{21072011-11},\rf{21072011-13} is given by
\beq \label{manold-11112011-05}
R^a & = & r_\smzero^a + r_\smone^a + R_\smG^a \,,
\\
r_\smzero^a & = & \rb_{0,1} (\alpha^a - \frac{1}{d-2} \alpha^2 \bar\alpha^a)
+ r_{0,1} \bar\alpha^a + \rb_{0,6}\alpha^2\bar\alpha^a +
r_{0,6}\alpha^a\bar\alpha^2  \,,
\\
r_\smone^a &= & (r_{1,1} + r_{1,2}\alpha^2\bar\alpha^2 + r_{1,3}\alpha^2 +
r_{1,4}\bar\alpha^2)\partial^a
\nonumber\\
& + & r_{1,5} (\alpha^a - \frac{1}{d-2} \alpha^2 \bar\alpha^a)\albpar +
r_{1,6} \bar\alpha^a \albpar + r_{1,10} \alpha^2 \bar\alpha^a \albpar \,,
\\
R_\smG^a & =  & G r_\smG^a\,,\qquad G \equiv G_\smone + G_\smzero\,,
\\
&& r_\smG^a = r_{\smG,1} \bar\alpha^a  + r_{\smG,2} \alpha^a + r_{\smG,3}
\alpha^a \bar\alpha^2\,,
\\
&& r_{0,n} = \zeta \rwt_{0,n} \bar\upsilon^\oplussm\,,\qquad  \rb_{0,n} =
\upsilon^\oplussm \rbwt_{0,n} \bar\zeta\,, \qquad n=1,6\,;
\\
&& r_{1,n} = \upsilon^\oplussm  \rwt_{1,n} \bar\upsilon^\oplussm\,,\qquad
n=1,2,5,10\,;
\\
&& r_{1,3} = \upsilon^\oplussm \upsilon^\oplussm  \rwt_{1,3} \bar\zeta^2 \,,
\\
&& r_{1,n} =  \zeta^2 \rwt_{1,n} \bar\upsilon^\oplussm
\bar\upsilon^\oplussm\,, \qquad n= 4,6\,;
\\
&& r_{\smG,n} =  \upsilon^\oplussm  \rwt_{\smG,n} \bar\upsilon^\oplussm \,,
\qquad n= 1,3\,;
\\
&& r_{\smG,2} =  \upsilon^\oplussm \upsilon^\oplussm \rwt_{\smG,2}
\bar\zeta^2 \,,
\\
\label{21072011-19} && \rwt_{0,1} =\rwt_{0,1}(N_\zeta, \Delta')\,, \qquad
\rbwt_{0,1} =\rbwt_{0,1}(N_\zeta, \Delta')\,,
\\
&& \rwt_{0,6} =\rbwt_{0,6}(\Delta')\,, \hspace{1.5cm} \rbwt_{0,6} =
\rbwt_{0,6}(\Delta')\,,
\\
&& \rwt_{1,n} =\rwt_{1,n}(N_\zeta,\Delta')\,, \ \quad n=1,5\,;\qquad
\\
&& \rwt_{1,n} =\rwt_{1,n}(\Delta')\,, \qquad \quad n=2,3,4,6,10\,;
\\
\label{21072011-20} && \rwt_{\smG,1} = \rwt_{\smG,1}(N_\zeta,\Delta')\,,
\qquad \rwt_{\smG,n} = \rwt_{\smG,n}(\Delta')\,,\qquad n=2,3\,,
\eeq
where $G_\smone$, $G_\smzero$ are given in \rf{21072011-01}. We see that in
order to fix $R^a$ we have to find quantities $\rwt$ in
\rf{21072011-19}-\rf{21072011-20}. Note that because $R^a$
\rf{manold-11112011-05} can be presented as in \rf{2kapkap00kpa01} with
\be
R_\smzero^a = r_\smzero^a + G_\smone r_\smG^a \,,
\qquad\quad R_\smone^a = r_\smone^a + G_\smzero r_\smG^a\,,
\ee
and view of the relation $E Gr_\smG^a\phik = 0$, all terms proportional to
$r_\smG^a$ cancel automatically in Eqs.\rf{2E2kap1eq01}-\rf{2E2kap1eq04}.
This implies that, in \rf{2kapkap00kpa01}, we can replace $r_\smzero^a$ for
$R_\smzero^a$ and $r_\smone^a$ for $R_\smone^a$ in analysis of
Eqs.\rf{2E2kap1eq01}-\rf{2E2kap1eq04}. We now analyze
Eqs.\rf{2E2kap1eq01}-\rf{2E2kap1eq04} in turn.

\noindent {\bf i)} Because analysis of Eqs.\rf{2E2kap1eq01} is straightforward
we just present our result. This is to say that Eqs.\rf{2E2kap1eq01} lead to
the following constraints:
\be r_{1,1}^\dagger = r_{1,1}\,,\quad r_{1,n} =  0\,,\qquad n=2,3,4,5,6,10\,.
\ee
{\bf ii)} Analysis of Eqs.\rf{2E2kap1eq02} is also straightforward. Therefore
we just present result of our analysis. Namely,  Eqs.\rf{2E2kap1eq02} amount
to the following relations:
\be
\label{k01l1k11} \rb_{0,1} = [\eb_1,r_{1,1}]\,,
\qquad r_{0,1} = - \rb_{0,1}^\dagger\,,
\qquad \rb_{0,6} = 0\,,
\qquad r_{0,6} = 0\,.
\ee
To summarize, Eqs.\rf{2E2kap1eq01},\rf{2E2kap1eq02} lead to the following
expressions for $r_\smzero^a$, $r_\smone^a$:
\beq
&& r_\smzero^a = \rb_{0,1}\bigl(\alpha^a - \frac{1}{d-2}\alpha^2 \bar\alpha^a
\bigr) + r_{0,1} \bar\alpha^a \,,
\qquad
r_\smone^a = r_{1,1}\partial^a \,,
\\
&& \rb_{0,1} = [\eb_1,r_{1,1}]\,, \qquad r_{1,1}^\dagger = r_{1,1}\,.
\eeq
\noindent {\bf iii)} We proceed with the detailed analysis of
Eqs.\rf{2E2kap1eq03}. Using the relations
\beq
\label{2sec150028} E_\smone r_\smzero^a & = &   e_1 \rb_{0,1}\partial^a   -
\eb_1 \rb_{0,1}\alpha^2 \partial^a
\nonumber\\
& + & \bigl(\frac{2}{d-2} e_1 \rb_{0,1}  + \eb_1 r_{0,1} \bigr) \alpar
\bar\alpha^a
+  e_1 \rb_{0,1} \alpha^a \albpar
\nonumber\\
& + & \eb_1 \rb_{0,1}\alpha^a \alpar  + e_1 r_{0,1}\albpar \bar\alpha^a -
\eb_1 r_{0,1} \alpha^2  \albpar \bar\alpha^a\,,
\\
\label{2sec150029} E_\smzero r_\smone^a & = &\bigl( m_1 r_{1,1} + m_2
r_{1,1}\alpha^2\bar\alpha^2 + \mb_3r_{1,1}\alpha^2  + m_3 r_{1,1}
\bar\alpha^2\bigr)\partial^a\,,
\eeq
we find that Eqs.\rf{2E2kap1eq03} amount to the following equations:
\beq
\label{2sec150030}
&& - r_{0,1} \eb_1 + e_1 \rb_{0,1} + [m_1,r_{1,1}] + 2\Delta'\Bigr|_{N_\zeta
=0,1,2}  =0\,,
\\
\label{2sec150030n01} && [m_2,r_{1,1}] - \Delta'\Bigr|_{N_\zeta =0} =0\,,
\\
\label{2sec150030n03}
&&  - \eb_1 \rb_{0,1} + [\mb_3, r_{1,1}]\Bigr|_{N_\zeta =0} =0\,,
\\
\label{2sec150030n03n1}
&&  r_{0,1} e_1  + [m_3, r_{1,1}]\Bigr|_{N_\zeta =0} =0\,,
\\
\label{2sec150031}
&& \frac{2}{d-2} e_1 \rb_{0,1} + \eb_1 r_{0,1} - r_{0,1} \eb_1 - \Delta' -
N_\zeta + \frac{d-2}{2}\Bigr|_{N_\zeta =0,1} =0 \,,
\\
\label{2sec150032} && - \eb_1 r_{0,1} + \Delta'  -
\frac{d-2}{2}\Bigr|_{N_\zeta =0 } =0 \,,
\\
\label{2sec150032n02} && [\eb_1, \rb_{0,1}]\Bigr|_{N_\zeta =0} =0 \,.
\eeq
In \rf{2sec150030}-\rf{2sec150032n02}, we use shortcut $A|_{N_\zeta=0,1,2}=0$
for the respective three equations $A|\phi_2\rangle=0$, $A|\phi_1\rangle=0$,
$A|\phi_0\rangle=0$, where ket-vectors $|\phi_2\rangle$,  $|\phi_1\rangle$,
and $|\phi_0\rangle$ are defined in \rf{phispin2def01x1},
\rf{phispin2def01x2}, and \rf{phispin2def04}. We proceed with analysis of
Eqs.\rf{2sec150031}. For $N_\zeta=0$, Eqs.\rf{2sec150031} lead to the
relation
\be
\label{2sec150034n1} \ebwt_1(0,\Delta') \rwt_{0,1}(0,\Delta') = - 2 \,,
\ee
which implies that $\ebwt_1(0,\Delta') \ne 0$. Taking this into account,
using \rf{21072011-04x1} and appropriate field redefinitions, we find the
solution
\be  \label{2sec150034n3} \ebwt_1(0,\Delta') = - 1 \,,\qquad
\ewt_1(0,\Delta') =  1 \,. \ee
Using $\rwt_{0,1}^\dagger = - \rbwt_{0,1}$ (see \rf{k01l1k11}) and
\rf{2sec150034n1},\rf{2sec150034n3}, we get
\be  \label{2sec150034n4} \rwt_{0,1}(0,\Delta') = 2 \,,\qquad
\rbwt_{0,1}(0,\Delta') = -2 \,.\ee
Using \rf{2sec150034n3},\rf{2sec150034n4}, we find that, for $N_\zeta=1$,
Eqs.\rf{2sec150031} lead to
\be
\label{2sec150034n5} \ebwt_1(1,\Delta') \rwt_{0,1}(1,\Delta') = -
2\frac{d-1}{d-2}\,,
\ee
while Eqs.\rf{2sec150032n02} lead to
\be \label{2sec150034n6} \rbwt_{0,1}(1,\Delta') = 2\ebwt_1(1,\Delta') \,,
\qquad  \rwt_{0,1}(1,\Delta') = 2\ewt_1(1,\Delta') \,.\ee
Plugging \rf{2sec150034n6} into \rf{2sec150034n5}, we obtain
\be \label{2sec150034n7} \ebwt_1(1,\Delta') \ewt_1(1,\Delta') = - u_\f^2\,,
\ee
where $u_\f$ is given in \rf{oldman-26012011-03}. From \rf{21072011-04x1},
\rf{2sec150034n7}, we see that $\ewt_1(1,\Delta')$ is fixed uniquely up to a
phase. Using field redefinitions, the phase of $\ewt_1(1,\Delta')$ can be set
equal to zero. We obtain then
\be \label{l11finsol01}  \ewt_1(1,\Delta') = u_\f\,,\qquad \ebwt_1(1,\Delta')
= - u_\f\,. \ee
Using \rf{2sec150034n3},\rf{l11finsol01} and \rf{spin2m101}-\rf{spin2m103},
we find
\be \label{m1m2m3sol01}
\mwt_1(0,\Delta') = -1\,,\quad   \mwt_1(1,\Delta') = 0,\quad \mwt_2 =
\half\,,\quad \mwt_3 = \half u_\f\,,\quad \mbwt_3 = \half u_\f\,.
\ee
Eq.\rf{spin2m104} and relations \rf{spin2m101},\rf{spin2m102} allow us to fix
$\mwt_1(2,\Delta')$:
\be  \label{m1m2m3sol02} \mwt_1(2,\Delta') = \frac{d}{d-2}\,.\ee
Using the results above-obtained, we find that Eqs.\rf{2sec150030} amount to
$\rwt_{1,1} = - 2$.

\noindent {\bf iv)}  Finally, using the results above-obtained, we check that
Eqs.\rf{2E2kap1eq04} are satisfied automatically.

\appendix{ Counting of on-shell D.o.F for spin-$\frac{3}{2}$ conformal field }

We analyze on-shell D.o.F for spin-$\frac{3}{2}$ conformal field that is
described by the standard higher-derivative Lagrangian given in
\rf{3-2counDoFt03}. Lagrangian \rf{3-2counDoFt03} leads to the equations of
motion given by
\be\label{3-2counDoFt08}
\Box^k P_{3/2}^{ab} \parline \psi^b  = 0\,.
\ee
We analyze equations of motion \rf{3-2counDoFt08} by using light-cone gauge
condition and gamma-transversality condition to fix the respective $\xi$- and
$\lambda$-gauge symmetries \rf{3-2gaugsym01},
\be \label{3-2lcgaug01}
\psi^+  = 0 \,,
\qquad
\gamma^a \psi^a = 0 \,.
\ee
Using \rf{3-2lcgaug01}, we prove that Eqs.\rf{3-2counDoFt08} amount to
the following equations (for details, see below):
\be
\label{3-2counDoFt09} \Box^k \parline \psi^a =  0 \,,
\qquad\qquad
\Box^{k-1} \parline \partial^a\psi^a =0 \,.
\ee
Introducing fields $\psi_{k'}^a$, $k'\in [k]_1$, $\psi_{k'}$, $k'\in
[k-1]_1$, and making use of the identification $\psi_{-k}^a \equiv \psi^a$,
we rewrite Eqs.\rf{3-2counDoFt09} into the ordinary-derivative form,
\beq \label{sysequ001}
&& \parline \psi_{k'}^a  +  \psi_{k'+1}^a = 0\,,
\quad\qquad k' \in [k]_1;
\\
\label{sysequ002} && \parline \psi_{k'} + \psi_{k'+1} = 0\,,
\quad\qquad \ \ k' \in [k-1]_1;
\\
\label{sysequ003} && \gamma^a\psi_{k'}^a =0 \,,
\quad\qquad \qquad \quad \ \ \ k' \in [k]_2;
\\
\label{sysequ004} && \gamma^a\psi_{k'}^a + \psi_{k'} = 0\,,
\quad\qquad \quad \ k' \in [k-1]_2\,,
\eeq
where we use the conventions $\psi_{k+1}^a \equiv 0$, $\psi_{- k} \equiv 0$,
$\psi_k \equiv 0$. Simple way to derive \rf{sysequ001}-\rf{sysequ004} is to
note the representation for the fields $\psi_{k'}^a$, $\psi_{k'}$ in terms of
the generic field $\psi^a\equiv \psi_{-k}^a$,
\beq
&& \psi_{k'}^a = \Box^{^{\frac{k+k'}{2}}} \psi_{-k}^a\,,
\quad \ \quad \qquad\qquad k' \in [k]_2;
\\
&& \psi_{k'}^a = - \Box^{^{\frac{k-1+k'}{2}}} \parline \psi_{-k}^a\,, \quad \
\ \ \qquad k' = [k-1]_2;
\\
&& \psi_{k'} = 2 \Box^{^{\frac{k-1+k'}{2}}} \partial^a\psi_{-k}^a\,,
\qquad\qquad k' \in [k-1]_2;
\\
&& \psi_{k'} = -2\Box^{^{\frac{k-2+k'}{2}}} \parline
\partial^a\psi_{-k}^a\,, \qquad \ k' \in [k-2]_2.
\eeq
Note that light-cone gauge $\psi^+=0$ and Eqs.\rf{sysequ001} imply light-cone
gauge for vector-spinor fields,
\beq \label{sysequ006}
&& \psi_{k'}^+ = 0\,,
\quad\qquad k'\in [k]_1.
\eeq
Introducing the notation
\be \label{defplhatminhatfields01} \psi_{k'}^{a\plushat } \equiv
\Pi^{\plushat}\psi_{k'}^a\,,\quad \psi_{k'}^{a\minushat } \equiv
\Pi^{\minushat}\psi_{k'}^a\,,\quad \psi_{k'}^{\plushat} \equiv
\Pi^{\plushat}\psi_{k'}\,,\quad \psi_{k'}^{\minushat} \equiv
\Pi^{\minushat}\psi_{k'}\,, \ee \be \Pi^{\plushat} \equiv \half
\gamma^-\gamma^+\,,\qquad \Pi^{\minushat} \equiv \half \gamma^+\gamma^-\,,
\ee
we note that the following fields in \rf{defplhatminhatfields01}
\beq \label{appdofdervec01}
&& \psi_{k'}^{i\plushat }\,,\qquad k' \in [k]_1\,;
\\
\label{appdofdervec02} && \psi_{k'}^{\plushat }\,, \qquad k' \in [k-1]_1,
\eeq
turn out to be dynamical on-shell D.o.F. This to say that using
Eqs.\rf{sysequ001},\rf{sysequ002}, constraints \rf{sysequ003},
\rf{sysequ004}, and light-cone gauge \rf{sysequ006}, we can solve the
remaining fields in \rf{defplhatminhatfields01} in terms of on-shell
dynamical D.o.F. \rf{appdofdervec01},\rf{appdofdervec02},
\beq
\label{lcfresol001} && \psi_{k'}^{i \minushat } =
-\frac{\gamma^+}{2\partial^+} \gamma^j\partial^j \psi_{k'}^{i\plushat} -
\frac{\gamma^+}{2\partial^+} \psi_{k'+1}^{i\plushat}\,,
\\
\label{lcfresol002} && \psi_{k'}^{-\plushat} = -\frac{\partial^j}{\partial^+}
\psi_{k'}^{j \plushat}  + \frac{1}{2\partial^+}  \psi_{k'+1}^{ \plushat}\,,
\\
\label{lcfresol003} && \psi_{k'}^{- \minushat } =
-\frac{\gamma^+}{2\partial^+} \gamma^j\partial^j \psi_{k'}^{-\plushat} -
\frac{\gamma^+}{2\partial^+} \psi_{k'+1}^{-\plushat}\,,
\\
\label{lcfresol004} && \psi_{k'}^{\minushat } = -\frac{\gamma^+}{2\partial^+}
\gamma^j\partial^j \psi_{k'}^{\plushat} - \frac{\gamma^+}{2\partial^+}
\psi_{k'+1}^{\plushat}\,.
\eeq
Relations \rf{lcfresol001}, \rf{lcfresol002}, and \rf{lcfresol004} are
obtained from \rf{sysequ001}, \rf{sysequ002}, while relation \rf{lcfresol003}
is obtained from \rf{lcfresol001} and the following relations:
\beq
\label{lcfresol005} && \hspace{-0.5cm}
\psi_{k'}^{- \plushat } = -\half \gamma^- \gamma^i
\psi_{k'}^{i\minushat}\,,\hspace{3cm} k' \in [k]_2;
\\
\label{lcfresol006} && \hspace{-0.5cm} \gamma^i \psi_{k'}^{i\plushat} =0\,,
\hspace{4.6cm} k' \in [k]_2;
\\
\label{lcfresol007}
&&\hspace{-0.5cm} \psi_{k'}^{- \plushat } = -\half \gamma^-\gamma^i
\psi_{k'}^{i\minushat} - \half \gamma^- \psi_{k'}^{\minushat}\,,  \qquad k'
\in [k-1]_2; \  \ \ \ \
\\
\label{lcfresol008} && \hspace{-0.5cm}
\gamma^i \psi_{k'}^{i\plushat} + \psi_{k'}^{\plushat} =0\,, \quad \qquad
\quad \qquad \qquad k' \in [k-1]_2\,,
\eeq
which are obtainable from algebraic constraints
\rf{sysequ003},\rf{sysequ004}.

Thus, we get on-shell D.o.F \rf{appdofdervec01},\rf{appdofdervec02} subject
to constraints \rf{lcfresol006},\rf{lcfresol008} and the constraints
$\Pi^{\minushat}\psi_{k'}^{i\plushat}=0$,
$\Pi^{\minushat}\psi_{k'}^\plushat=0$. Obviously, such on-shell D.o.F are in
one-to-one correspondence with on-shell D.o.F
\rf{ferspin2DoF01}-\rf{ferspin2DoF04} subject to the constraints $\gamma^i
\psi_{k'}^i=0$, $\Pi^{\minushat}\psi_{k'}^i=0$, $\Pi^{\minushat}\psi_{k'}=0$.

We finish with the derivation of Eqs.\rf{3-2counDoFt09}. Using gauge
\rf{3-2lcgaug01}, we obtain the relation
\be \label{manold-08112011-02} (1-d)\Box^k P_{3/2}^{+b} \psi^b =
\Box^{k-1}\left( (d-2)
\partial^+ \parline \partial^b\psi^b + \gamma^+ \Box \partial^b\psi^b\right)
\,.
\ee
Use of relation \rf{manold-08112011-02} and Eq.\rf{3-2counDoFt08} leads to
the equation
\be \label{3-2counDoFt20} \Box^{k-1}\left( (d-2) \partial^+ \parline
\partial^b\psi^b + \gamma^+ \Box \partial^b\psi^b\right) = 0  \,. \ee
Multiplying this equation by $\gamma^+$ gives
\be \label{3-2counDoFt21} \Box^{k-1} \gamma^+  \partial^+ \parline
\partial^b\psi^b =0 \,. \ee
Eq.\rf{3-2counDoFt21} implies the relations
\be \label{3-2counDoFt22} \Box^{k-1} \gamma^+ \Box \partial^b\psi^b =
\Box^{k-1} \gamma^+
\parline \parline \partial^b\psi^b =   \Box^{k-1} 2 \partial^+  \parline
\partial^b\psi^b\,. \ee
Plugging \rf{3-2counDoFt22} into \rf{3-2counDoFt20} and assuming that kernel
of the operator $\partial^+$ is trivial, we obtain the second equation in
\rf{3-2counDoFt09}. Finally, using the second equation given in
\rf{3-2counDoFt09} and the gauge condition $\gamma^a\psi^a=0$ in equations
\rf{3-2counDoFt08}, we obtain the first equation in \rf{3-2counDoFt09}.

\appendix{ Spin-$\frac{3}{2}$ field: Derivation of Lagrangian, gauge transformations,
and operator $R^a$ }

We begin with the study of restrictions imposed by gauge symmetries. For the
spin-$\frac{3}{2}$ conformal field, general ordinary-derivative Lagrangian
and gauge transformations are given by
\beq
\label{21072011-21} {\rm i}\LL &  = &  \psibr E \psik\,,
\\
\label{20072011-01} && E  = E_\smone + E_\smzero\,,
\\
&& E_\smone  \equiv
\parline
- \alpha\partial\gamma\bar\alpha
- \gamma\alpha\bar\alpha\partial
+ \gamma\alpha
\parline\gamma\bar\alpha
\,,
\\
&& E_\smzero =  (1-\gamma\alpha\gamma\bar\alpha)e_1^\smGamma + \gamma\alpha
\eb_1 + e_1 \gaalb\,,
\\
\label{21072011-22} \delta\psik & = & (G_\smone + G_\smzero)\xik\,,\qquad
G_\smone \equiv \alpar\,, \qquad G_\smzero \equiv f_1 + \gaal f_2\,,
\\
\label{22072011-01} && e_n^\smGamma = \ewt_{n-}^\smGamma
\bar\upsilon^\ominussm \sigma_- + \upsilon^\ominussm \ewt_{n+}^\smGamma
\sigma_+\,, \qquad n=1,2,
\\
\label{22072011-02} && e_1 =  \zeta  (\ewt_{1-} \pi_- +  \ewt_{1+}\pi_+)
\bar\upsilon^\ominussm \,,
\\
\label{22072011-03} && \eb_1 = \upsilon^\ominussm ( \ebwt_{1-} \pi_- +
\ebwt_{1+}\pi_+) \bar\zeta\,,
\\
\label{22072011-04} && f_1 =  \zeta  ( \fwt_{1-}\pi_- +  \fwt_{1+}\pi_+)
\bar\upsilon^\ominussm \,,
\\
\label{22072011-05} && f_2 = \upsilon^\ominussm ( \fwt_{2-}\pi_- +
\fwt_{2+}\pi_+) \bar\zeta\,,
\\
\label{21072011-23} && \ewt_{1,\pm}^\smGamma =
\ewt_{1,\pm}^\smGamma(N_\zeta,\Delta')\,,\qquad \ewt_{2,\pm}^\smGamma =
\ewt_{2,\pm}^\smGamma(\Delta')\,,
\\
\label{21072011-24} && \ewt_{1,\pm}= \ewt_{1,\pm}(\Delta')\,,\hspace{1.5cm}
\ebwt_{1,\pm} = \ebwt_{1,\pm}(\Delta')\,,
\\
\label{21072011-25} && \fwt_{1,\pm}= \fwt_{1,\pm}(\Delta')\,,\hspace{1.5cm}
\fwt_{2,\pm} = \fwt_{2,\pm}(\Delta')\,,
\\
&& \ewt_{1\pm}^{\smGamma \dagger}=\ewt_{1\mp}^\smGamma\,, \qquad
\ewt_{2\pm}^{\smGamma \dagger}=\ewt_{2\mp}^\smGamma\,,\qquad
\ewt_{1\pm}^\dagger = -\ebwt_{1\pm}\,.
\eeq
We note that:\\
{\bf i)} The Rarita-Schwinger operator $E_\smone$ \rf{20072011-01} is fixed by
requiring the $\phibr E_\smone\phik$ part of the Lagrangian to be invariant
under standard gradient $\alpar\xik$ part of gauge
transformations \rf{21072011-22};\\
{\bf ii)} dependence of operators $E$ \rf{20072011-01}, $G_\smone$,
$G_\smzero$ \rf{21072011-22} on the oscillators in
\rf{22072011-01}-\rf{21072011-25} is fixed by requiring the ket-vectors
$E\psik$, $(G_\smone + G_\smzero)\xik$ to
satisfy constraints in \rf{3-2Nalzet01}-\rf{NzetNups02};\\
{\bf iii)} Quantities $\ewt^\smGamma$, $\ewt$, $\fwt$ in
\rf{21072011-23}-\rf{21072011-25} depend $N_\zeta$ and $\Delta'$.  Because $\psik$
\rf{3-2psidef0001} is polynomial of degree-1 in $\zeta$, the quantities
$\ewt_{1\pm}^\smGamma$ should be determined only for $N_\zeta=0,1$.

Thus all that is required is to find dependence of the quantities
$\ewt^\smGamma$, $\ewt$, $\fwt$ on $N_\zeta$ and $\Delta'$. To this end we
study restrictions imposed by the gauge symmetries. Variation of Lagrangian
\rf{21072011-21} under gauge transformations \rf{21072011-22} takes the form
\beq
\delta \LL  & = & {-\rm i}
\psibr ( \VV_1 + \VV_0 )\xik  +  h.c. \,,
\\
&& \VV_1  = \parline X_{\smone,1} + \alpha\partial X_{\smone,3}
+ \gamma\alpha\parline X_{\smone,4}\,,
\nonumber\\
&& \VV_0 = X_{\smzero,1}  + \gamma\alpha X_{\smzero,2}\,,
\\
&& \qquad \ \ X_{\smone, 1} = f_1 + e_1\,,
\nonumber\\
&& \qquad \ \ X_{\smone ,3} = -(d-2)f_2 + e_1^\smGamma\,,
\qquad\quad
X_{\smone,4} = (d-2)f_2 +  e_2^\smGamma\,,
\nonumber\\
&& \qquad \ \ X_{\smzero,1}  =  e_1^\smGamma f_1 + d e_1 f_2\,,
\hspace{1.9cm}
X_{\smzero,2}  =  e_1^\smGamma f_2 +  d e_2^\smGamma f_2 + \eb_1 f_1\,.\qquad
\eeq
Requiring the variation to vanish amounts to the equations $ \VV_\smone\xik =
0$, $\VV_\smzero\xik = 0$. The equation $\VV_\smone\xik = 0$ amounts to the
relations
\beq
\label{15072011-01} && f_1  = -  e_1\,, \qquad
f_2  = \frac{1}{d-2} e_1^\smGamma\,, \qquad
f_2 = - \frac{1}{d-2} e_2^\smGamma\,,
\eeq
which should be considered on space of $\xik$ \rf{oldman-31012011-02}. Using
\rf{15072011-01}, we find that the equation $\VV_\smzero\xik = 0$ amounts to
the relations
\beq \label{23072011-01}
&& - e_1^\smGamma e_1 + \frac{d}{d-2} e_1 e_1^\smGamma = 0 \,, \qquad\quad
\eb_1 e_1 + \frac{d-1}{d-2} e_1^\smGamma e_1^\smGamma = 0\,,
\eeq
which should also be considered on space of $\xik$. Equations
\rf{15072011-01},\rf{23072011-01} summarize restrictions imposed by the gauge
symmetries. It is seen, that these equations alone are not enough to
determine the quantities $\ewt^\smGamma$, $\ewt$, $\fwt$ uniquely. In other
words, restrictions imposed by the gauge symmetries alone are not enough to
determine the Lagrangian and gauge transformations uniquely. Therefore we
proceed with analysis of restrictions imposed by conformal boost symmetries.

{\bf Analysis of restrictions imposed by conformal boost symmetries}.
Requiring Lagrangian \rf{21072011-21} to be invariant (up to total
derivative) under conformal boost transformations given in \rf{20072011-03},
\rf{conalggenlis04},  we find the equations
\beq
\label{3-2E2kap1eq02nn01} && \hspace{-1.8cm} (E R^a + \hcwh) + E_\smzero^a
\approx  0\,,
\\
E_\smzero^a & \equiv & \Delta' (\gamma^a + \gaal \gamma^a \gaalb)
 -   \alpha^a (\Delta' + \frac{d-2}{2}) \gaalb - \gaal ( \Delta' -
\frac{d-2}{2}) \bar\alpha^a\,.
\eeq
In \rf{3-2E2kap1eq02nn01} and below, relations like $A+ \hcwh$
stand for $A - \gamma^0 A^\dagger \gamma^0$, where the hermitian
conjugation rules are given in \rf{03082011-01}. Also, in
\rf{3-2E2kap1eq02nn01} and below, we simplify our formulas as follows. Let
$A$ be some operator. We use the relation $A \approx 0$ in place of
$A\psik=0$, where $\psik$ is defined in \rf{3-2psidef0001}.

The operator $R^a$ turns out to be degree-1 polynomial in the derivative.
Using power series expansion of the operator $E$ given in \rf{20072011-01}
and power series expansion of the operator $R^a$ given by
\be R^a = R_\smzero^a + R_\smone^a\,, \ee
where $R_\smn^a$ stands for the contribution involving $n$ derivatives, we
represent Eqs.\rf{3-2E2kap1eq02nn01} as
\beq
\label{3-2E2kap1eq02}
&& E_\smone R_\smone^a + \hcwh \approx 0 \,,
\\
\label{3-2E2kap1eq03}
&& E_\smone R_\smzero^a + E_\smzero R_\smone^a + \hcwh \approx 0 \,,
\\
\label{3-2E2kap1eq04}
&& ( E_\smzero R_\smzero^a + \hcwh )  + E_\smzero^a \approx 0 \,.
\eeq
Most general form of the operator $R^a$ that respects constraints
\rf{3-2Nalzet01}-\rf{NzetNups02} is given by
\beq
\label{16072011-04} R^a & = & r_\smzero^a + r_\smone^a + R_\smG^a +
R_\smE^a\,,
\\
r_\smzero^a & = & r_{0,1}^\smGamma \gamma^a  + r_{0,2}^\smGamma
\alpha^a\gaalb + r_{0,3}^\smGamma \gaal \bar\alpha^a
\nonumber\\
& +  & r_{0,4}^\smGamma \gaal \gamma^a \gaalb  + r_{0,5}^\smGamma \gaal
\gamma^a + r_{0,6}^\smGamma \gamma^a \gaalb
\nonumber\\
& + &  \rb_{0,1} \alpha^a + r_{0,1} \bar\alpha^a\,,
\\
r_\smone^a &= & r_{1,1}\partial^a  + r_{1,1}^\smGamma \gaal
\partial^a \gaalb  + r_{1,2}^\smGamma \alpha^a\parline \gaalb  +
r_{1,3}^\smGamma \gaal \gamma^a \parline \gaalb
\nonumber\\
& + & r_{1,4}^\smGamma \gaal \partial^a +  r_{1,5}^\smGamma
\partial^a \gaalb + r_{1,6}^\smGamma \gamma^a \parline \gaalb\,,
\nonumber\\
\label{kappasmE001def} R_\smG^a &  = & G r_\smG^a\,,
\qquad
R_\smE^a = r_\smE^a E\,,
\\
r_\smG^a  & =  &   r_{\smG,1} \bar\alpha^a  + r_{\smG,2} \gamma^a  +
r_{\smG,3} \gamma^a \gaalb \,,
\\
r_\smE^a & = &  r_{\smE,1} \gamma^a + r_{\smE,2} \alpha^a \gaalb + r_{\smE,3}
\gaal\bar\alpha^a + r_{\smE,4} \gaal \gamma^a \gaalb
\nonumber\\
& + &  r_{\smE,5} \alpha^a +  r_{\smE,6}\bar\alpha^a
+  r_{\smE,7} \gaal \gamma^a  + r_{\smE,8}\gamma^a \gaalb
+   r_{\smE,9} \gamma^a \,,
\eeq

\vspace{-0.5cm} \beq
&& r_{0,n}^\smGamma =  \rwt_{0,n-}^\smGamma \bar\upsilon^\oplussm \sigma_- +
\upsilon^\oplussm \rwt_{0,n+}^\smGamma  \sigma_+\,,\qquad n=1,2,3,4;
\\
&& r_{0,5}^\smGamma =  \upsilon^\oplussm (\rwt_{0,5-}^\smGamma  \pi_- +
\rwt_{0,5+}^\smGamma  \pi_+)\bar\zeta\,,
\\
&& r_{0,6}^\smGamma =  \zeta (\rwt_{0,6-}^\smGamma  \pi_- +
\rwt_{0,6+}^\smGamma \pi_+)\bar\upsilon^\oplussm\,,
\\
&& r_{0,1} = \zeta(\rwt_{0,1-}\pi_- +
\rwt_{0,1+}\pi_+)\bar\upsilon^\oplussm\,,
\\
&& \rb_{0,1} = \upsilon^\oplussm (\rbwt_{0,1-}\pi_- +
\rbwt_{0,1+}\pi_+)\bar\zeta\,,
\\
&& r_{1,1} = \upsilon^\oplussm (\rwt_{1,1-}\pi_- +
\rwt_{1,1+}\pi_+)\bar\upsilon^\oplussm\,,
\\
&& r_{1,n}^\smGamma = \upsilon^\oplussm (\rwt_{1,n-}^\smGamma\pi_- +
\rwt_{1,n+}^\smGamma\pi_+)\bar\upsilon^\oplussm\,,\qquad n=1,2,3,
\\
&& r_{1,4}^\smGamma = (\upsilon^\oplussm \rwt_{1,4-} \bar\upsilon^\oplussm
\sigma_- + \upsilon^\oplussm \upsilon^\oplussm \rwt_{1,4+}\sigma_+)\bar\zeta
\,,
\\
&&  r_{1,n}^\smGamma = \zeta (\rwt_{1,n-}^\smGamma \bar\upsilon^\oplussm
\bar\upsilon^\oplussm \sigma_- + \upsilon^\oplussm \rwt_{1,n+}^\smGamma
\bar\upsilon^\oplussm \sigma_+) \,, \hspace{2cm} n=5,6\,;
\eeq

\vspace{-0.5cm}
\beq \label{21072011-26}
&& \rwt_{0,1\pm}^\smGamma = \rwt_{0,1\pm}^\smGamma(N_\zeta,\Delta')\,, \qquad
\rwt_{0,n\pm}^\smGamma = \rwt_{0,n\pm}^\smGamma(\Delta')\,,\qquad
n=2,3,\ldots ,6\,,
\\
\label{21072011-27} && \rwt_{0,1\pm} = \rwt_{0,1\pm}(\Delta')\,,
\hspace{1.5cm} \rbwt_{0,1\pm} = \rbwt_{0,1\pm}^\smGamma(\Delta')\,,
\\
\label{21072011-28} && \rwt_{1,1\pm} = \rwt_{1,1\pm}(N_\zeta,\Delta')\,, \qquad
\rwt_{1,n\pm}^\smGamma = \rwt_{1,n\pm}^\smGamma(\Delta')\,,\qquad n=1,\ldots
,6\,.
\eeq
From these relations, we see that in order to fix $R^a$ we have to find
quantities $\rwt$ in \rf{21072011-26}-\rf{21072011-28}. Because of the
relation $E Gr_\smG^a\psik = 0$, we note that the $Gr_\smG^a$ contribution to
$R^a$ satisfies equations \rf{3-2E2kap1eq02nn01} automatically without any
restrictions on $r_{\smG,n}$. Therefore in what follows we ignore all
$r_{\smG,n}$ contributions in our analysis of
Eqs.\rf{3-2E2kap1eq02}-\rf{3-2E2kap1eq04}. We now analyze
Eqs.\rf{3-2E2kap1eq02}-\rf{3-2E2kap1eq04} in turn.

\noindent {\bf i)} Using Eqs.\rf{3-2E2kap1eq02}, we find the following
relations:
\beq
 \label{kEsool001}  && r_{1,1}^\dagger = r_{1,1}\,,
\\
&& r_{1,n}^\smGamma =  0\,, \qquad n=1,\ldots, 6\,,
\\
\label{kEsool004} && r_{\smE,n}^\dagger = r_{\smE,n}\,,\qquad \qquad
n=1,4,9\,;
\\
\label{kEsool005} && r_{\smE,2}^\dagger = r_{\smE,3}\,,\qquad
r_{\smE,5}^\dagger = - r_{\smE,6}\,,\qquad r_{\smE,7}^\dagger =-
r_{\smE,8}\,.
\eeq
Note that operator $R_\smE^a$ \rf{kappasmE001def}, with $r_{\smE,n}$ subject
to constraints \rf{kEsool004},\rf{kEsool005}, satisfies the equations
\be \label{manold-08112011-01} E R_\smE^a + \hcwh \approx 0\,. \ee
Comparing \rf{manold-08112011-01} and \rf{3-2E2kap1eq02nn01}, we see that the
operator $R_\smE^a$ does not contribute to Eqs.\rf{3-2E2kap1eq03},
\rf{3-2E2kap1eq04}. This implies that Eqs.\rf{3-2E2kap1eq03},\rf{3-2E2kap1eq04}
do not impose additional restrictions on the $R_\smE^a$ and we can
ignore the operator $R_\smE^a$ in analysis of
Eqs.\rf{3-2E2kap1eq03},\rf{3-2E2kap1eq04}. Therefore in the subsequent
analysis of Eqs.\rf{3-2E2kap1eq03},\rf{3-2E2kap1eq04} we replace
$r_\smzero^a$ for $R_\smzero^a$ and $r_\smone^a$ for $R_\smone^a$.

\noindent {\bf ii)} We now analyze Eqs.\rf{3-2E2kap1eq03}. From these
equations, we find the following relations:
\beq
&& r_{0,2}^\smGamma =0\,, \quad r_{0,4}^\smGamma =0\,,\quad r_{0,6}^\smGamma
=0 \,,
\\
&& r_{0,1}^{\smGamma \dagger} =  - r_{0,1}^\smGamma  \,,
\\
&& r_{0,3}^\smGamma = - \frac{2}{d-2} r_{0,1}^\smGamma \,,
\qquad r_{0,5}^\smGamma = - \frac{1}{d-2} \rb_{0,1} \,,
\\
&&  r_{0,1} = - \rb_{0,1}^\dagger\,,
\\
\label{3-2k01k11001} &&  r_{0,1}^\smGamma = \half [r_{1,1},e_1^\smGamma]\,,
\qquad
\rb_{0,1} =   [\eb_1,r_{1,1}]\,.
\eeq
These relations lead to the following representation for the operator $R^a$:
\beq
\label{3-2k01k11003} r_\smzero^a & = & r_{0,1}^\smGamma (\gamma^a
-\frac{2}{d-2} \gaal \bar\alpha^a)
+  \rb_{0,1} (\alpha^a - \frac{1}{d-2}\gaal \gamma^a) + r_{0,1}
\bar\alpha^a\,,
\\
\label{3-2k01k11004} r_\smone^a &= & r_{1,1}\partial^a\,.
\eeq
\noindent{\bf iii)} Using \rf{3-2k01k11003}, we find that
Eqs.\rf{3-2E2kap1eq04} amount to the following equations:%
\footnote{ In \rf{3-2k01k11005} and below, we use shortcut
$A|_{N_\zeta=0,1}=0$ for the respective two equations $A|\psi_1\rangle=0$,
$A|\psi_0\rangle=0$, where ket-vectors $|\psi_1\rangle$ and $|\psi_0\rangle$
are defined in \rf{3-2psidef0001}-\rf{psi10def04}.}
\beq
\label{3-2k01k11005} && \{ e_1^\smGamma,r_{0,1}^\smGamma\} + \frac{2}{d-2}
(-e_1 \rb_{0,1} + r_{0,1} \eb_1) + \Delta'\Bigr|_{N_\zeta=0,1} =0 \,,
\\
\label{3-2k01k11006} && \{ e_1^\smGamma,r_{0,1}^\smGamma\} +
\Delta'\Bigr|_{N_\zeta=0} =0 \,,
\\
\label{3-2k01k11007} && \frac{2}{d-2} e_1^\smGamma r_{0,1}^\smGamma + \eb_1
r_{0,1} -  \Delta' + \frac{d-2}{2}\Bigr|_{N_\zeta=0} =0 \,,
\\
\label{3-2k01k11008} && [e_1^\smGamma, \rb_{0,1}] - \frac{4}{d-2}
r_{0,1}^\smGamma \eb_1\Bigr|_{N_\zeta = 0}= 0 \,,
\\
\label{3-2k01k11009} && [\eb_1, r_{0,1}^\smGamma] +  \frac{1}{d-2}
e_1^\smGamma \rb_{0,1}\Bigr|_{N_\zeta = 0 }= 0 \,.
\eeq
Plugging \rf{3-2k01k11001} into \rf{3-2k01k11005}, we get
\beq \label{3-2k01k11010}
[r_{1,1},\half e_1^\smGamma e_1^\smGamma + \frac{2}{d-2}  e_1 \eb_1 ] +
\Delta'\Bigr|_{N_\zeta =0,1} =0 \,.
\eeq
Using Eq.\rf{3-2k01k11010} when $N_\zeta=0$, we make sure that
Eq.\rf{3-2k01k11010} implies $\ewt_{1\pm}^\smGamma(0,\Delta') \ne 0$. Taking
this into account and using appropriate field redefinitions, we find the
solution
\be \label{3-2k01k11012}
\ewt_{1\pm}^\smGamma(0,\Delta') =1\,. \ee
Using \rf{3-2k01k11012} in \rf{3-2k01k11010} when $N_\zeta = 0$, we find
\be  \label{3-2k01k11013}
\rwt_{1,1\pm}(0,\Delta') =-2\,.
\ee
From these relations and restrictions imposed by gauge symmetries
\rf{23072011-01}, we obtain
\beq
\label{3-2k01k11014} && \ewt_{1\pm} = u_\f\,,\qquad \ebwt_{1\pm} = - u_\f\,,
\qquad
\ewt_{1\pm}^\smGamma(1,\Delta') = \frac{d}{d-2}\,,
\eeq
where $u_\f$ is given in \rf{oldman-26012011-03}. Note that, from the 2nd
equation in \rf{23072011-01}, the quantities $\ewt_{1\pm}$ are determined up
to phase. Using field redefinitions, the phase can be set equal to zero.

Using Eq.\rf{3-2k01k11010} when $N_\zeta=1$, we find
$\rwt_{1,1\pm}(1,\Delta') = -2$. We make sure that
Eqs.\rf{3-2k01k11006}-\rf{3-2k01k11009} are satisfied automatically. To
summarize, we finished analysis of Eqs.\rf{3-2E2kap1eq03}.

\noindent {\bf iii)} Finally, we analyze Eqs.\rf{3-2E2kap1eq04}. Using the
results above-obtained, we check that these equations are satisfied
automatically.

\end{document}